\documentclass[apj]{emulateapj}
\usepackage{hyperref,graphicx,amssymb}
\usepackage{subfigure}
\usepackage{natbib,url,xcolor}
\usepackage{enumerate,appendix}

\long\def\symbolfootnote[#1]#2{\begingroup%
\def\thefootnote{\fnsymbol{footnote}}\footnote[#1]{#2}\endgroup}

\hypersetup{colorlinks=true, urlcolor=blue}

\definecolor{dark-green}{rgb}{0.1,0.49,0.4}
\hypersetup{colorlinks=true, urlcolor=blue, citecolor=dark-green}

\begin{document}
\shorttitle{\textit{Swift} black hole accretion survey}
\shortauthors{Reynolds et al.}

\title
{A \textit{Swift} survey of accretion onto stellar-mass black holes}
\author{Mark T. Reynolds\altaffilmark{1}, Jon M. Miller\altaffilmark{1}}  
\email{markrey@umich.edu}

\altaffiltext{1}{Department of Astronomy, University of Michigan, 500 Church
  Street, Ann Arbor, MI 48109} 

\begin{abstract}
We present a systemic analysis of all of the stellar mass black hole binaries
(confirmed \& candidate) observed by the \textit{Swift} observatory up to June 2010.
The broad \textit{Swift} bandpass enables a trace of disk evolution over an
unprecedented range in flux and temperature. The final data sample consists of 476
X-ray spectra containing greater than 100 counts, in the 0.6 -- 10 keV band. This is
the largest sample of high quality CCD spectra of accreting black holes published to
date. In addition, strictly simultaneous data at optical/UV wavelengths are
available for 255 (54\%) of these observations.
The data are modelled with a combination of an accretion disk and a hard spectral
component. For the hard component we consider both a simple power-law and a thermal
Comptonization model. An accretion disk is detected at greater than the 5$\sigma$
confidence level in 61\% of the observations.  Lightcurves and color-color diagrams
are constructed for each system. Hardness luminosity and disk fraction luminosity
diagrams are constructed and are observed to be consistent with those typically
observed by \textit{RXTE}, noting the sensitivity below 2 keV provided by
\textit{Swift}. The observed spectra have an average luminosity of $\sim$ 1\%
Eddington, though we are sensitive to accretion disks down to a luminosity of $\rm
10^{-3}~L_{Edd}$. Thus, this is also the largest sample of such cool accretion disks
studied to date.
The accretion disk temperature distribution displays two peaks consistent with the
classical hard and soft spectral states, with a smaller number of disks distributed
between these.  The distribution of inner disk radii is observed to be continuous
regardless of which model is used to fit the hard continua.  There is no evidence for
large scale truncation of the accretion disk in the hard state ($\rm at~least~for~L_x
\gtrsim 10^{-3}~L_{Edd}$), with all of the accretion disks having radii $\rm \lesssim
40~R_g$. Plots of the accretion disk inner radius versus hardness ratio reveal the
disk radius to be decreasing as the spectrum hardens, i.e., enters the hard
state. This is in contrast to expectations from the standard disk truncation paradigm
and points towards a contribution from spectral hardening.
The availability of simultaneous X-ray and optical/UV data for a subset of
observations, facilitates a critical examination of the role of disk irradiation via a
modified disk model with a variable emissivity profile (i.e., $\rm T(r) \propto
r^{-p}$). The broadband spectra (X-ray -- Opt/UV) reveal irradiation of the accretion
disk to be an important effect at all luminosities sampled herein, i.e., $\rm p
\lesssim 0.75~for~luminosities \gtrsim 10^{-3}~L_{Edd}$. The accretion disk is found
to dominate the UV emission irrespective of the assumed hard spectral
component. Overall, we find the broadband soft state spectra to be consistent with an
irradiated accretion disk plus a corona, but we are unable to make conclusive
statements regarding the nature of the hard state accretion flow (e.g., ADAF/corona vs
jet).
Finally, the \textit{Swift} data reveal a relation between the flux emitted by the
accretion disk and that emitted by the corona, for this sample of stellar mass black
holes, that is found to be in broad agreement with the observed disk -- corona
relationship in Seyfert galaxies, suggesting a scale invariant coupling between the
accretion disk and the corona.
\end{abstract}
 

\maketitle
\section{Introduction}
Accretion onto a black hole is a fundamental astrophysical process capable of
producing copious amounts of energy from radio -- $\gamma$-ray frequencies. This
radiation in turn will have a profound impact on its surroundings. In the local
Universe this is illustrated by the discovery of large jet blown cavities around
stellar mass black holes \citep{gallo04,pakull10}. When integrated over the lifetime
of the black hole, this energy will have a significant impact on its environment,
e.g., feedback from supermassive black holes (SMBH) in the quiescent state ($\rm
\lesssim 10^{-6}~L_{Edd}$)\footnote{$\rm L_{Edd} = 1.3 \times
  10^{39}~(M_x/10~M_{\sun})~erg~s^{-1}$} is required to reproduce the observed
properties of galaxies \citep{croton06}.  The observation of tight correlations
between the SMBH and properties of the host galaxy are further evidence for the
importance of feedback from the black hole, and understanding this will provide
insight into the process of Galaxy formation and evolution
\citep{ferraresemerritt00,gebhardt00,gultekin09}. At higher redshifts still, the
accretion power of early HMXBs could have an impact on the formation of the earliest
stars and galaxies \citep{mirabel11}.

There are over 500 accreting binaries currently known in the Milky
Way, though this sample is dominated by systems containing neutron star accretors.
Observations of other galaxies have also revealed large populations of accreting
objects \citep{fabbiano06}, which can be used to reveal information on the star
formation history \citep{grimm03} and stellar content of these galaxies
\citep{gilfanov04}.  The X-ray luminosity function of the Milky Way can be used to
constrain the population and evolution of X-ray emitting sources in the Galaxy
\citep{grimm02}, which is found to be dominated by accreting binary systems with
compact object accretors (BH, NS \& WD, \citealt{revnivtsev11a,revnivtsev11b}).

Detailed study of the accretion flow in Galactic X-ray binaries can provide
constraints on the nature of accretion in the presence of a large gravitational
potential, and the theory of general relativity itself \citep{psaltis08}. In
particular, they provide a nearby window on black hole spin
\citep{miller07,mcclintock11}, the production mechanism of the ubiquitous relativistic
jet emission \citep{markoff01,markoff05,fender06}, and even the nature of supernovae
(SNe) and gamma-ray bursts (GRBs; \citealt{miller11}).  To date, there are 20
confirmed stellar mass black holes (i.e., with dynamical constraints on the mass of
the BH) and greater than 20 strong candidate systems, e.g., see
\citet{mcclintockremillard06}.

BH X-ray binaries (XRBs) are typically transient systems that are observed to brighten
from $\rm \lesssim 10^{-6}~L_{Edd}~- L_{Edd}$ on timescales of weeks to months before
returning to a low luminosity state.  The Galactic black hole binaries have been
classified based on their behavior in the soft X-ray band (0.1 -- 10 keV). This
resulted in the definition of a number of apparently discrete spectral states
\citep{tanakashibazaki96,mcclintockremillard06}. When a black hole is accreting at an
appreciable percentage of its Eddington luminosity ($\rm L_x \gtrsim
10^{-3}~L_{Edd}$), 2 primary spectral forms are observed.  The first of these is the
low-hard state, characterized by a power-law continuum in the 2 -- 10 keV band with a
spectral index $\rm \Gamma \lesssim 1.8$, and a spectral cut-off at energies $\sim$
100 keV. The RMS spectral variability is observed to be high in this state. The second
state is the so-called high-soft state. The X-ray spectrum is dominated by emission in
the soft band ($\rm \lesssim 2~keV$) consistent with a thermal blackbody component, a
softer low luminosity power-law component at higher energies ($\Gamma \gtrsim 2.2$),
and low RMS noise \citep{tanakashibazaki96,grove98}. The launch of the \textit{RXTE}
mission \citep{bradt93} resulted in a refinement in our understanding of these
spectral states, and the transitions between them, which have resulted in the
definition of a number of intermediate spectral states based on both the spectral
state and the timing properties of the source
\citep{homanbelloni05,mcclintockremillard06,vanderklis06}.

The broadband emission from accreting black holes is correlated with the behavior at
X-ray wavelengths, as clearly illustrated by the observed phenomenology of the radio
jet \citep{fender06,gallo10}\footnote{Jets in X-ray binaries are typically observed in
  2 distinct forms: (i) transient ejections occurring close to a state transition, and
  (ii) steady state jets observed in the low-hard state, see \citet{fender06,gallo10}
  for details. Hereafter, when we discuss `jets', we are referring to the steady state
  form unless specifically stated otherwise.}. Studies of jets from accreting black
holes on all mass scales have led to the realization that the accretion disk (inflow)
and the jet (outflow) are closely related, i.e., `fundamental plane of black hole
activity' \citep{merloni03,falcke04} and `disk-jet coupling' \citep{fender04}. In
particular it has been noted that (i) the low-hard state ($\rm L_x \lesssim
0.05~L_{Edd}$ ) is typically associated with the presence of a compact quasi-steady
jet (ii) radio/jet emission appears to cease when the system enters the high-soft
state ($\rm L_x \gtrsim 0.1~L_{Edd}$). This suggests that the accretion disk inflow is
intimately involved in the process of launching a jet, e.g., \citet{fender06,gallo10}.

At higher energies ($\rm > 10~keV$), the large archive of high energy observations by
\textit{RXTE} \citep{bradt93}, observations by \textit{Integral} through the bulge
monitoring program \citep{kuulkers07}, and pointed observations are helping to
elucidate the nature of the high energy X-ray component. \textit{Suzaku} observations
have demonstrated the importance of having broadband spectral coverage, where the
broad bandpass (0.6 -- 300 keV) has proven to be crucial to simultaneously constrain
the soft accretion disk component, the hard X-ray emission and the reflection
features, i.e., Fe K line, Compton hump e.g., see
\citet{makishima08,reynolds10a,reynolds10b} for high energy spectra and
\citet{tomsick09,reis11a,reis11b} for constraints on the reflection component.

\begin{figure}[t]
\begin{center}
\includegraphics[height=0.27\textheight]{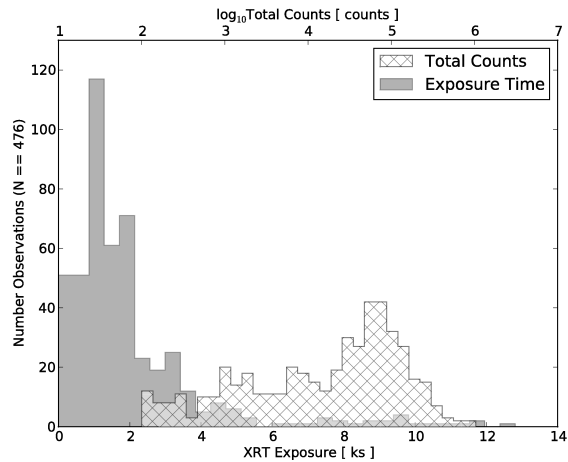}
\caption{Histogram illustrating the observational characteristics of the data
  sample. The filled histogram plots the exposure time for all of the systems, where
  they have been grouped in 0.5 ks bins (see x1 axis). The exposure peaks at
  approximately 1 ks, with $\sim$ 70\% of the observations having an exposure time
  less than 2 ks and 90\% less than 4 ks. The total exposure time is 940 ks. The
  hatched histogram shows the total number of counts for the spectra in our sample
  (see x2 axis). The peak is at $\sim$ 60k counts, while 70\% and 90\% of the spectra
  had greater than 3.5k and 500 total counts respectively. The number of counts in the
  background relative to the source is negligible in all cases.}
\label{histogram_expos}
\end{center}
\end{figure}

\begin{figure*}[t]
\begin{center}
\includegraphics[height=0.40\textheight]{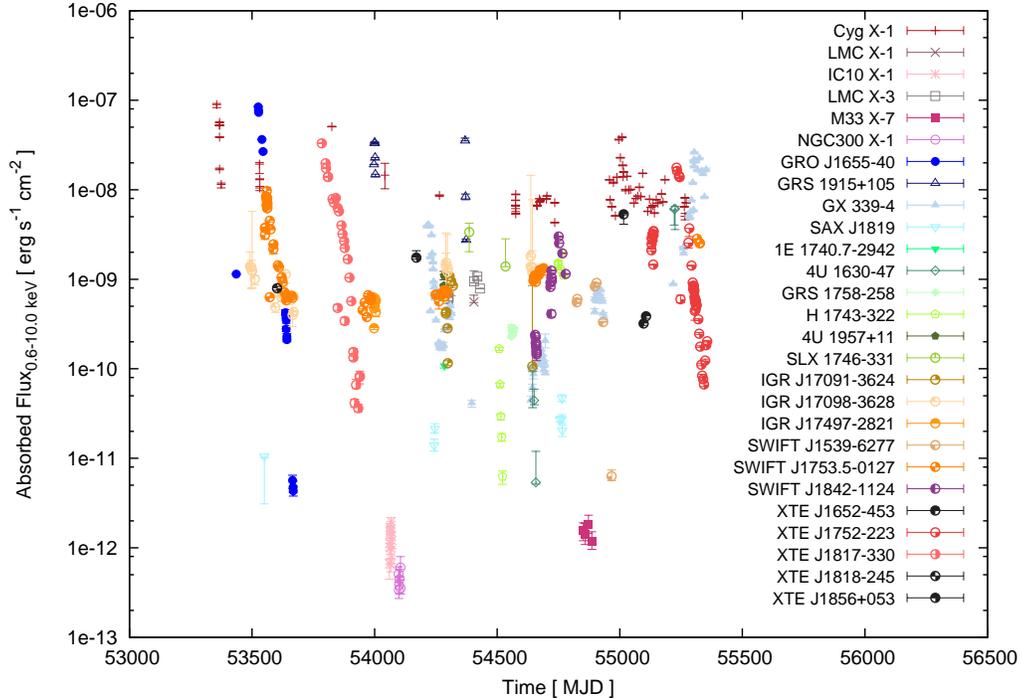}
\caption{Absorbed flux versus time for all of the black holes observed by
  \textit{Swift} up to May 2010, see the symbols indicated on the right of the plot
  and Table \ref{BHC_table}. All fluxes are measured in the 0.6 -- 10 keV band via the
  \texttt{flux} command in \textsc{xspec}. The outbursts of these black hole binaries
  are observed to exhibit a dynamic range of $\rm 10^3 - 10^5$ over the 0.6 -- 10 keV
  energy range.}
\label{BH_all_systems}
\end{center}
\end{figure*}

\subsection{Specific issues}\label{spec_issues}
Understanding the relationship between the accretion disk inflow and the jet outflow
is one of the outstanding goals of modern astrophysics. The anti-correlation between
the accretion state of the black hole and the presence or absence of a relativistic
jet indicates that understanding the geometry of the accretion flow in each of the
primary accretion states will be key, i.e., the low-hard state and the high-soft
state. The current paradigm for the configuration of these accretion states is
consistent with a hot geometrically thin optically thick multi-temperature blackbody
accretion disk \citep{ss73} extending to the innermost stable circular orbit (ISCO) in
the soft state. As the accretion rate and hence disk temperature drop, the inner edge
of the accretion disk recedes from the ISCO and the inner region is filled with a
geometrically thick ADAF (advection dominated accretion flow, \citealt{I17}), which
produces the observed hard spectral component that dominates the low-hard state X-ray
spectrum \citep{mcclintockremillard06,done07}.

However, a number of recent observations contradict this picture and reveal evidence
for the persistent of the accretion disk at the ISCO in the low-hard state, e.g.,
\citet{miller06a,miller06b,rykoff07,tomsick08,wilkinson09,reis09,reynolds10a,reynolds10b,reis10}. These
observations suggest that the accretion disk is not the primary driver of the observed
low-hard $\Leftrightarrow$ high-soft state transition. Indeed, the latest observations
suggest that the disk does not recede from the ISCO until luminosities of $\rm
\lesssim 10^{-3}~L_{Edd}$ \citep{tomsick09}. A number of alternative scenarios are
consistent with the accretion disk remaining at the ISCO, e.g., the hard X-ray flux
may originate in the relativistic jet \citep{markoff05} or from a hot corona
\citep{belo99,diskpn_gierlinski99,merloni02}, and it is the turning on/off of the jet
or evolution of the corona that drives the observed state transitions.

Questions also persist at longer wavelengths. Recent observations have highlighted a
problem in our understanding of the UV emission mechanism in X-ray binaries. Standard
theory explains the UV emission as direct emission from the accretion disk and
observations of the classical $\rm \nu^{1/3}$ spectrum in early IUE observations
supported this picture \citep{cheng92,schrader93}. It was soon realized that
the large flux of X-rays present in an X-ray binary should modify the standard disk
spectrum, with the X-ray emission irradiating the disk, where it is subsequently
reprocessed -- disk reprocessing
(e.g. \citealt{vanparadijs94,kingritter98,hynes98,dubus01,russell06}). 

A standard steady state accretion disk will be observed to exhibit a radial
temperature dependence of the form $\rm T(r) \propto r^{-0.75}$ \citep{ss73}.  It is
inherently difficult to constrain the temperature profile of the accretion disk at
X-ray wavelengths as the primary modification to the accretion disk spectrum occurs at
wavelengths to the red of the typical X-ray detector, i.e., UV/optical or $\rm E
\lesssim 0.1~keV$. This is illustrated in \citet{kubota05} who studied a sample of
\textit{ASCA} observations of black holes in outburst and found the temperature
profile of the measured spectra to be consistent with the standard $\rm T \propto
r^{-0.75}$ accretion disk at all times, though there was evidence, of low statistical
significance, for a deviation towards a more irradiated disk in some of the
observations. In contrast, observations at optical and UV wavelengths have revealed
clear signatures for disk irradiation
\citep{schrader94,vanparadijs94,russell06,rykoff07}. Indeed, the addition of X-ray
irradiation of the accretion disk to the disk instability model is required in order
to reproduce the observed outburst lightcurves of black holes XRBs
\citep{dubus01,lasota01}.

Over the last decade the picture has been further complicated by the realization that
the radio jet may contribute significantly across the entire spectral energy
distribution, e.g., \citet{markoff01,russell06}. Furthermore, simultaneous
observations at optical and X-ray wavelengths have revealed complex correlation
functions, hinting at the presence of a non-thermal component in addition to the
accretion disk emission, e.g., \citet{kanbach01,durant08,gandhi08}.

\citet{dunn10a,dunn10b} has recently carried out a comprehensive study of black hole
X-ray binaries using over 13 years of archival \textit{RXTE} observations. In
\citet{dunn10a}, hardness intensity (HID) and disk fraction luminosity diagrams (DFLD)
were constructed for 25 black holes.  \citet{dunn10b} focused on the behavior of the
accretion disk component in each of the outbursts studied in the previous paper. In
particular, they analyzed the observed distribution of temperature versus luminosity
for each outburst and found it to be consistent with $\rm L_{Disk} \tilde{\propto}
T_{Disk}^4$, and with previous studies, which found the inner radius of the accretion
disk to be constant in disk dominated states. They do note a deviation from this
relation at the entry to/exit from the disk dominated state, which they suggest is
consistent with a varying color correction factor (e.g., \citealt{shimura95}, see
\S\ref{accretion_disk}iii for further details). However, due to the 3 keV lower energy
limit of \textit{RXTE}, this study was necessarily restricted to relatively hot
accretion disks ($\rm T \gtrsim 0.5~keV$).

Despite the significant progress made over recent years, a number of unanswered
questions remain. In this study, we focus our attention on the following issues:
\vspace{-2mm}
\begin{enumerate}[(i)]
\item{Does the accretion disk recede from the ISCO during the soft to hard state
  transition?}
\vspace{-2mm}
\item{What is the nature of the accretion geometry in the low hard state, i.e.,
  Is the accretion disk truncated? What is the source of the observed hard X-ray flux,
  a jet or a corona?}
\vspace{-2mm}
\item{What produces the UV emission, does it originate from the accretion disk
  or is there a non-thermal contribution?}
\end{enumerate} 
\vspace{-2mm}
The \textit{Swift} observatory was designed to discover and provide rapid
multi-wavelength follow-up of gamma ray bursts (GRBs). These same attributes allow
detailed study of many other time variable phenomena, in particular accreting stellar
mass black holes. Due to the rapid pointing capability and low pointing overhead
\textit{Swift} provides the first detailed CCD resolution monitoring of black hole
binaries in outburst. Combined with the strictly simultaneous observations at
optical/UV wavelengths provided by UVOT, \textit{Swift} is the ideal platform
to study the accretion process in detail. In comparison to \textit{RXTE}, the low
energy X-ray cutoff allows us to study the accretion disk across the entire outburst
cycle, whereas previously the cool disk would have been outside the RXTE low energy
cutoff ($\sim$ 3 keV). 

In this paper, we describe observations of black hole binaries undertaken with the
\textit{Swift} observatory prior to June 2010.  In \S2, we describe the observations
and extraction of source spectra and lightcurves. We proceed to analyze the data in
\S3 and the results are presented in \S4. In particular, we find that there is no
evidence for significant truncation of the accretion disk, at least to luminosities
$\rm \gtrsim 10^{-3}~L_{Edd}$. In the broadband data, we find evidence for an
irradiated accretion disk at all luminosities probed, but increasing as we enter the
low-hard state. These results are discussed in the context of models for the accretion
flow in \S5, and finally our conclusions are presented in \S6.

\section{Observations}
Our data sample consists of \textit{Swift} observations of all known and suspected
stellar mass black hole candidates observed through August 2010 (see Table
\ref{BHC_table}). Data for each system was obtained via the \textit{HEASARC} archive
service. 

\subsection{\textit{Swift}}
The \textit{Swift} observatory \citep{gehrels04} is a highly versatile
multi-wavelength platform, providing rapid pointing capabilities and simultaneous
broad wavelength coverage from optical to hard X-ray energies.  The \textit{Swift}
observatory contains three instruments: 

\noindent \textit{(i) Burst alert telescope (BAT, \citealt{barthelmy05}):}
A coded mask hard X-ray ($\rm > 10~keV$) imaging telescope, covering the energy band
from 15 -- 150 keV. Low resolution spectra are also provided at higher fluxes.

\noindent\textit{(ii) X-ray telescope (XRT, \citealt{burrows05}):} A 3.5m focal length
Wolter I type grazing incidence X-ray telescope, containing a 600 pix$^2$ CCD ($\sim$
2.35$\arcsec$ pix$^{-1}$, i.e., 23 arcmin$^2$), sensitive over the energy range 0.2 --
10 keV at the telescope focus, with an effective area of $\sim$ 135, 65, 20 cm$^{-2}$
at 1.5, 6.4 and 8 keV respectively. Bright sources are typically problematic to
observe with CCD based detectors (see \S\ref{pile_up}).  The XRT uses a number of
different read-out modes to avoid/mitigate this problem, facilitating observations of
sources with fluxes in excess of $\sim$ 1 Crab ($\rm \gtrsim 300~ct~s^{-1}~or~f_{2 -
  10~keV} \sim 2.4\times10^{-8}~erg~s^{-1}~cm^{-2}$). These may be divided into
windowed timing (WT) and photon counting modes. In windowed timing mode only spectral
information is provided as the central 8 arc-minutes of the CCD is collapsed into a
single spatial direction, providing 1.8 ms timing resolution. This is the primary mode
used at high fluxes, with the observations presented herein primarily utilizing WTW2
mode. Photon counting mode (PC) is the standard method of reading out a CCD providing
both spectral and 2 dimensional spatial information but with a limited time resolution
of only 2.5 s. The majority of photon counting mode observations herein are in PCW2
mode.

\noindent\textit{(iii) Ultra-violet/optical telescope (UVOT,
  \citealt{roming05,uvot_calI}):} Simultaneous observations are obtained at optical/UV
energies with the UVOT instrument (30 cm Ritchie Chretien telescope). The detector is
a complex system consisting of a photodiode, micro-channel plates and a CCD (see the
instrument paper for details). The CCD provides a 17 arcmin$^2$ FoV and a frame time
of 10.8 ms. Coincidence loss occurs at high count rates, typically those in excess of
20 ct s$^{-1}$, but this is not an issue for the optical/UV counterparts to the black
hole binaries in our sample. Observations are normally obtained in only an single
filter; however, there are some sources with multi-wavelength lightcurves. The systems
in our sample contain data in one or more of the 6 UVOT photometric filters V, B ,U,
W1, M2, W2 ($\rm \lambda_c \approx$ 5468 \AA, 4392 \AA, 3465 \AA, 2600 \AA, 2246 \AA,
1928 \AA).

\begin{table}[t]
\begin{center}
\caption{Exclusion radius as a function of count rate}\label{pile_up_table}
\begin{tabular}{cc|cc}
\tableline\\[-2.0ex]
\multicolumn{2}{c}{Photon Counting} & \multicolumn{2}{c}{Windowed Timing}\\[0.5ex]
\tableline\\[-2.0ex]
\tableline\\[-2.0ex]
0.5 -- 1 ct s$^{-1}$ & 2 pix & 150 -- 200 ct s$^{-1}$ & 3 pix\\[0.5ex] 
1 -- 3 ct s$^{-1}$ & 5 pix & 200 -- 300 ct s$^{-1}$ & 5 pix\\[0.5ex]
3 -- 6 ct s$^{-1}$ & 7.5 pix & 300 -- 500 ct s$^{-1}$ & 10 pix\\[0.5ex]
6 -- 9 ct s$^{-1}$ & 10 pix & $\gtrsim$ 500 ct s$^{-1}$ & 15 pix\\ [0.5ex]
\tableline\\[-2.0ex]
\end{tabular}
\tablecomments{Radius of the inner excluded region as function of count rate for all
  spectra considered herein, see also \citet{romano06,rykoff07}. For comparison, in
  windowed timing mode 200 $\rm ct~s^{-1} \sim 600~mCrab$.}
\end{center}
\end{table}

\subsection{Sample characteristics}\label{sample_characteristics}
Our sample comprises 27 stellar mass black holes. These range from those systems with
dynamical constraints on the mass of the compact object, i.e., $\rm M_x \gtrsim
3~M_{\sun}$ (e.g., Cyg X-1, GRO J1655-40), to a large number of systems that are more
correctly labeled as candidate black holes. These are systems that have displayed
spectral/photometric attributes consistent with those of the known black hole systems,
e.g., soft spectrum, timing properties, lack of pulsations or X-ray bursts, but lack a
dynamical constraint on the mass of the compact object, e.g., 4U 1957+11, Swift
J1753.5-0127.  The results presented herein are necessarily biased by those objects
with the majority of observations, i.e., Cyg X-1, GX 339-4, XTE J1817-330, XTE
J1752-223 and SWIFT J1753.5-0127. However, these object are distributed relatively
uniformly over the relevant binary parameters, e.g., HMXB vs LMXB, $\rm P_{orb}$ etc.,
and hence do not overly bias our results. The known system parameters for each of the
black hole binaries in our sample are listed in Table \ref{BHC_table}.

A typical \textit{Swift} pointing is short with an exposure time less than 2 ks. For
the purposes of this study, we ignore all pointings for which the total number of
counts detected by the XRT is less than 100. Thus our sample consists of the following:
\begin{itemize}
\item{476 observations of 27 BHs with $\geq$ 100 counts in the 0.6 --
10 keV band.}
\item{255 observations for which we have simultaneous detections at optical/UV
  wavelengths, in at least one band (V, B, U, W1, W2, M2).} 
\end{itemize}
The lower number of optical/UV observations reflects the high column density through
which some of these systems are observed, with many systems having $\rm N_H \gtrsim
10^{22}~cm^{-2}$. In Fig. \ref{histogram_expos}, we plot the histogram of exposure
times and total number of detected counts for all of the observations considered
herein. The exposure time distribution peaks at an exposure time of $\sim$ 1 ks, with
$\gtrsim$ 70\% and 90\% of the exposures being less than 2 ks and 4 ks in length
respectively. The total counts distribution peaks at $\sim$ 60k counts, with 70\% and
90\% of the spectra containing greater than 3.5k and 500 total counts respectively.
In Fig. \ref{BH_all_systems}, we plot the absorbed flux vs time for all of the
systems in our sample. We are sensitive to fluxes over a dynamic range of $\rm 10^6$.

\section{Analysis}
We aim to use the data sample described herein (see \S\ref{sample_characteristics}) to
probe the physics of accretion onto black holes. To begin, we outline the 
steps taken to extract the final spectral and photometric data products.

\subsection{Reduction procedure: XRT}
All data reduction and analysis was performed using the \textsc{heasoft 6.9}
suite, which includes \textsc{ftools 6.9, swift 3.5} and \textsc{xspec
  12.6.0k}. The latest versions of the relevant \textit{Swift} \textsc{caldb} files
are also used.

Data were downloaded from the HEASARC archive using the coordinates listed in Table
\ref{BHC_table}, where the search radius was restricted to 10$\arcmin$. All raw-data
were reprocessed using the \texttt{xrtpipeline} to ensure usage of the latest
\textit{Swift} \textsc{caldb} when producing the relevant observation event
files. Spectra were then extracted using \texttt{xselect} from circular and annular
regions in the photon counting mode data, while rectangular regions were used to
extract spectra from the windowed timing mode data. Exposure maps were generated for
each observation using the \texttt{xrtexpomap} task. These are particularly important
due to the presence of a number of dead columns near the center of the XRT CCD due to
a likely micro-meteorite impact in May 2005 \citep{abbey06}.

Background spectra were subsequently extracted from neighbouring regions. Ancillary
response function files (.arf) were created with the \texttt{xrtmkarf} task, while the
appropriate response matrix (.rmf) was sourced from the calibration database. These
files were then grouped together with \texttt{grppha} and exported to \textsc{xspec}
for spectral fitting.

\begin{figure}[t]
\begin{center}
\includegraphics[height=0.27\textheight]{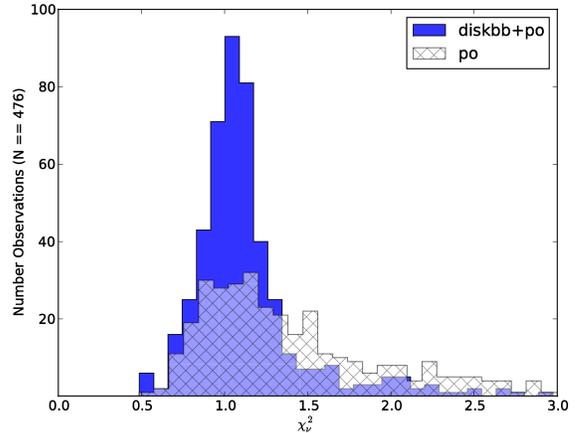}
\caption{Histogram of $\rm \chi^2_{\nu}$ for our best fit \texttt{pha(diskbb+po)}
  model. The dashed histogram is the model containing the power-law component alone,
  while the blue histogram denotes the best fit model after the addition of the
  \texttt{diskbb} component. An accretion disk is required at greater than the
  5$\sigma$ confidence level in $\gtrsim$ 61\% of the observations. Fits with
  alternative continuum components reveal similar results, see Table
  \ref{chi2_table}.}
\label{chi_diskbb_po}
\end{center}
\end{figure}

\subsubsection{pile-up}\label{pile_up}
Due to the photon counting nature of modern X-ray CCDs, they are strongly susceptible
to a phenomenon known as `pile-up'. This occurs at high fluxes, where multiple photons
can impact a detector pixel at approximately the same time (i.e., less than the
detector frame time), and be recorded as a single higher energy event. This will
produce an excess of hard photons in the measured spectrum, which as a result will
appear harder than is actually the case. Proper interpretation of the observed spectra
requires a strategy to mitigate the adverse pile-up related effects.  For further
details see \citet{miller10} for a detailed discussion of pile-up in X-ray CCDs,
including \textit{Chandra, XMM-Newton and Suzaku} in addition to the \textit{Swift}
XRT.

In our analysis, pile-up is corrected in the standard manner, i.e., by the extraction
of annular regions centered on the source position. The excluded region radius was
chosen to be conservative and agrees with those used by other groups, e.g.,
\citet{romano06,rykoff07}\footnote{\citet{done10} claim that such a strategy will not
  successfully retrieve the original spectral form; however, it has been subsequently
  demonstrated that the conclusions of this paper are incorrect. (i) \citet{miller10}
  clearly show how pile-up actually effects the line shape and demonstrate that the
  timing spectrum utilized by \citet{done10} is quite clearly piled-up, e.g., see the
  similarity to Fig. 5 in the Miller paper. (ii) There are also unknown calibration
  issues with the PN timing mode data utilized in this paper, for example,
  \citet{walton12} demonstrate previously unrecognized problems with the SAS tool
  \textsc{epfast}.}. The exclusion regions utilized are listed in Table
\ref{pile_up_table} along with the corresponding count rates. In order to confirm the
effectiveness of this strategy, we ensured that the chosen exclusion radius was such
that the best fit spectral parameters did not exhibit significant variance in
comparison to the next largest exclusion region. The sample we present here is
dominated by observations of black holes in the spectrally hard state, which typically
results in lower observed fluxes in comparison to the soft state. In total,
approximately 20\% of our spectra are effected by pile-up of some form when classified
using the count rates in Table \ref{pile_up_table}. The majority of this occurs at the
lower count rate levels, e.g., only 2.5\% \& 4.5\% of our observations required the
largest and second largest exclusion regions respectively.  The final spectra that we
choose to analyze are those where the quantity of interest is robustly on the
conservative side, for example, \citet{romano06} exclude only the inner pixel for
count rates in the range 100 -- 300 $\rm ct~s^{-1}$ and 2 pixels for 300 -- 400 $\rm
ct~s^{-1}$ and 4 pixels for counts rates $\rm >~400~ct~s^{-1}$.

\subsection{Reduction procedure: UVOT}\label{uvot_reduction}
The optical \& UV photometry required minimal additional processing
\citep{uvot_calI,uvot_calII}.  We begin by utilizing the level II pipeline processed
image file (sk.img). Images are aspect corrected via the \texttt{uvotskycorr} tool,
where the images are registered to the USNO-B1 catalogue \citep{usnob1}. Source flux
is extracted from an aperture of 5$\arcsec$ radius and the background is extracted
from a neighbouring source free sky position. The counterparts to the X-ray sources
are easily identified, primarily due to the transient nature of the optical/UV
counterpart in combination with the accurate X-ray position provided by the XRT. This
also allows us to detect any contamination from possible line of sight stars
contributing to the 5$\arcsec$ source aperture. Source flux and magnitudes are
calculated using \texttt{uvotmaghist} tool. These fluxes are converted to
\textsc{xspec} readable `.pha' files with \texttt{uvot2pha}, which are then exported
for spectral fitting.

\subsection{Spectral models}\label{models}
The are a large number of models available in \textsc{xspec} for modelling the X-ray
emission from black hole binary systems. As outlined in the introduction, the standard
model for black hole X-ray binaries involves the combination of soft emission
($\lesssim$ 2 keV) from a thermal accretion disk and hard X-ray ($\gtrsim$ 2 keV)
emission from power-law like component, typically attributed to Compton scattering of
the thermal photons from the accretion disk to higher energies
\citep{mcclintockremillard06}, or perhaps emission from a jet
\citep{markoff01,markoff05}. In an effort to model the data in a manner that minimizes
bias, we choose two complimentary approaches:

\noindent (i) Phenomenological -- Here we aim to determine the empirical relationships
displayed in the data, which can be used to inform/constrain theoretical models and
characterise source types. For example, is the system spectrally hard or soft? A very
soft X-ray spectrum is one of the characteristic signatures of an accreting black hole
in the soft state. Additionally, we can create color-color and hardness intensity
diagrams to search for characteristic variability and search for correlation at other
frequencies, i.e., optical/UV.

\noindent (ii) Physical -- Using these models we will attempt to constrain the
intrinsic properties of the accretion flow, e.g., accretion disk temperature and inner
radius. For the soft component, we assume a standard steady state, geometrically thin,
optically thick accretion disk \citep{ss73}, described by the \texttt{diskbb} model
\citep{diskbb_mitsuda84,diskbb_makishima86}.  Extensions to this base model are also
considered, e.g., corrections for GR effects in the inner disk region close to the
black hole (\citealt{zhang97}, \citealt{kubota98}, \citealt{makishima00}, or via the
\texttt{diskpn} model \citealt{diskpn_gierlinski99}) and accounting for possible
changes in the temperature profile of the disk due to, for example, irradiation by the
corona or flux generated otherwise close to the black hole (\texttt{diskpbb},
\citealt{mineshige94}). These models have the advantage of simplicity while
incorporating the important physics, and a large archive of published observations
with which the resulted presented herein may be readily compared.  Detailed state of
the art models of accretion disks have been developed, e.g., \texttt{bhspec}
\citep{davis05} and \texttt{kerrbb} \citep{li05}; however, systematic uncertainties
remain (e.g., \citealt{kubota10}) and a detailed investigation of these issues is
beyond the scope of the present work.

The hard X-ray component, i.e., that which typically dominates above $\sim$ 2 keV, is
modelled using 2 distinct prescriptions. The most basic model for the hard X-ray flux
is a simple power-law, characterized by a spectral index, $\Gamma$. While this model
offers the simplest method to characterize the hard X-ray flux, at low energy it does
not exhibit a cut-off as one would expect if the hard X-rays are generated via Compton
scattering of the soft X-ray photons from the accretion disk. This is an important
issue in the current analysis due to the sensitivity of the \textit{Swift}/XRT to
energies as low as 0.5 keV.  In this case, the hard component may be modelled assuming
it originates via scattering of the low temperature seed photons from the accretion
disk by a high temperature corona.  Detailed Comptonization models including
reflection \texttt{compps} \citep{ps96} and non-thermal electron contributions
\texttt{eqpair} \citep{coppi99} are available; however, given the upper energy bound
of the XRT (10 keV), we choose to use the most computationally efficient thermal
Comptonization model (\texttt{comptt}, \citealt{titarchuk94}). In the interest of
completeness, we also consider the \texttt{simpl} Comptonization model of
\citet{steiner09}. This model differs from the \texttt{comptt} model in that it allows
one to specifically choose the input model for the seed photons, whereas the seed
photons are assumed to originate from the Wein tail of a blackbody in the
\texttt{comptt} case.

Where simultaneous UVOT/XRT data are available, joint fits are undertaken in an effort
to constrain the nature of the broadband X-ray to opt/UV spectrum. The extinction in
this case is modelled using the \textsc{xspec} \texttt{phabs} \& \texttt{redden}
models for the X-ray and opt/UV data respectively. The standard Galactic extinction
curve was assumed \citep{cardelli89} along with the gas to dust ratio from
\citet{predehl95}, i.e., the reddening was set to $\rm E(B-V) =
N_H/5.3\times10^{21}~cm^{-2}$. We note that in sightlines towards the Galactic center
the column density is highly variable and as such the standard extinction curve may
not be appropriate for all sources in our sample.

\begin{figure}[t]
\begin{center}
\includegraphics[height=0.27\textheight]{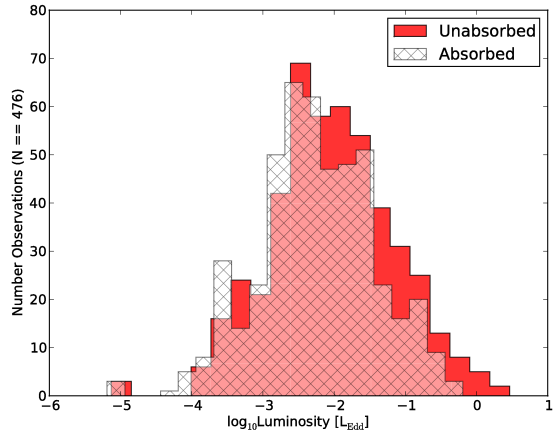}
\caption{Histogram of the observed luminosities for all of the observations in our
  sample, where the distances and masses listed in Table \ref{BHC_table} are
  assumed. The unabsorbed luminosity is consistent with a normal distribution peaking
  at $\sim$ 1\% Eddington.}
\label{histogram_luminosity}
\end{center}
\end{figure}

\subsection{Analysis procedure}
For our baseline model, we utilize a blackbody accretion disk in addition to a
power-law to model the X-ray spectrum (i.e., \texttt{diskbb+po}). This model will also
aid in the comparison of our results to the significant archive of \textit{RXTE}
observations, e.g., \citep{dunn10a}. To model the absorption by intervening neutral
hydrogen, we use \texttt{phabs}, where throughout this paper the abundances and
cross-sections assumed are \texttt{bcmc} \citep{bcmc92} and \texttt{angr}
\citep{angr89} respectively. The column density is held fixed at a value consistent
with previous observations, see Table \ref{BHC_table} for details. This is in
agreement with the observed absence of significant local absorption in these systems
\citep{miller09}, with obvious exceptions for the case of the wind-accreting HMXBs.
 
Spectra with less than 100 total counts were ignored in all further spectral
fitting. Such spectra are typically those obtained with the XRT operating in
auto-mode, i.e., the first observation takes place in windowed timing mode and if the
count rate is safe the observation is stopped and switched to photon counting mode,
sometimes resulting in an under-exposed or piled-up spectrum. The best fit statistic is
chosen based on the total number of counts in the spectrum. Standard chi-squared
fitting is used for observations with more than 500 counts. All spectra with
sufficient counts to meet our $\chi^2$ fitting criteria were binned so as to have a
minimum of 20 counts in each bin via the \texttt{grppha} task thus ensuring that use
of this statistic is valid. The Cash-statistic was utilized for those observations
with less than 500 total counts \citep{cash79}.  Finally, the spectral fitting is
restricted to the energy range 0.6 -- 10 keV.

\begin{figure}[t]
\begin{center}
\includegraphics[height=0.27\textheight]{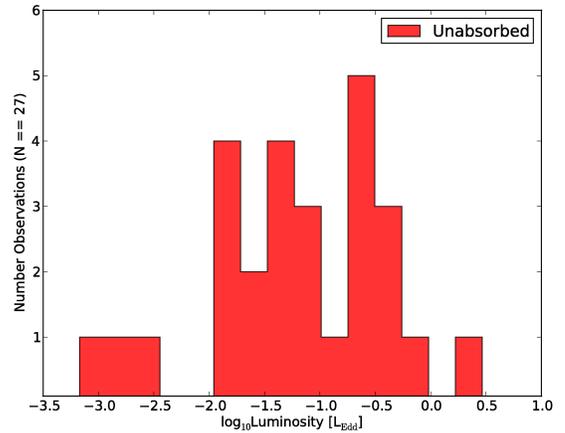}
\caption{Histogram of the peak luminosity observed from each of the systems in our
  sample, where the distances and masses listed in Table \ref{BHC_table} are
  assumed. Although the overall number is small, this distribution appears to peak
  above 10\% Eddington with a significant tail to lower luminosities consistent with
  previous observations \citep{chen97,dunn10a}.}
\label{histogram_peak_luminosity}
\end{center}
\end{figure}

\begin{table}[b]
\begin{center}
\caption{Best fit model comparison}\label{chi2_table}
\begin{tabular}{lcc}
\tableline\\ [-2.0ex]
Model & $\rm \chi^2_{\nu} \leq 1.5$ & $\rm \chi^2_{\nu} \leq 2.0$ \\ [0.5ex]
\tableline\tableline\\ [-2.0ex] 
\texttt{pha*(po)}            & 0.47 & 0.58 \\   [0.5ex]
\texttt{pha*(comptt)}        & 0.53 & 0.68 \\   [0.5ex]
\texttt{pha*(diskbb+comptt)} & 0.83 & 0.90 \\   [0.5ex]
\texttt{pha*(diskbb+po)}     & 0.88 & 0.93 \\   [0.5ex]
\texttt{pha*(diskpn+po)}     & 0.85 & 0.94 \\   [0.5ex]
\tableline
\end{tabular}
\tablecomments{Results of the best fit models for our sample, where a number of
  components are considered in addition to those plotted in
  Fig. \ref{chi_diskbb_po}. The final two columns indicate the percentage of systems
  with a best fit statistic less than indicated.}
\end{center}
\end{table}

\begin{table}[t]
\begin{center}
\caption{lightcurve morphology}\label{lc_morph}
\begin{tabular}{lcccc}
\tableline\\ [-2.0ex]
System & $\rm t_b$ & $\tau_1$ & $\tau_2$ & Comment\\ [0.5ex]
\tableline\tableline\\ [-2.0ex] 
GRO J1655-40         & 115 & 20 & 6  & Largest amplitude\\   [0.5ex]
Swift J1842.5-1124   & -- & 30 & --  & Slow decay\\   [0.5ex]
XTE J1817-330        & 119 & 32 & 5  & Brightens  $\sim$ 60 days\\   [0.5ex]
                     &     &    &    & Const after $\sim$ 130 days\\   [0.5ex]
XTE J1752-223        & -- & 20 & --  & Highly variable\\   [0.5ex]
H1743-322            & -- & -- & 4   & Fast decay \\   [0.5ex]
Swift J1539.2-6277   & 26 & 35 & 7.5 & Slow decay for $\sim$ 20 days\\ [0.5ex]
GX 339-4             & 15 & 35 & 7.5 & Flattens at $\sim$ 30 days\\   [0.5ex]
Cyg X-1              & 5  & 5  & 120 & Const after $\sim$ 5 days\\   [0.5ex]
Swift J1753.5-0127   & 28 & 20 & 40  & Const after $\sim$ 80 days\\   [0.5ex]
GRS 1915+105         & -- & -- & 1   & Very fast decay \\   [0.5ex]
\tableline
\end{tabular}
\tablecomments{Parameters of the decay lightcurves displayed in
  Fig. \ref{lc_plot}. All times are measured in days elapsed from the time of maximum
  flux $\rm t_0$, where $\rm \tau_1$ and $\rm \tau_2$ are the decay timescales before
  and after the break in the lightcurve, occurring at $\rm t_b$ days after maximum
  flux, see Eqn. 1.}
\end{center}
\end{table}

We then proceed to fit each spectrum as follows: (i) the spectrum is loaded in
\textsc{xspec}, the number of counts are determined and if this is greater than 100
the fitting statistic is then chosen. (ii) the initial continuum model is fit to the
data assuming a constant value for the interstellar column density as discussed
previously, e.g., a power-law -- \texttt{pha*po}. Best fit parameters and the
associated errors are determined in addition to the model flux (iii) an accretion disk
component is then added to the model in the previous step, the best fit parameters of
this new model are determined, e.g., \texttt{pha*(diskbb+po)}. (iv) an F-test is
carried out to determine if the new 2 component model is a statistically significant
improvement on the previous single component model, if the second component is
required error scans are undertaken and fluxes are calculated. 

The above fitting procedure was then repeated, replacing the continuum components
above with those indicated earlier, i.e., \texttt{diskpn} or \texttt{diskpbb} in place
of the \texttt{diskbb} and/or \texttt{comptt} to replace the power-law
component. The seed photon temperature for the \texttt{comptt} component is set to
  equal the temperature of the accretion disk, i.e., $\rm T_0 \equiv T_{in}$. A
subset of these fits were also repeated with the column density as a free parameter to
check the sensitivity of our fitting methodology to this parameter. As the XRT
bandpass is limited to energies below 10 keV, we are insensitive to the presence of a
spectral cut-off at higher energies. As such, we are unable to simultaneously
constrain both the electron temperature ($\rm kT_e$) and optical depth ($\tau$) of the
\texttt{comptt} model. We therefore choose to freeze the electron temperature to 50
keV in all fits using the \texttt{comptt} component unless otherwise indicated. This
will result in the production of a power-law like spectrum in the 2 -- 10 keV
range. Additional fits were repeated at other electron temperatures to confirm the
robustness of the fits to changes in this parameter. In the case of the opt/UV + X-ray
fits, the basic procedure was similar. The X-ray and UVOT data are loaded into
\textsc{xspec} and the model is initialized at the relevant best fit model from the
X-ray fits alone. The extinction at optical and ultra-violet wavelengths is modelled
using the \texttt{redden} component, i.e, \texttt{pha(po)+redden(po)}, and we relate
this to the X-ray column density following \citet{predehl95}, i.e., $\rm E(B-V) =
N_H/5.3\times10^{21}~cm^{-2}$.

All of the above models were defined and fit using custom \texttt{tcl} scripts. Errors
where quoted are calculated via the \texttt{error} command and are equivalent to the
90\% confidence interval unless otherwise explicitly stated.

\begin{figure}[t]
\begin{center}
\includegraphics[height=0.25\textheight]{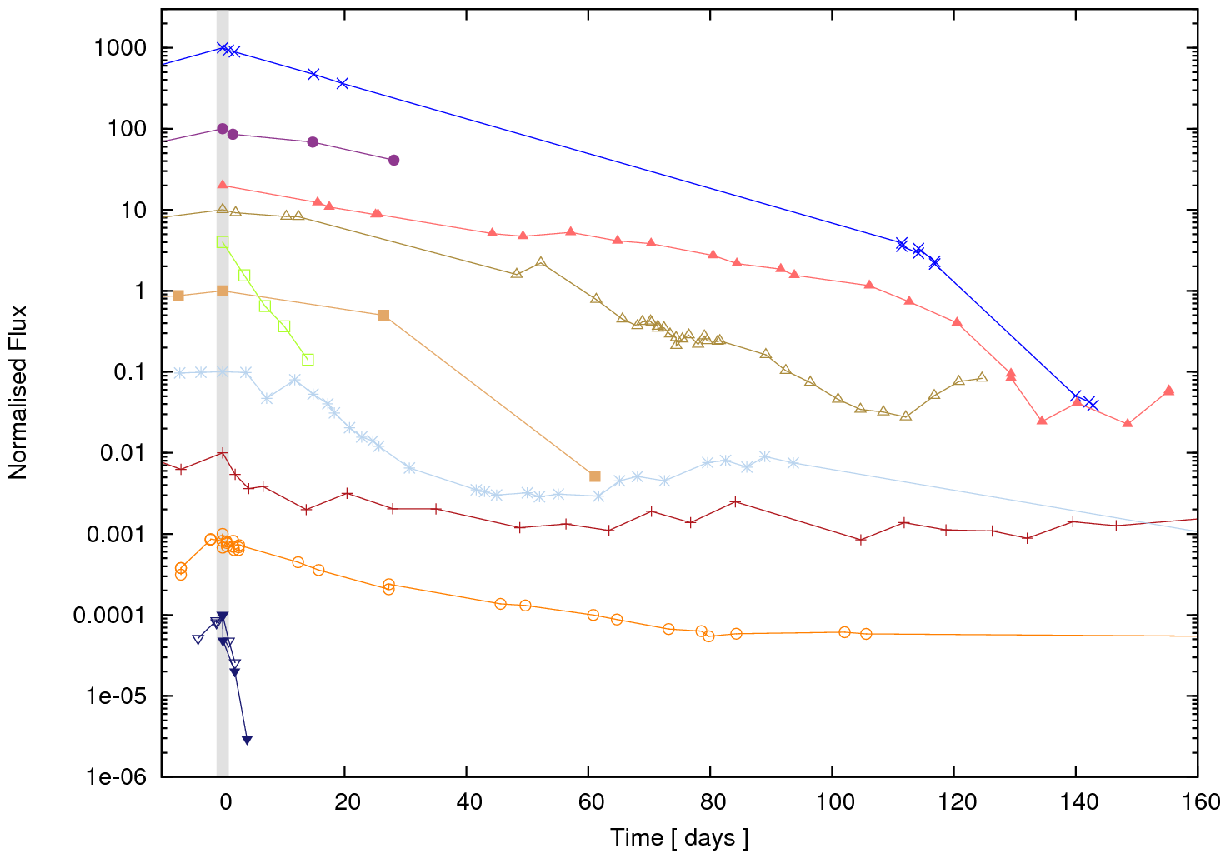}
\caption{Detailed zoom in on the outburst decay lightcurve for the best sampled
  systems, where $\rm t_0$ has been set to the time of maximum flux indicated by the
  shaded region (see also Fig. \ref{BH_all_systems}). Each system has been offset by
  an arbitrary amount for clarity. The symbol/color are as defined in
  Fig. \ref{BH_all_systems}.}
\label{lc_plot}
\end{center}
\end{figure}

\section{Results}\label{results}
The primary results from this study relate to the analysis of the X-ray and UV spectra
of each of the black hole systems in our sample. We briefly discuss the results of the
X-ray spectral fits here.

\begin{figure*}[t]
\begin{center}
\subfigure[\texttt{diskbb+po}]{\includegraphics[height=0.26\textheight]{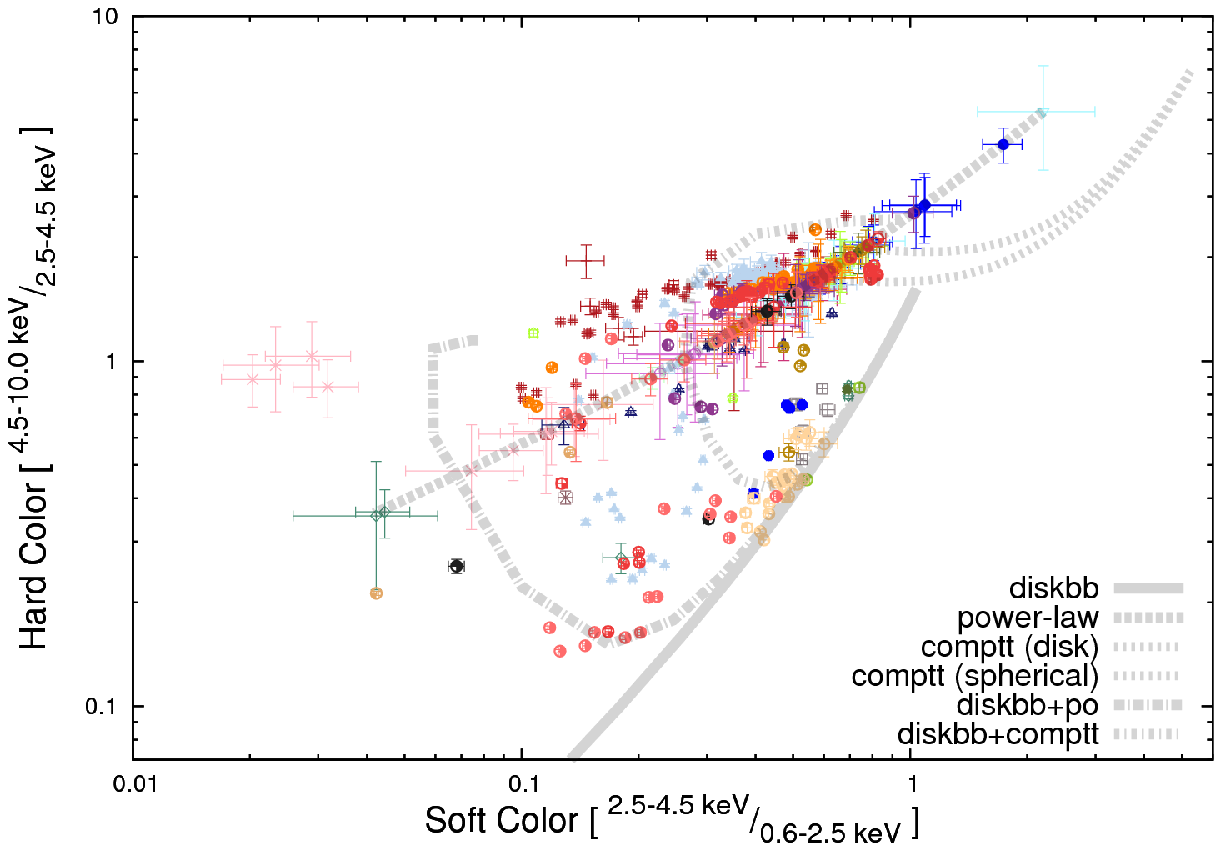}}
\subfigure[\texttt{diskbb+comptt}]{\includegraphics[height=0.26\textheight]{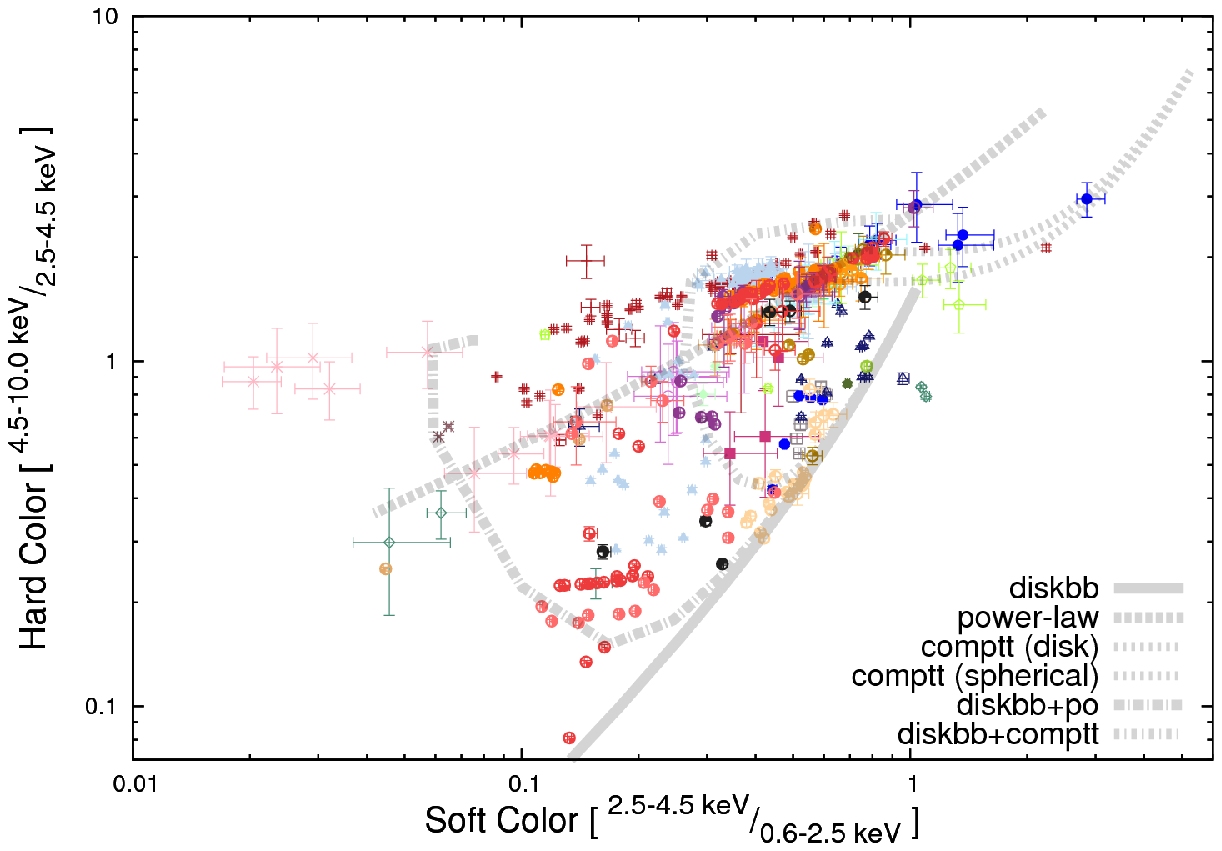}}
\caption{Color-color plots for the systems in our sample. \textbf{(a)} Tracks traced
  assuming the best fit \texttt{diskbb+po} model. \textbf{(b)} Tracks traced assuming
  the best fit \texttt{diskbb+comptt} model. Unabsorbed fluxes have been measured from
  the best fit models with the \texttt{cflux} command in \textsc{xspec}. Only those
  observations that returned a best fit $\rm \chi^2_{\nu} \leq 2$ are plotted. The
  thick grey lines denote the tracks traced by various model components, i.e.,
  accretion disk, power-law and Comptonization. The symbol/color are as defined in
  Fig. \ref{BH_all_systems}.}
\label{color_color_plot}
\end{center}
\end{figure*}

The spectra were fit with 2 families of models, one with a power-law representing the
hard spectral component and the other with the power-law component replaced with a
Comptonization model. In Fig. \ref{chi_diskbb_po}, we plot the $\rm \chi^2_{\nu}$
distribution for the spectra when fit with a power-law only, and after the addition of
a disk component, i.e., \texttt{po} $\rightarrow$ \texttt{diskbb+po}. It is
immediately apparent that a disk component is statistically required in a large number
of observations. In total, we had 476 observations with $\geq$ 100 counts that were
fit with the above model, and of these 445/476 ($\sim$ 94\%) returned a best fit $\rm
\chi^2_{\nu} \leq 2$. An accretion disk is required at greater than the 5$\sigma$
confidence level, as measured by an \texttt{ftest}, in 282/445 (63\%) of the
observations with a best fit $\rm \chi^2_{\nu} \leq 2$. A further 30 observations also
required a disk, but in this case the ``best'' fit is still poor, i.e., $\rm
\chi^2_{\nu} \geq 2$.

In Table \ref{chi2_table}, we list the percentage of spectra below best-fit cut-offs
of $\chi^2_{\nu} \leq$ 1.5 and 2 for our sample of spectral models. We find that fits
with the hard component represented by the \texttt{comptt} spectral model return best
fit values consistent with that returned by those modelled with a \texttt{po}
component. This is to be expected, as we are unable to place constraints on the
electron temperature of the Comptonizing corona due to the lack of data above 10 keV.
Nonetheless, the accretion disk component is still strongly required in this case,
though the disk properties are slightly different to those found with the power-law
component (see \S\ref{accretion_disk}). Likewise, replacing the \texttt{diskbb}
component above with a \texttt{diskpn} model does not provide a statistically
significant improvement in the best fit results.

Before proceeding to a detailed discussion of the results of our spectral analysis, we
first present the photometric properties of our sample.

\subsection{Phenomenology}

\subsubsection{Light-curve morphology}\label{lc_morph_text}
The morphology of a black hole binary X-ray lightcurve contains information which may
be used to constrain the outburst, e.g., outburst energy, $\rm \dot{M}$ etc. In
Fig. \ref{BH_all_systems}, we plot the absorbed flux for each of the 476 pointed
observations in our sample containing greater than 100 total counts. We are sensitive
to outbursts over a dynamic range of greater than 4 orders of magnitude, and including
the extra-galactic systems (NGC 300 X-1, IC10 X-1, M33 X-7), we are sensitive to
accreting stellar mass black holes over 6 orders of magnitude in flux.

In Fig. \ref{histogram_luminosity} the 0.6 -- 10 keV luminosity histogram of all
observations in our sample is plotted, in both absorbed and unabsorbed form. There is
a clear peak in the distribution at a luminosity of $\sim$ 1\% Eddington. Our limiting
luminosity is $\rm \sim 10^{-4}~L_{Edd}$, though there are a number of isolated
observations an order of magnitude fainter (due to GRO J1655-40, which experienced a
large amplitude outburst, see Fig. \ref{BH_all_systems}). The unabsorbed luminosity
equals or marginally exceeds the Eddington limit for a small sample of observations,
which correspond to the known Eddington limited persistent source GRS 1915+105
\citep{vierdayanti10}, while the 2005 outburst from the transient black hole GRO
J1655-40 also approached this limit.  In terms of flux, the distribution is broader
owing to the varying distances of the detected sources. This peak corresponds to an
unabsorbed flux of $\rm \sim 5 \times 10^{-9}~erg~s^{-1}~cm^{-2}$. The distribution of
the maximum luminosity observed during each outburst is displayed in
Fig. \ref{histogram_peak_luminosity}. A peak is observed at a luminosity of $\sim$
20\% $\rm L_{Edd}$, this is consistent with the earlier observations ($\rm \sim
0.2~L_{Edd}$ \citealt{chen97}, $\rm \sim 0.12~L_{Edd}$ \citealt{dunn10a}); however,
small number statistics preclude any deeper analysis.

In Fig. \ref{lc_plot}, we plot the lightcurves for 11 outbursts in our sample, where
we have set $\rm t_0$ to be the time of observed outburst maximum. Due to the variable
and transient nature of the black hole binaries in our sample, we note that our
estimate of $\rm t_0$ is inherently uncertain for those sources without observations
prior to the peak, i.e., XTE J1817-330, H1743-322, GRS 1915+105.  In addition, due to
uncertainty in assignment of $\rm t_0$, our interpretation of the lightcurves is also
biased by the sampling and the duration of the outburst for which the system continued
to be observed. This is particularly noteworthy for the cases of Swift J1842.5-1124,
Swift J1539.2-6277, H1743-322, GRS1915+105, and to a lesser extent GRS J1655-40. The
lightcurves are characterized by fitting with a model assuming an exponential decay
with timescale $\tau_1$ \citep{chen97}. A break is added at late times ($t_b,~\tau_2$)
where necessary, i.e.,
\begin{equation}
f(t) = \left\{
\begin{array}{l l}
a_1e^{-(t)/\tau_1} & t < t_{b}\\
a_2e^{-(t-t_{b})/\tau_2} & t > t_{b}\\
\end{array}
\right.
\end{equation}
The results of these fits are listed in Table \ref{lc_morph}. Our sample of
lightcurves includes 2 persistent systems that display plateau type lightcurves (Cyg
X-1, Swift J1753.5-0127). Three systems display an exponential decay type lightcurve
followed by a break a later times (GRO J1655-40, XTE J1817-330, XTE J1752-223). Swift
J1842.5-1124 \& Swift J1539.2-6277 also display exponential decay type lightcurves but
with a characteristic decay timescale much shorter than the 3 previous systems.
H1743-322 was likely observed late in the decay phase of the outburst as the
lightcurve is consistent with that of GRO J1655-40 after the late time break. Finally,
we display 2 `outbursts' from GRS1915+105. These events display a much shorter
timescale than that observed from the other systems, although the event is roughly
consistent with that seen from Cyg X-1 during its transition to the soft
state. Perhaps such flaring type variability is a characteristic of persistent
accretion flows.

The characteristic exponential decay timescales is similar across a number of systems,
lying in the 20 -- 40 day range. This is consistent with findings of the study of XRB
lightcurves by \citet{chen97}.  The typical outburst durations of $\sim$ 50 -- 100
days (Fig. \ref{lc_plot}), imply a total outburst fluence in the range $\rm 10^{44}
- 10^{45}~erg$ for the galactic stellar mass black hole binary
systems. Assuming the standard accretion efficiency ($\eta = 0.1$), we find the
typical amount of matter accreted per outburst to be $\rm \sim 10^{-10} -
10^{-11}~M_{\sun}$. Though the actual amount may be lower than this as the accretion
efficiency is expected to be lower than 0.1 in the hard state, e.g., due to the
presence of outflow/jets and/or an ADAF. Taking this into account, the measurement of
an accretion rate onto the outer disk in quiescent black holes of $\rm \dot{M}_x
\sim 10^{-10}~M_{\sun}~yr^{-1}$ would support this \citep{mcclintock95,orosz11a}.

\subsubsection{Diagnostic diagrams}
In this section, we create a number of diagrams that provide significant insight into
the behavior of the accreting black holes in our sample. We begin by considering the
source evolution in color-color diagrams.

\vspace{5mm}
\noindent \textit{i. Color color diagrams}\label{cc_diagram}\\
Color-color diagrams \citep{hasinger89} are a useful means by which to characterise
the spectral behavior and evolution of accreting black holes and neutron stars in a
model independent manner. Though, we note that the results are detector and bandpass
dependent, which demands care when making inter-mission comparisons, e.g.,
\textit{Swift} vs \textit{RXTE}.

\begin{figure}[t]
\begin{center}
\includegraphics[height=0.25\textheight]{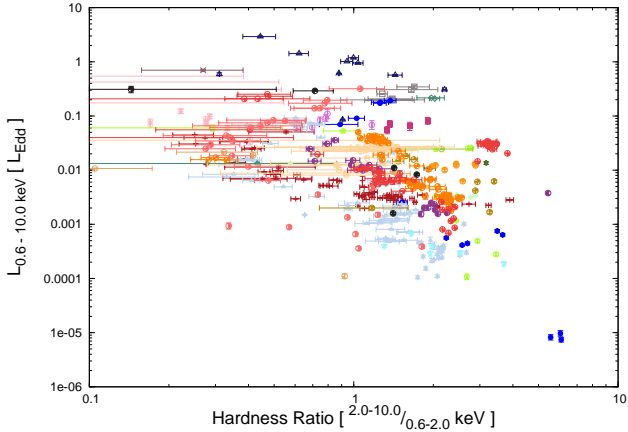}
\caption{Hardness luminosity diagram for the black holes in our sample. The hardness
  ratio is calculated from the ratio of the unabsorbed hard to soft X-ray fluxes,
  calculated from the best fit model \texttt{pha(diskbb+po)}. While there is a hint of
the characteristic `q'-pattern visible in \textit{RXTE} observations, e.g.,
\citet{dunn10a}, there is considerable scatter. The symbol/color are as defined in
  Fig. \ref{BH_all_systems}.}
\label{hld}
\end{center}
\end{figure}

In Fig. \ref{color_color_plot}, we plot color-color diagrams for the systems in our
sample, based on the results of the \texttt{diskbb+po} model and the
\texttt{diskbb+comptt} model. The soft colour is defined as (2.5 keV -- 4.5 keV)/(0.6
keV -- 2.5 keV), while the hard color is defined as (4.5 keV -- 10.0 keV)/(2.5 keV --
4.5 keV).  We plot unabsorbed fluxes as we want to avoid the detector dependent
characteristics introduced by using the measured count rate. A number of our spectra
contain some pile-up (\S\ref{pile_up}), and correcting for this to obtain the true
count rate is not straightforward.  The colors are also impacted by the column density
in the direction of each object, which varies considerably for the sources in out
sample, i.e., $\rm N_H = 10^{21} - 10^{23}~cm^{-2}$. The expected behavior for a
number of different continuum components is plotted in the center panel, e.g.,
power-law, accretion disk etc. The large number of low-hard state observations in our
data sample is immediately apparent with the power-law dominated spine
dominating the color-color diagrams. These \textit{Swift} diagrams are qualitatively
similar to those constructed by \citet{done03} for a sample of black holes observed by
\textit{RXTE}.

\begin{figure}[t]
\begin{center}
\includegraphics[height=0.25\textheight]{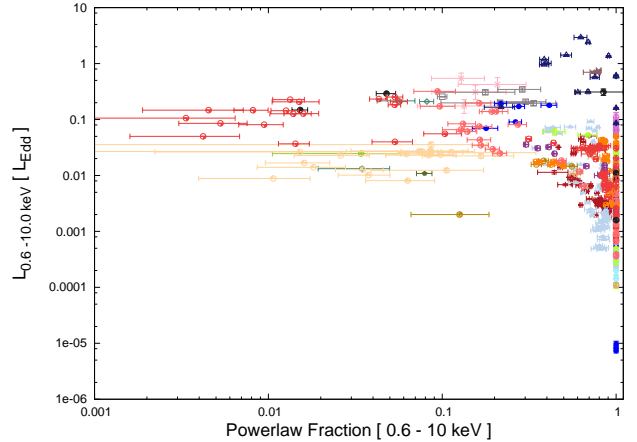}
\caption{Disk fraction luminosity diagram for the black holes in our sample. The
  dominance of hard state sources in our sample is clearly visible. Those systems with
  a power-law fraction less than 0.7 typically require the presence of a low
  temperature disk component. The spectra for which no disk component is required can
  be seen with a power-law fraction of 1. The symbol/color are as defined in
  Fig. \ref{BH_all_systems}.}
\label{dfld}
\end{center}
\end{figure}

\vspace{5mm}
\noindent \textit{ii. Hardness luminosity diagram}\label{hld_diagram}\\
The hardness luminosity diagram (HLD; \citealt{homan01,belloni04,fender04}) for the
sources in our sample is plotted in Fig. \ref{hld}.  This diagram is limited due to
its reliance on count rates to calculate the hardness ratio. There will additionally
be errors due to uncertainty in the column density, black hole mass, and distance
estimates ($\rm N_H,~M_x,~d$), which make it difficult to separate the different
spectral states.  If simultaneous timing information is available, then the HLD is
very useful to characterise spectral evolution during an outburst. Nonetheless, we
include it for comparison with previous work, e.g., \citet{dunn10a}. As we utilize a
CCD detector, we choose not to calculate a hardness ratio from the spectral counts due
to possible pile-up related effects. Hence, we define the hardness ratio as the ratio
of the hard to the soft flux, i.e., $\rm F_{2.0 - 10.0 keV}/F_{0.6 - 2.0 keV}$.  While
the overall shape of the pattern traced by the objects is consistent with the
`q'-pattern observed with \textit{RXTE}, there is considerable scatter. In particular,
the low-hard state dominated vertical branch see in \textit{RXTE} HLDs is not present
here due to the superior ability of \textit{Swift} to detect low temperature accretion
disks. These differences are to be expected given the caveats outlined above and the
differing bandpasses between the 2 instruments.

We can also ask if defining the hardness ratio in terms of energy ranges will mask
changes in the source properties. This is most likely an issue in the energy range 0.6
-- 3 keV, where there will be significant overlap between the hard (\texttt{po,
  comptt}) \& soft (\texttt{diskbb}) spectral components. This has been investigated
by comparing the number of soft/hard sources when we define the hardness ratio as the
ratio of the flux in each spectral component ($\rm f_{comptt}/f_{disk}$) vs the the
flux in the spectral bands above and below 2 keV. We find that for both definitions,
the percentage of sources for which the above ratio is less than 1 is approximately
70\%, i.e., both methods find approximately 30\% of the observations to be dominated
by the soft component.

\vspace{5mm}
\noindent \textit{iii. Disk fraction luminosity diagram}\label{dfld_diagram}\\
In Fig. \ref{dfld}, we display the disk fraction luminosity diagram (DFLD)
\citep{kalemci04,kalemci06,tomsick05,kording06} for the black holes in our
\textit{Swift} sample. The plot is generated from the best fit model
\texttt{diskbb+po} over the spectral range 0.6 -- 10.0 keV. Creating this plot with
the Comptonization model does not appreciably modify the resulting pattern traced by
the observations in the DFLD.  The sample used by \citet{dunn10a} is biased to higher
luminosities in comparison to the current sample presented herein, which is dominated
by hard state spectra.  \citet{dunn10a} calculated the luminosity in the 1 -- 100 keV
and 0.001 -- 100 keV bands for the power-law and disk components respectively. This
requires an extrapolation to energies well below the \textit{RXTE/PCA} bandpass (i.e.,
below 3 keV), which will lead to errors as the accretion disk cools. Nonetheless, the
DFLD presented herein is qualitatively similar to that measured by \textit{RXTE} for a
similar sample of black hole binaries \citet{dunn10a}.

\begin{figure}[t]
\begin{center}
\includegraphics[height=0.26\textheight]{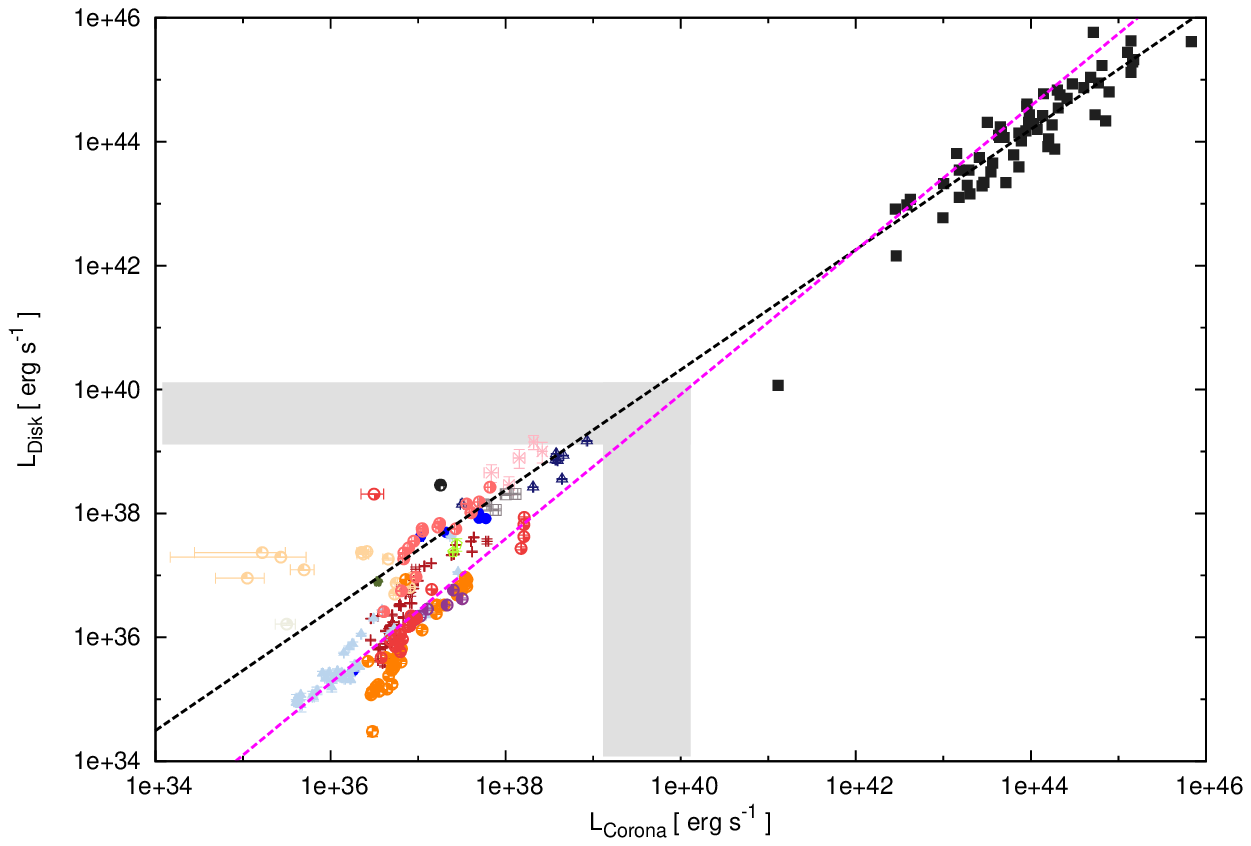}
\caption{Relationship between the unabsorbed luminosity from the corona and that
  emitted by the accretion disk for the sample of stellar mass BHs considered herein,
  where the fluxes have been measured assuming the \texttt{diskbb+comptt} model. We
  also plot the hard X-ray selected sample of AGN (black squares) from
  \citet{sazonov12}. The gray shaded region denotes the region from $\rm L_{Edd} -
  10~L_{Edd}$ assuming a 10 M$_{\sun}$ black hole. The symbol/color are as defined in
  Fig. \ref{BH_all_systems}. The lone AGN at a luminosity of $\rm \sim
  10^{41}~erg~s^{-1}$ is the low luminosity AGN NGC 4395.  The black dashed line
  denotes the best fit relationship between the corona and the disk ($\rm L_{disk}
  \propto L_{corona}^{0.97}$) for the AGN sample, while the magenta line plots the
  inverse relationship corrected for Malmquist bias ($\rm L_{corona} \propto
  L_{disk}^{0.85}$), see equation \#15 \& \#B4 in \citet{sazonov12}.}
\label{ldisk_lcomptt}
\end{center}
\end{figure}

\begin{figure*}[t]
\begin{center}
\subfigure[Disk temperature]{\includegraphics[height=0.27\textheight]{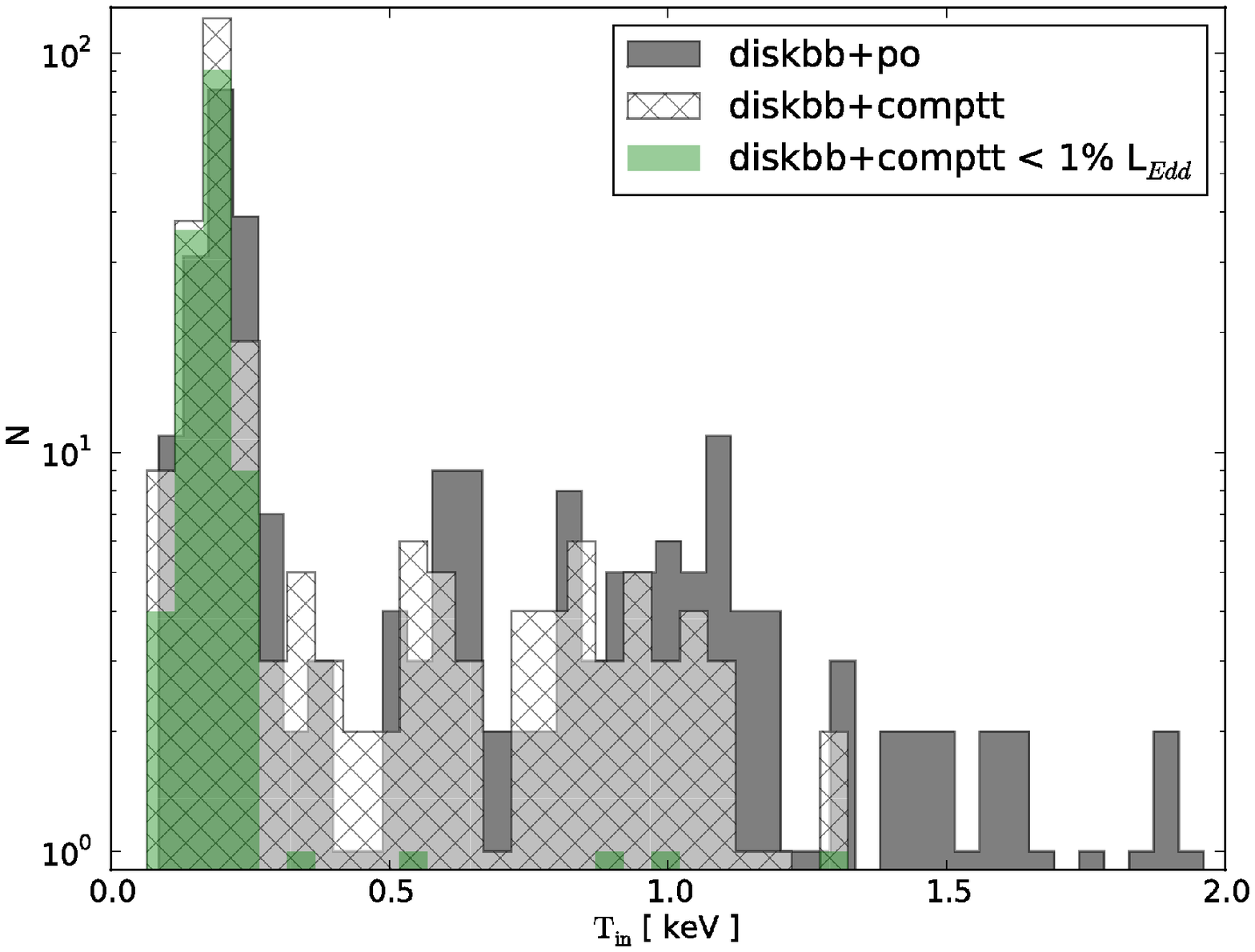}}
\subfigure[Disk inner radius]{\includegraphics[height=0.27\textheight]{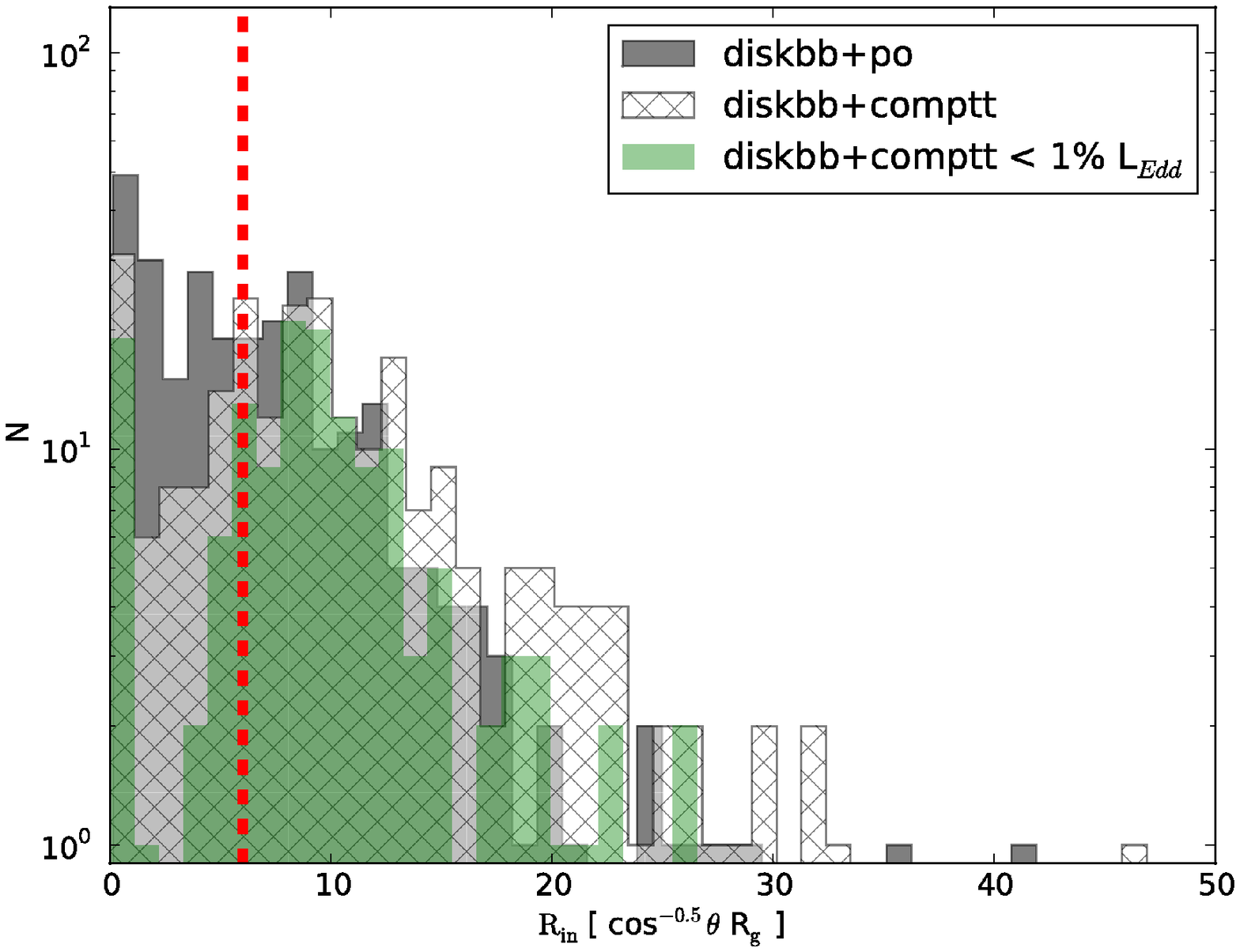}}
\caption{\textbf{Left:} Histogram of the accretion disk temperature for the disks
  detected in our sample. Two peaks are observed consistent with the classical hard
  and soft spectral states with a number of disks detected at temperatures between
  these. \textbf{Right:} Histogram of the accretion disk inner radius measured for the
  disks detected in our sample. The dashed vertical line denotes a radius of $\rm
  6~R_g$ consistent with the ISCO for a Schwarzschild black hole. The hard continuum
  component has been modelled in 2 separate ways (i) a power-law -- \texttt{po}, and
  (ii) Comptonization -- \texttt{comptt}. While both models are similar in the hard
  spectral region ($\gtrsim$ 2 keV), below this the Comptonization has a low energy
  roll over whereas the power-law does not. Nonetheless, there is considerable
  agreement between the derived disk properties in each case. The green histograms
  denote those disks detected when the luminosity is lower than 1\% Eddington,
  assuming the masses and distances in Table \ref{BHC_table}.}
\label{disk_temp_histogram}
\end{center}
\end{figure*}

\subsection{Physical Interpretation}

\subsubsection{The inner accretion flow}\label{accretion_disk}
In this section, we focus on the 0.6 -- 10 keV X-ray properties of the accretion disk
as measured in our sample, before also considering the available UV data. As outlined
earlier (\S\ref{results}), an accretion disk is statistically required at greater than
the 5$\sigma$ level in $\sim$ 56\% of our observations. These disks are detected at
varying levels of dominance (relative to the hard component), which can be seen in the
disk fraction luminosity diagram (Fig. \ref{dfld}). Specifically, in the broad 0.6 --
10 keV bandpass, 95\% (68\%) of the detected disks contribute greater than 7\% (17\%)
of the total unabsorbed flux. If we consider only the soft bandpass (0.6 -- 2.0 keV),
then 95\% (68\%) of the detected accretion disks contribute greater than 16\% (40\%)
of the total unabsorbed flux. These disk fraction percentages correspond to the
spectra when the hard component is modelled using a Comptonization model. Utilizing a
power-law model results in slightly lower disk fractions as expected, i.e., 95\%
(68\%) of the disks contribute greater than 14\% (34\%) of the total unabsorbed flux
in the 0.6 -- 2.0 keV band. 

\vspace{5mm}
\noindent\textit{i. Accretion disk -- corona relationship}\label{disk_vs_corona}\\ In
Fig. \ref{ldisk_lcomptt}, we plot the relationship between the unabsorbed luminosity
in the corona versus that generated by the accretion disk, where the fluxes have been
calculated from the \texttt{diskbb+comptt} model. We note that when the hard component
in our sample is instead modelled with a simple powerlaw component, the distribution
of sources in this plot is similar albeit with somewhat larger scatter.

\citet{sazonov12} have investigated this relationship in a sample of 61 hard X-ray
selected luminous Seyfert galaxies, where it was found that the disk flux is
approximately proportional to the coronal flux. These systems are also plotted in
Fig. \ref{ldisk_lcomptt} along with the best fitting relationship from
\citet{sazonov12}. The black dashed line denotes the relation between the disk flux as
a function of the coronal flux ($\rm L_{disk} \propto L_{corona}^{0.97}$), while the
magenta line denotes the inverse relation corrected for Malmquist bias ($\rm
L_{corona} \propto L_{disk}^{0.85}$), see the discussion in \citet{sazonov12} for
details.  It is immediately apparent that an extrapolation of these relations to the
stellar mass systems reveals an intriguing correspondence.\footnote{ When comparing
  the luminosities of systems at such vast distances, one must always worry that the
  common dependence of $\rm L_{disk},~L_{corona}$ on the distance introduces a
  spurious correlation. A partial correlation (PC) test can be used to reveal this
  effect, e.g., \citet{akritas96,merloni03}. For the data presented in
  Fig. \ref{ldisk_lcomptt}, a PC test reveals this correlation to be robust and not
  simply a spurious distance driven correlation. This is true for the entire sample
  and for the AGN/stellar systems when considered on their own. We defer a detailed
  discussion of this subject to a future publication.}

In contrast to AGN where the accretion disk emission is difficult to constrain, for
example, requiring the use of mid-IR emission as a proxy in the \citet{sazonov12}
study, the accretion disk flux in stellar mass systems peaks in the standard X-ray
bandpass (0.2 -- 10.0 keV) allowing simultaneous constraints to be placed on both the
disk and coronal emission. It is apparent from the stellar systems that the use of a
single relation is unlikely to provide an accurate description of the observed
disk -- coronal behavior for the entire sample. This is not surprising given that the
stellar mass sample studied herein contains a large number of transient systems,
dominated by accretion rates varying by many orders of magnitude, in contrast to the
relatively constant accretion rates exhibited by the AGN studied in \citet{sazonov12}.
In addition, the stellar mass systems display apparently discreet spectral states,
while Seyfert galaxies are more likely to correspond to stellar mass systems in a Cyg
X-1-like soft state, e.g., \citet{mchardy04}.

Although there is considerable scatter, extrapolation of the Seyfert galaxy disk --
corona relationship to the stellar mass black holes suggests that at least a subset of
the stellar mass BHs exhibit a disk -- corona relationship that is consistent with
that of their super-massive counterparts (e.g., see Fig. \ref{ldisk_lcomptt}). Further
study of this relation is clearly warranted, as it would add evidence pointing to the
scale invariant nature of the accretion process, e.g., fundamental plane
\citep{merloni03,falcke04}. Currently, the fundamental plane relates radio flux to
hard X-ray flux and the black hole mass. While the 2 -- 10 keV flux may be a good
proxy for the true disk luminosity, it is necessarily an indirect measure. The
relation presented in Fig. \ref{ldisk_lcomptt} may provide a means to finally link the
true disk and jet emission across the mass scale for accreting black holes. We defer a
detailed consideration of this subject to a future publication.

\begin{table}[b]
\begin{center}
\caption{Disk Inner Radius}\label{disk_rin_table}
\begin{tabular}{lccc}
\tableline\\ [-2.0ex]
Radius [ $\rm R_g$ ] & $\rm (d,~M_x)$ & $\rm  (1.2d,~0.8M_x)$ & $\rm (0.8d,~1.2M_x)$\\ [0.5ex]
\tableline\tableline\\ [-2.0ex] 
6    & 0.57 (0.36) & 0.43 (0.21) & 0.74 (0.59) \\   [0.5ex]
10   & 0.78 (0.59) & 0.57 (0.36) & 0.91 (0.81) \\   [0.5ex]
33   & 0.97 (0.97) & 0.95 (0.97) & 0.98 (0.97) \\   [0.5ex]
\tableline
\end{tabular}
\tablecomments{The cumulative number of spectra with an inner radius less than the
  values listed in column one. Column 2 lists the inner radius calculated from the
  \texttt{diskbb+po} (\texttt{diskbb+comptt}) model assuming the values for mass and
  distance listed in Table \ref{BHC_table} (where $\rm R_{in}[R_g] \propto d/M_x$, see
  appendix A\ref{appendix_param} for a discussion of the system parameters and their
  related uncertainties). Columns 3 \& 4 list the values assuming 20\% uncertainty in
  the mass and distance such that the inner radius is minimized and maximized
  respectively. Six gravitational radii is consistent with the ISCO for a
  Schwarzschild black hole. We see that essentially all detected disks have radii less
  than $\rm \sim 40~R_g$.}
\end{center}
\end{table}

\begin{figure*}[t]
\begin{center}
\subfigure[\texttt{diskbb+po}]{\includegraphics[height=0.25\textheight]{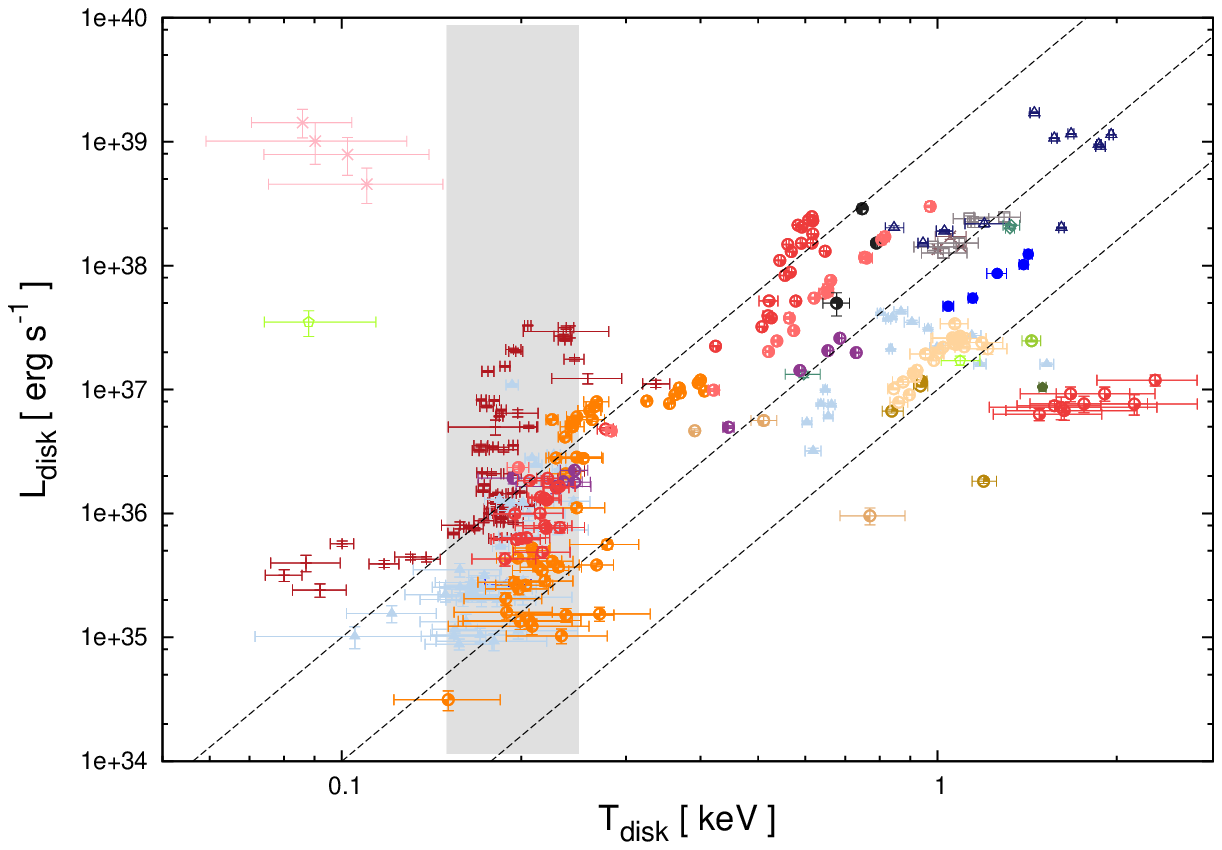}}
\subfigure[\texttt{diskbb+comptt}]{\includegraphics[height=0.25\textheight]{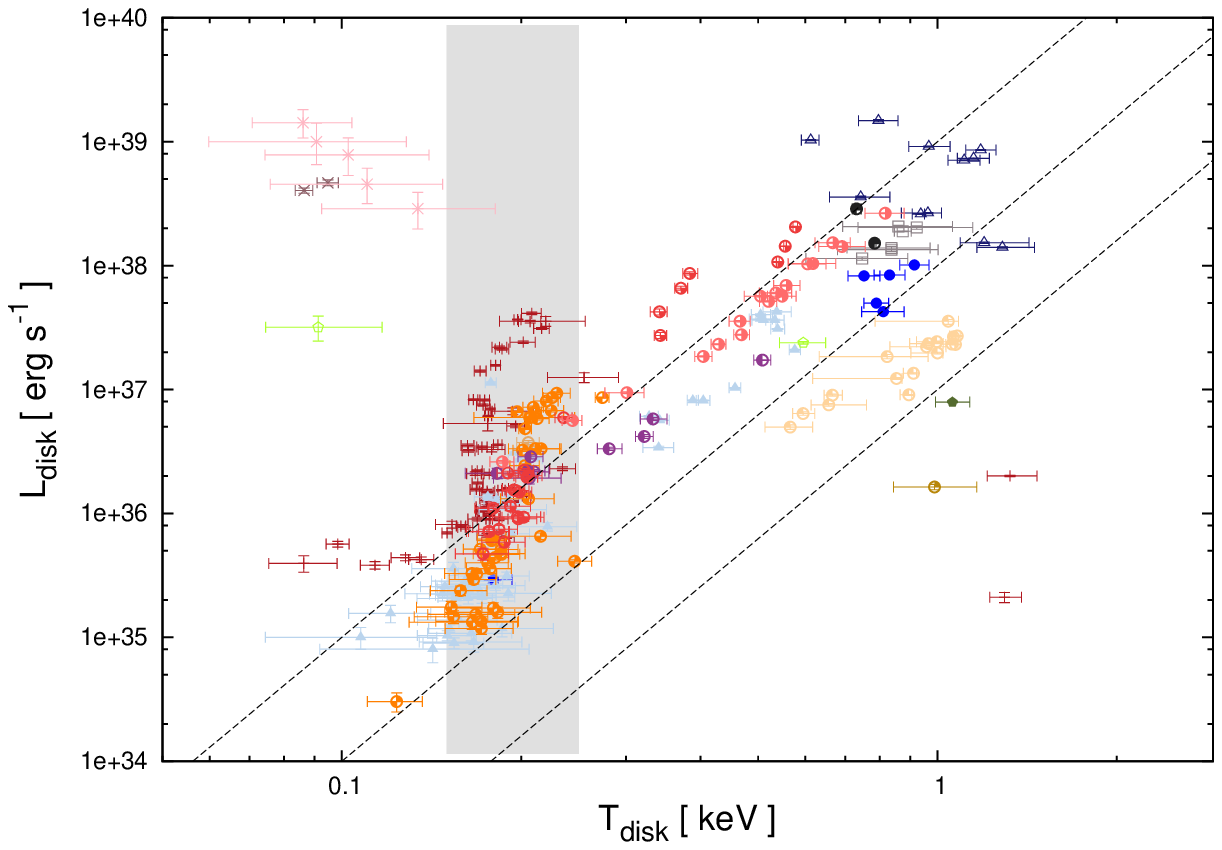}}
\vspace{-3mm}
\caption{Luminosity of the accretion disk component versus the disk
  temperature. Dashed lines indicate the relation $\rm L_x \propto T^4$, as would be
  expected from a disk of constant emitting area. At large temperatures/fluxes this
  relation is followed by all of the systems with a small amount of scatter. At low
  luminosities, this is not the case with the disk temperature appearing to remain
  constant with changing flux. We note, at this disk temperature we are dominated by
  observations of the persistent systems Cyg X-1 and SWIFT J1753.5-0127. The
  symbol/color are as defined in Fig. \ref{BH_all_systems}.}
\label{lx_t4}
\end{center}
\end{figure*}

\vspace{5mm}
\noindent\textit{ii. Disk temperature \& radius}\\ The primary observational
quantities providing us with information on the configuration of the accretion disk
are the inner disk color temperature ($\rm T_{col}$) and the disk normalization, which
is proportional to the distance and the disk color radius, i.e., $\rm norm \propto
(r_{col}/d)^2$. The best fit values for the parameters are dependent on how the
spectra are modelled. In particular, the choice of model for the hard X-ray component
can modify the measured disk flux, e.g., using a power-law (\texttt{po}) or thermal
Comptonization (\texttt{comptt} -- \citealt{titarchuk94}) because the Comptonization
cuts off at low energies consistent with the input seed photon temperature.  We
compare the 2 models above with the goal of determining, which model is a better
description of the spectra. This will also allow us to probe the effect of incorrect
modelling on the accretion paradigm (see \S\ref{spec_issues}). In the appendix
(A\ref{appendix_cyg}), we discuss the evolution of the persistent BH Cyg X-1 \& the
transient system GX 339-4 in detail, and compare the behavior of the key observational
quantities as a function of time for both the \texttt{diskbb+po} and
\texttt{diskbb+comptt} models.

In Fig. \ref{disk_temp_histogram} (left panel), we plot the measured accretion disk
temperature histogram for both of the \texttt{diskbb+po} and \texttt{diskbb+comptt}
continuum models.  The primary difference between these models is that fewer
observations require the addition of an accretion disk component at greater than the
5$\sigma$ significance level, i.e., 282 and 254 observations require a disk at the
5$\sigma$ level for the \texttt{diskbb+po} and \texttt{diskbb+comptt}
respectively. This is due to the differing nature of the hard components at low
energies, which also modifies the disks parameters, i.e., the accretion disks detected
when the hard component is modelled with \texttt{comptt} are slightly cooler and have
a slightly larger inner radius as is evident in the figure. 

The inner radius of the accretion disk may be determined from the normalization of the
\texttt{diskbb} model, i.e., $\rm norm = (r_{in} [km]/d [10 kpc])^2 cos\theta$. The
disk radius thus measured is subject to a number of additional uncertainties, which
bias the measured radius. The most important are (i) to account for spectral
hardening, a multiplicative correction factor of 1.7 is typically assumed
(\citealt{shimura95}, though see below), and (ii) to account for the fact that the
disk temperature does not peak at the inner radius \citep{kubota98,makishima00}, which
combined give us the corrected inner radius\footnote{The radius measured in this
  manner is an approximation for the actual radius (see \citealt{kubota98}),
  determination of which requires detailed relativistic spectral modelling, which is
  outside the scope of the current project. Nonetheless, within the limits of the
  current data, the calculated radii are consistent with more detailed modelling. For
  example, if we compare the value we measure for the inner radius of the accretion
  disk in LMC X-3 with that determined via detailed relativistic models by
  \citet{steiner10}, our \texttt{diskbb} corrected inner radius is consistent with
  this within the error bars.}
\begin{equation}
R_{in} [ km ] \simeq \frac{1.18\sqrt{norm}}{\sqrt{cos\theta}}d_{10 kpc}
\end{equation}
In Fig. \ref{disk_temp_histogram} (right panel), we plot the inner disk radius
calculated for each of the accretion disks in the sample, assuming the distances and
black hole masses listed in Table \ref{BHC_table} (see appendix A\ref{appendix_param}
for a discussion of the system parameters and their related uncertainties). The
distribution peaks at small inner radii, $\rm R_{in} \lesssim 10~R_g$, with a tail to
larger radii\footnote{$\rm 1~R_g = 14.8~(M_x/10~M_{\sun})~km$}. Of particular note,
the largest radii measured in our sample are $\lesssim$ 40 $\rm R_g$. In Table
\ref{disk_rin_table}, we list the cumulative number of accretion disk with radii below
6, 10, 33 $\rm R_g$ respectively and these values taking into account uncertainties in
the mass and distance to each system respectively. The maximum inner radius is
determined to be robustly $\lesssim$ 40 $\rm R_g$. In Fig. \ref{disk_temp_histogram},
those spectra for which the source luminosity is less than 1\% Eddington are plotted
in green. It is clear that at low luminosities, the disk temperature is low $\sim$ 0.2
keV, but the inner radius of the accretion disk also remains low, exhibiting a clear
peak at a radius of $\sim$ 10 R$\rm _g$.

\begin{figure*}[t]
\begin{center}
\subfigure[$\rm T_{in}~vs~\Gamma$]{\includegraphics[height=0.17\textheight]{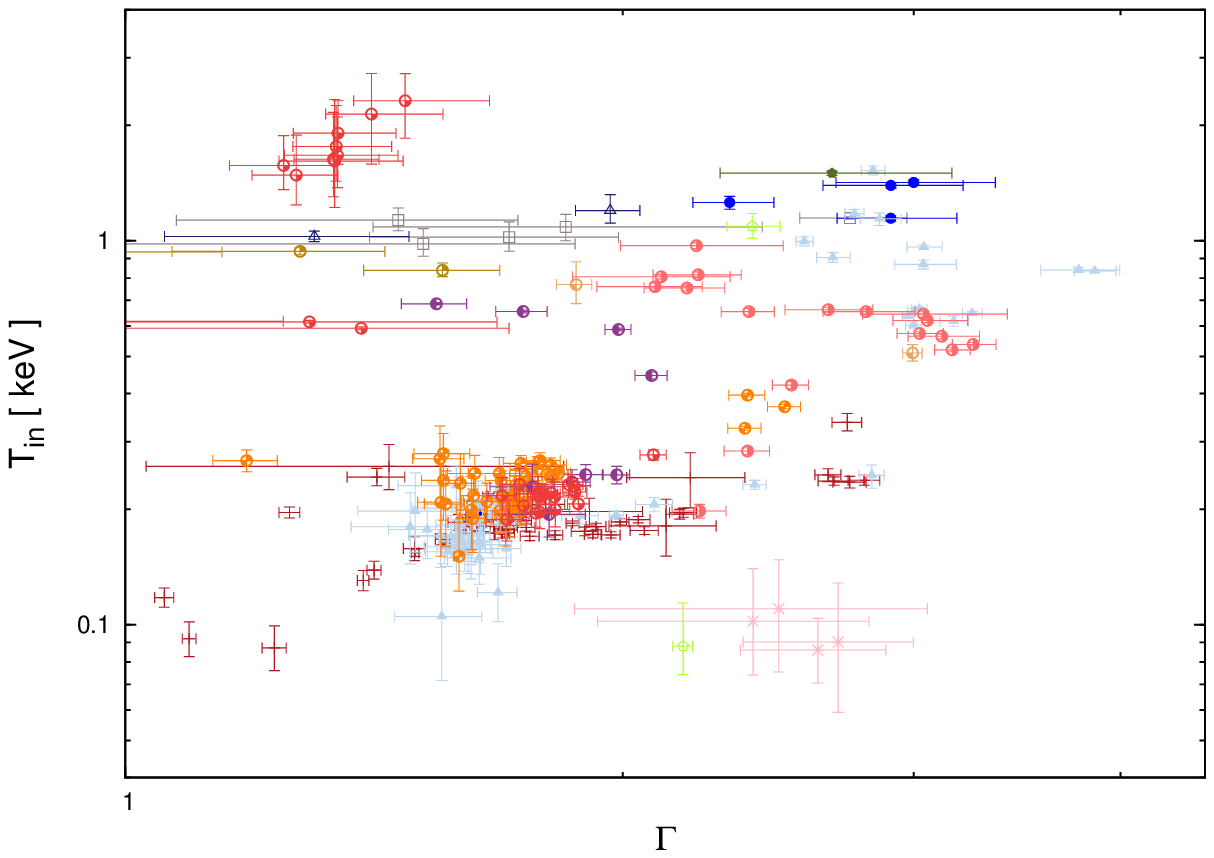}}
\subfigure[$\rm R_{in}~vs~\Gamma$]{\includegraphics[height=0.17\textheight]{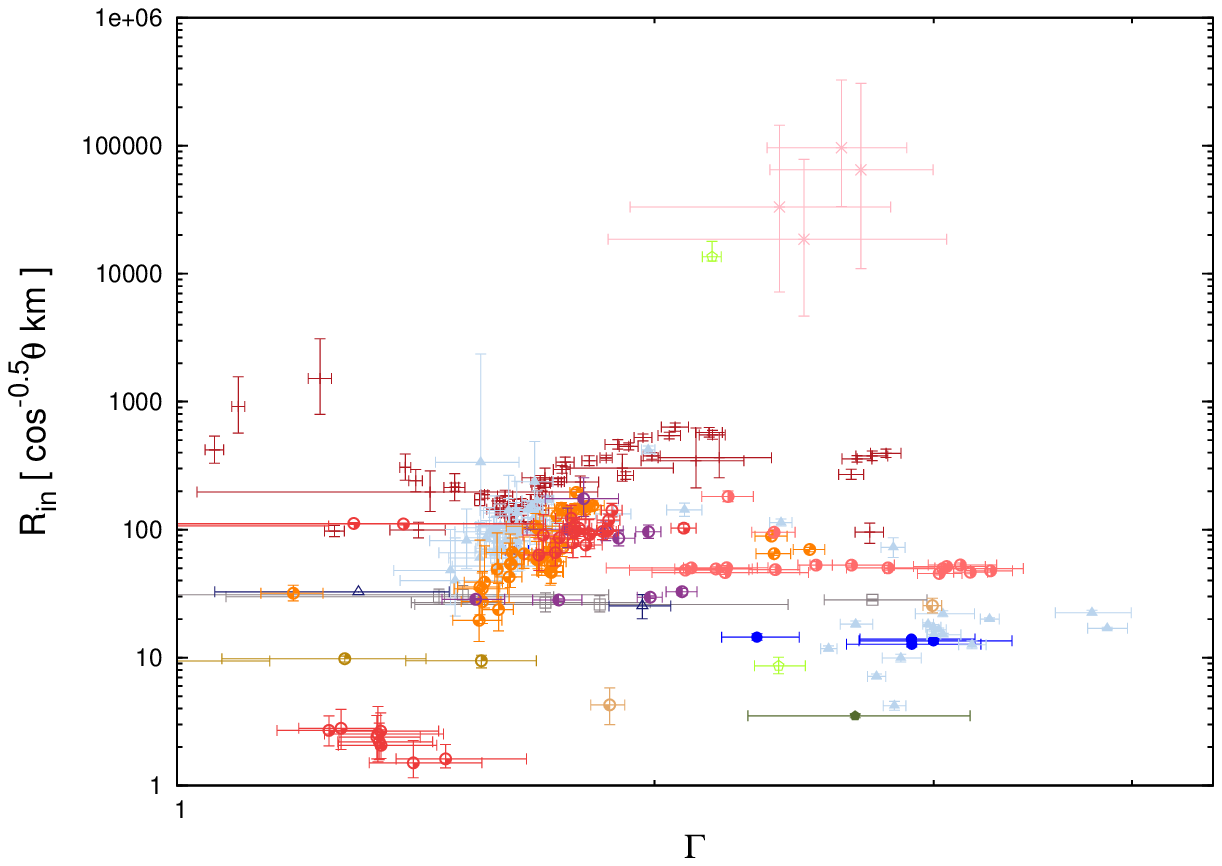}}
\subfigure[$\rm L_{Edd}~vs~\Gamma$]{\includegraphics[height=0.17\textheight]{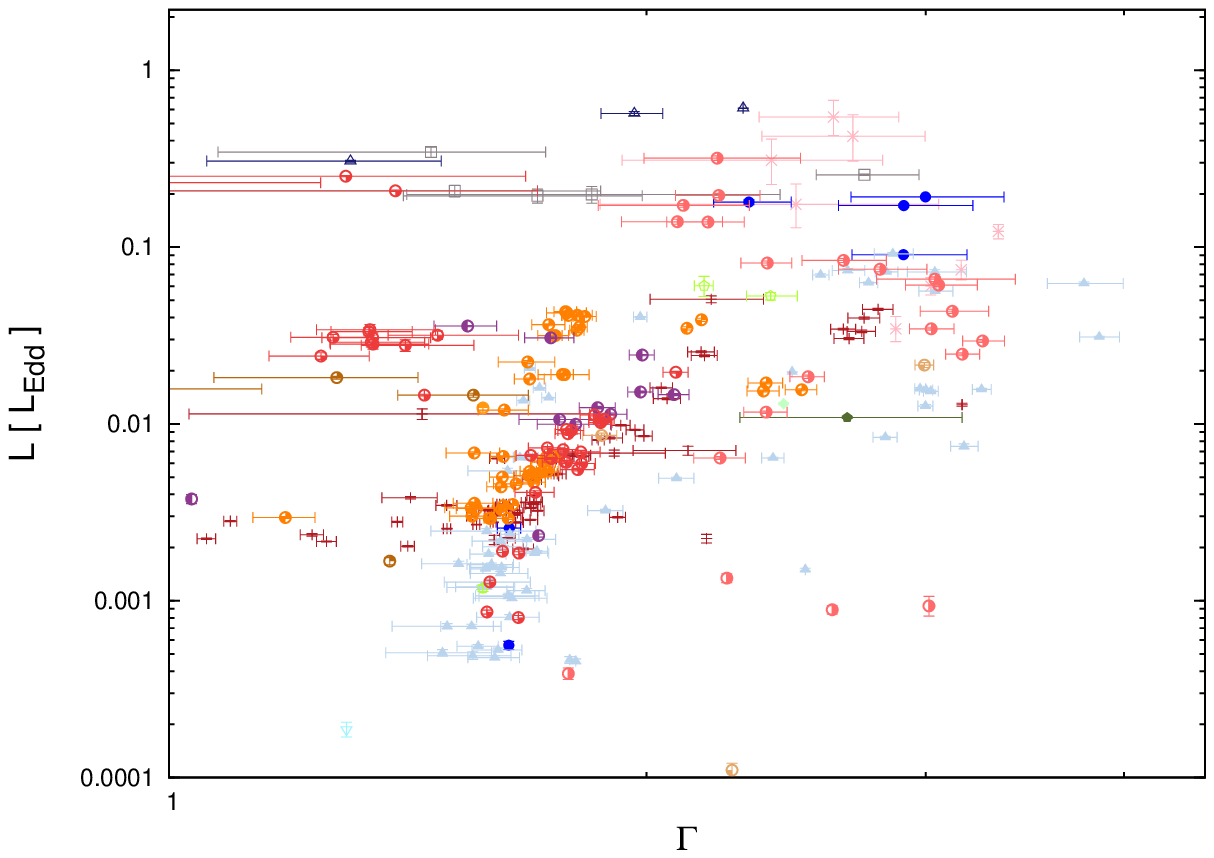}}
\subfigure[$\rm T_{in}~vs~\tau$]{\includegraphics[height=0.17\textheight]{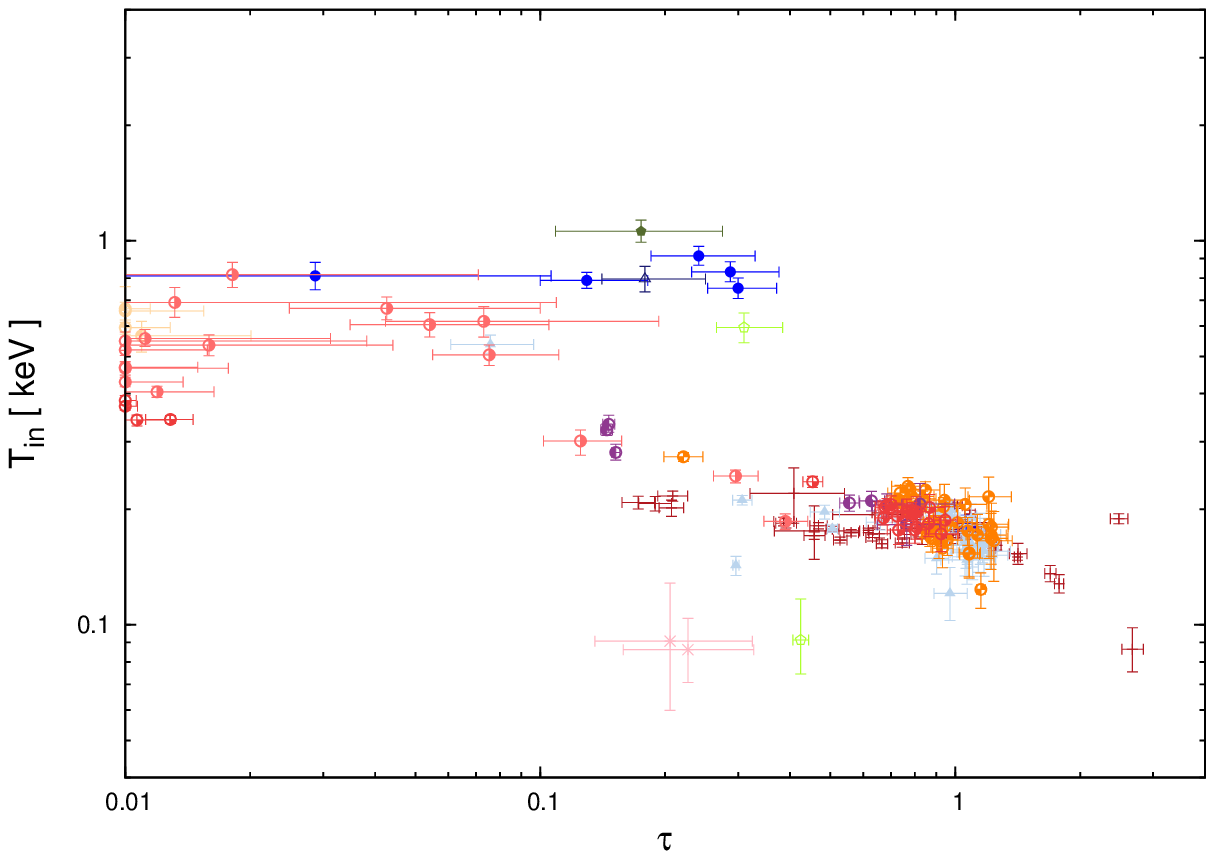}}
\subfigure[$\rm R_{in}~vs~\tau$]{\includegraphics[height=0.17\textheight]{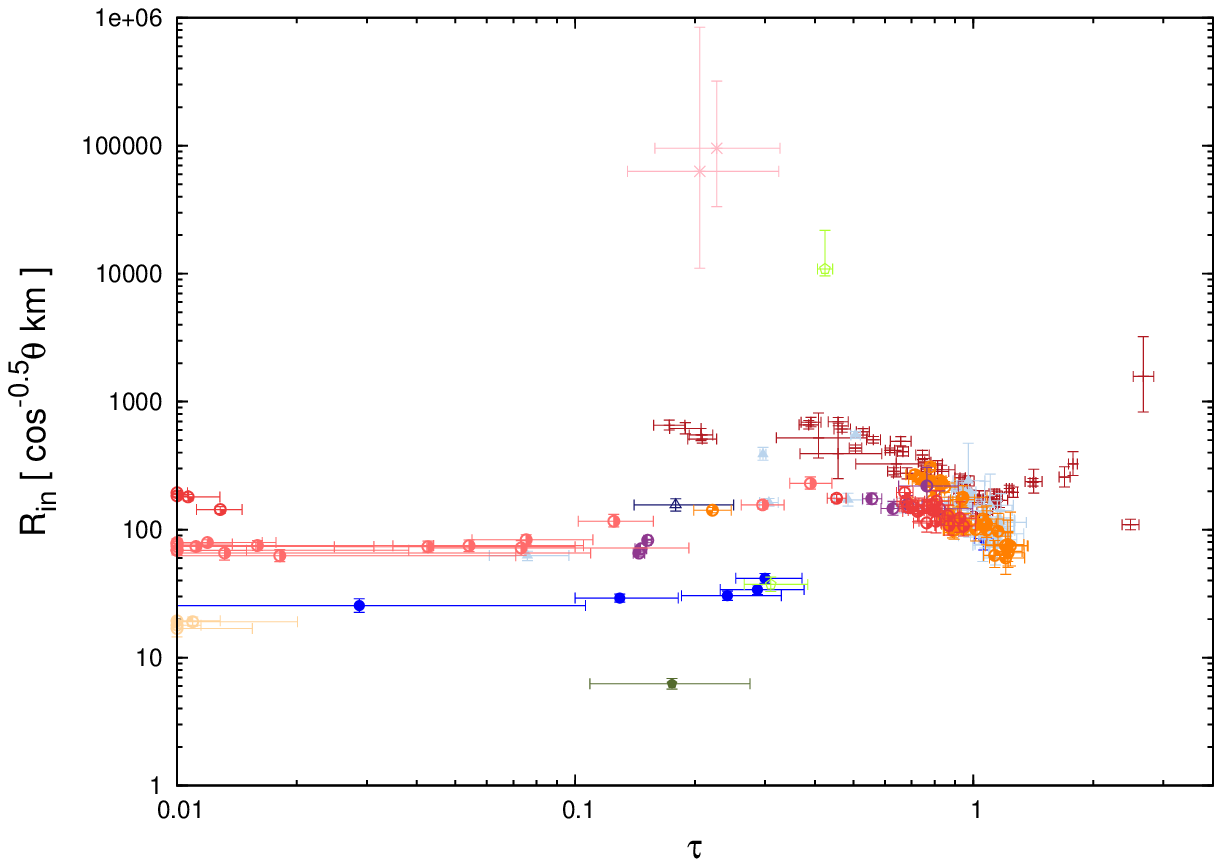}}
\subfigure[$\rm L_{Edd}~vs~\tau$]{\includegraphics[height=0.17\textheight]{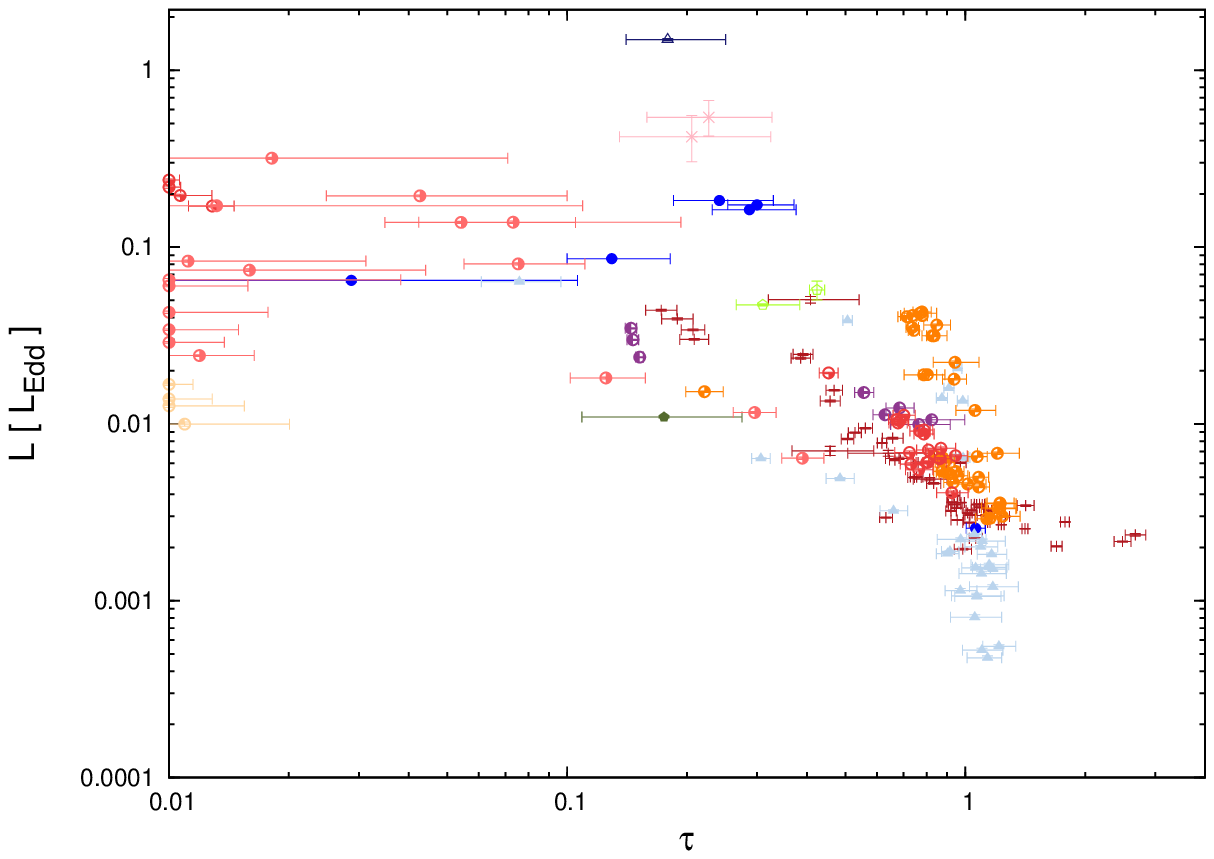}}
\caption{Behavior of the hard X-ray component as characterized by the power-law index
  $\Gamma$ (top), and the Comptonizing corona optical depth $\tau$ (bottom). The
  relationship between the above quantities and inner disk temperature, inner disk
  radius and Eddington luminosity are plotted from left to right respectively. We note
  that the electron temperature for the corona has been frozen at 50 keV. The
  symbol/color are as defined in Fig. \ref{BH_all_systems}.}
\label{gamma_tau_plot}
\end{center}
\end{figure*}

Examination of the temperature distribution in Fig. \ref{disk_temp_histogram} clearly
reveals 2 distinct regimes, i.e., those at $\sim$ 0.2 keV and those at a temperature
of $\gtrsim$ 0.5 keV. These are consistent with the primary active accretion states,
the hard state and the soft state respectively. We note that there is a deficit of
systems in the temperature range between these two. The so called intermediate states
are expected to reside here, i.e., those states during which the system transitions
from the hard state to the soft state and vice versa \citep{mcclintockremillard06}.
In contrast, the inner disk radius distribution does not display obvious evidence for
2 or more distinct spectral states, instead being consistent with a single continuous
distribution of low inner radii with a tail to higher radii, $\rm R_{in} \lesssim
40~R_g$. 

In the \texttt{diskbb+po} fits there are a number of high temperature disks that are
not required when the hard component is modelled using a Comptonization
component. These high temperature disks are seen from GRS 1915+105, GX 339-4 and XTE
J1752-223. In all of these cases, the spectrum is dominated by the hard component,
i.e., these spectra appear to be consistent with the systems being in the very high
state at the time of the observation \citep{mcclintockremillard06}. A number of high
temperature disks also persist at a luminosity less than 1\% Eddington (see
Fig. \ref{disk_temp_histogram}). These spectra belong to Cyg X-1, IGR J17091-3624 and
IGR J17098-3628. The IGR J17098-3628 spectra are only marginally below the nominal 1\%
selection cut, i.e., $\rm L_x \sim 0.0098~L_{Edd}$. In contrast, the spectra of the
other two systems are closer to 0.001~L$\rm _{Edd}$. The IGR J17091-3624 observation
is consistent with a soft state disk dominated spectrum, whereas the Cyg X-1 spectrum
is dominated by power-law emission and is more consistent with a very-high state like
spectrum. The system parameters for both of the IGR sources are poorly constrained,
unlike Cyg X-1, where both the mass and distance are accurately known
\citep{orosz11b}. However, we note that the simple spectral modelling used herein may
be a poor representation of the actual spectral form for these systems, e.g.,
\citet{miller12}.

In Fig. \ref{lx_t4}, the relationship between the disk luminosity and temperature is
plotted for both of the continuum models. If the emitting surface area of the
accretion disk is constant (i.e., constant inner radius), then we expect the systems
to follow a relation of the form $\rm L_{disk} \propto T^4$. At higher temperatures
and luminosities, a relation of this form (as indicated by the dashed lines) is
observed; however, at the lowest disk temperatures (indicated by the shaded region),
this no longer holds and instead the disk temperature is observed to remain
approximately constant while the luminosity changes by orders of magnitude. Modelling
the accretion disk emission with the \texttt{diskpn} model
\citep{diskpn_gierlinski99}, returns similar results. The deviation of the spectra at
low temperature from the expected relation implies a change in the conditions of the
accretion flow, for example, a changing inner accretion disk radius in the low-hard
state \citep{done07}.

\begin{figure*}[t]
\begin{center}
\subfigure{\includegraphics[height=0.25\textheight]{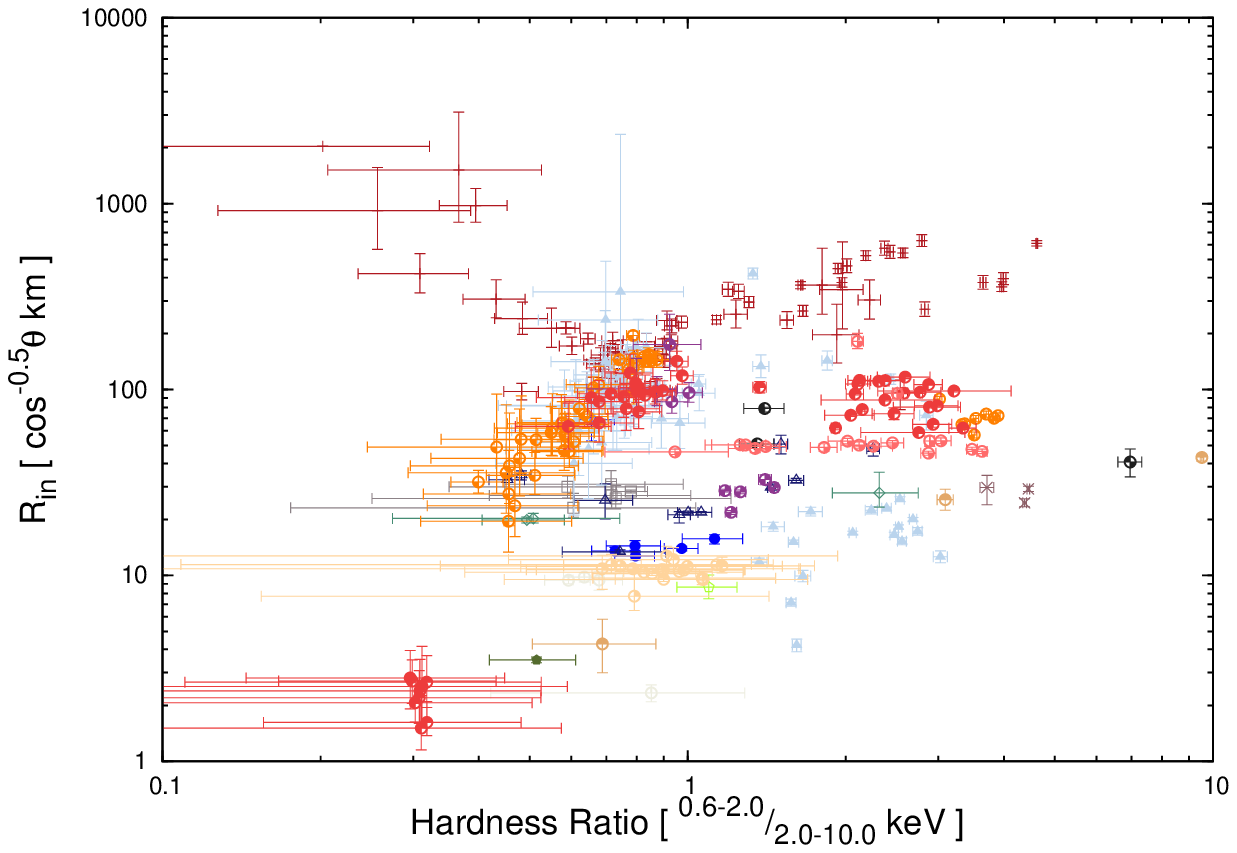}}
\subfigure{\includegraphics[height=0.27\textheight]{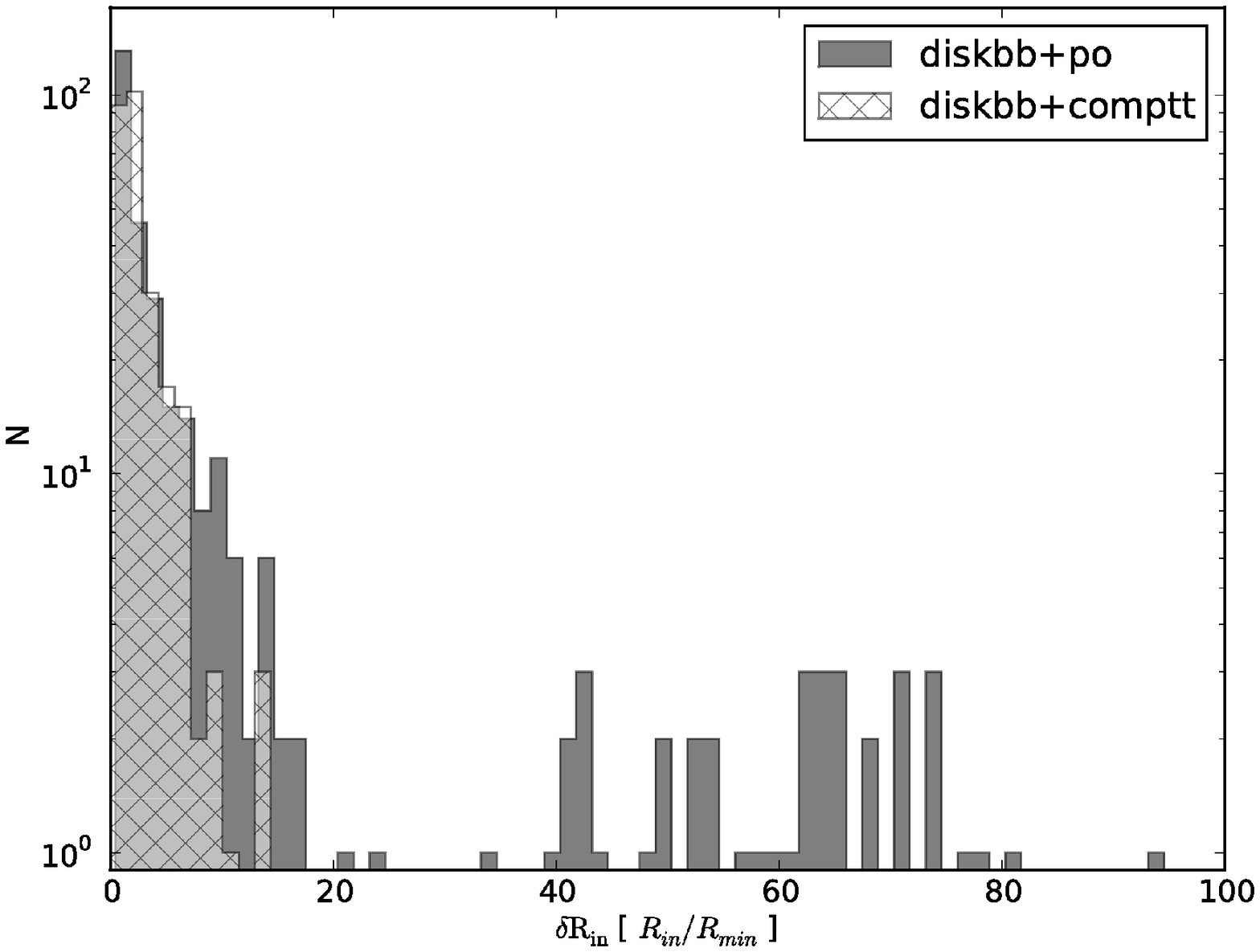}}
\vspace{-3mm}
\caption{\textbf{Left:} Normalization of the best fit accretion disk model
  (\texttt{diskbb}) versus hardness ratio, for the systems in our sample. The measured
  inner disk radii are low, consistent with $\rm R_{in} \lesssim 30~R_g$ at all
  times. The symbol/color are as defined in Fig. \ref{BH_all_systems}. \textbf{Right:}
  Histogram of the ratio of the measured inner radius of the accretion disk in each
  observation relative to the smallest radius measured in any observation for each
  system. There are a small number of systems that appear to require a large change in
  disk radius; however, modelling the hard component with a Comptonization model
  instead of a simple power-law reveals these to be a systematic artifact of using the
  power-law model (see text).}
\label{rin_histogram}
\end{center}
\end{figure*}

The behavior of the hard X-ray components and their relationship to the observed
behavior of the accretion disk was also examined, i.e., the power-law characterized by
spectral index, $\Gamma$, and the Comptonizing component characterized by the optical
depth, $\tau$, see Fig. \ref{gamma_tau_plot}. Focusing on the results of the
\texttt{diskbb+po} fit, we find from left to right (a) disk temperature: a clear
dichotomy exists where we see a population of low temperature disks consistent with
the hard state, and then a second population of hotter disks. Unfortunately due to our
relatively narrow X-ray bandpass, the power-law is not well constrained in those
spectra where the disk is dominant, i.e., $\rm T_{disk} \sim 1~ keV$. (b) The spectral
index is observed to increase (i.e., soften) as the radius of the disk increases until
a spectral index of approximately 2 is reached at which point the radius tends to be
constant. The decrease of the accretion disk radius as the spectrum hardens points
towards a changing spectral hardening factor, see \S\ref{spec_hard}iii.  (c) At the
lowest luminosities $\rm L_x \lesssim 10^{-3}~L_{Edd}$, the spectral index is observed
to remain constant, $\Gamma \sim 1.8$. At luminosities above this the spectral index
increases, though the rate at which it increases appears to vary from system to
system. Additionally, the spectral index may soften at the lowest luminosities (e.g.,
\citealt{corbel06}); however, the number of spectra at these luminosities is too few
to make any definitive statement.
 
As noted previously, the \texttt{diskbb+comptt} model exhibits similar behavior to the
\texttt{diskbb+po} model with the caveat that the optical depth, $\tau$, trends in the
opposite sense to the power-law spectral index. This effect is strikingly apparent in
Fig. \ref{gamma_tau_plot}. In the lower panels, we see that (d) the optical depth
varies significantly with high temperature disks mainly requiring optically thin
coronae whereas those low temperature disks in the hard state require higher optical
depths. In the hard state it is noticeable that while the disk temperature remains
approximately constant ($\sim$ 0.2 keV), the optical depth varies significantly
suggestive of coronally driven spectral evolution at this disk
temperature/luminosity. (e) the inner radius of the accretion disk increases as the
optical depth decreases, and (f) the optical depth of the corona decreases as the 0.6
-- 10 keV luminosity increases, i.e., as the spectrum becomes dominated by the soft
thermal emission from the accretion disk.  Similar behavior is observed when the hard
component is modeled via the \texttt{simpl} model, where the scattering fraction can
be understood as a proxy for the optical depth of the corona.

\vspace{5mm}
\noindent\textit{iii. Spectral Hardening}\label{spec_hard}\\ In the previous section,
the accretion disk emission is modelled as a standard multi-color blackbody accretion
disk using the \texttt{diskbb} model. However, the emission from the accretion disk is
not a true blackbody and instead will be modified as it scatters through the disk
atmosphere. This emission may nonetheless be approximated by a blackbody. The
temperature returned by the \texttt{diskbb} model, $\rm T_{in}$, actually represents
the temperature of this modified blackbody which approximates the emergent disk
spectrum. This temperature is related to the true effective disk temperature through
the relation
\begin{equation}
T_{col} \equiv f_cT_{eff}
\end{equation}
where $f_c$ is the spectral hardening factor \citep{shimura95}. In the standard
scenario $f_c$ is assumed to be approximately 1.7 based on the work of
\citet{shimura95}. \citet{merloni00} revisited this problem and found $\rm 1.7
\lesssim f_{col} \lesssim 3$, with the spectral hardening factor increasing as the
spectrum became less disk dominated. Further theoretical study of the behavior of
$f_c$ in the disk dominated state predicts $\rm f_{col} \sim 1.7~for~T \lesssim 1~keV$
\citep{davis05} and $\rm f_{col} \lesssim 2.2$ for disk temperatures greater than 1
keV \citep{davis06,done08}. With the exception of the study by \citet{merloni00}, each
of the other investigations focused on the soft state behavior of $f_c$.

In Fig. \ref{rin_histogram} (left), we plot the relationship between the hardness
ratio and the inner disk radius, as calculated from the best-fit \texttt{diskbb+po}
model. For the softer spectra (HR $\gtrsim$ 1), the radius is observed to be
approximately constant for each source, whereas for the harder spectra (HR $\lesssim$
1), the radius is observed to decrease. This is consistent with the behavior predicted
by the \citet{merloni00} spectral hardening model for coronally dominated spectra. In
Fig. \ref{rin_histogram} (right), we plot the histogram showing the inner disk radius
measured for each spectrum, normalised to the smallest radius measured from that
system. The relative radius change is observed to peak at $\lesssim$ 20x and 10x for
the \texttt{diskbb+po} and \texttt{diskbb+comptt} models respectively.

The apparent changes in the disk radius implied by the previous figures suggests that
spectral hardening may play an important role. In order to investigate this, the previous
continuum fits were repeated but the accretion disk model \texttt{diskbb} is replaced
with \texttt{diskpn} \citep{diskpn_gierlinski99}. In this model, the inner radius is a
free parameter (which we fix at $\rm R_{in} = 6~R_g$) and the normalization explicitly
includes the spectral hardening factor, i.e., 
\begin{equation}
norm = \frac{M_xcos(\theta)}{D^2}\frac{1}{f_c^4}
\end{equation}
The results of these spectral fits, where a constant inner radius is explicitly
assumed, are found to be consistent those using the variable radius \texttt{diskbb}
model, see Table \ref{chi2_table}. Restricting ourselves to the predicted values for
the spectral hardening factor mentioned earlier ($1.4 \leq f_c \leq 3.0$), we see that
a change in the disk normalization by a factor of 10 -- 20 is achievable through
variations in the spectral hardening alone.

\begin{figure}[t]
\begin{center}
\includegraphics[height=0.27\textheight]{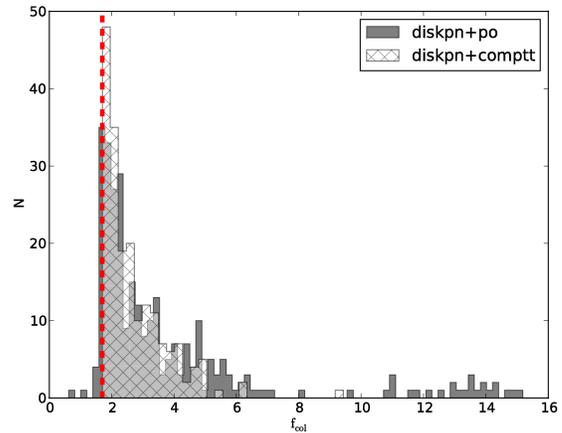}
\caption{Histogram of the color correction factor assuming 2 differing hard X-ray
  components, i.e., a power-law and Comptonization. The dashed grey line indicates a
  color correction factor of $\rm f_{col} = 1.7$.}
\label{fcol_histogram}
\end{center}
\end{figure}

Assuming that the radius variations observed in Fig. \ref{rin_histogram} are solely
due to variations in the spectral hardening factor, the \texttt{diskpn} model fits may
be used to calculate $f_c$. To do this, we assume that the minimum measured accretion
disk normalization corresponds to the canonical spectral hardening factor, i.e., $f_c$
= 1.7 \citep{gierlinski04}. This minimum normalization corresponds to an accretion
disk dominated observation, and as the disk temperature/luminosity decreases we expect
$f_c$ to increase as the continuum becomes dominated by emission from the hard X-ray
component \citep{merloni00}. As the normalization $\propto f_c^{-4}$, we scale the
other observations relative to this, hence allowing us to estimate the spectral
hardening factor for the observations in our sample. In Fig \ref{fcol_histogram}, we
plot the histogram of calculated spectral hardening factors. The vast majority of
observations are consistent with the theoretically expected bounds on $f_c$ ($1.4 \leq
f_c \leq 3.0$), with a tail of observations requiring a spectral hardening factor
outside this range, i.e., 76\% (88\%) and 60\% (73\%) of the observations require $f_c
\leq 3.0~(4.0)$ for the \texttt{comptt} and \texttt{po} models respectively.

\begin{figure*}[t]
\begin{center}
\subfigure{\includegraphics[height=0.17\textheight]{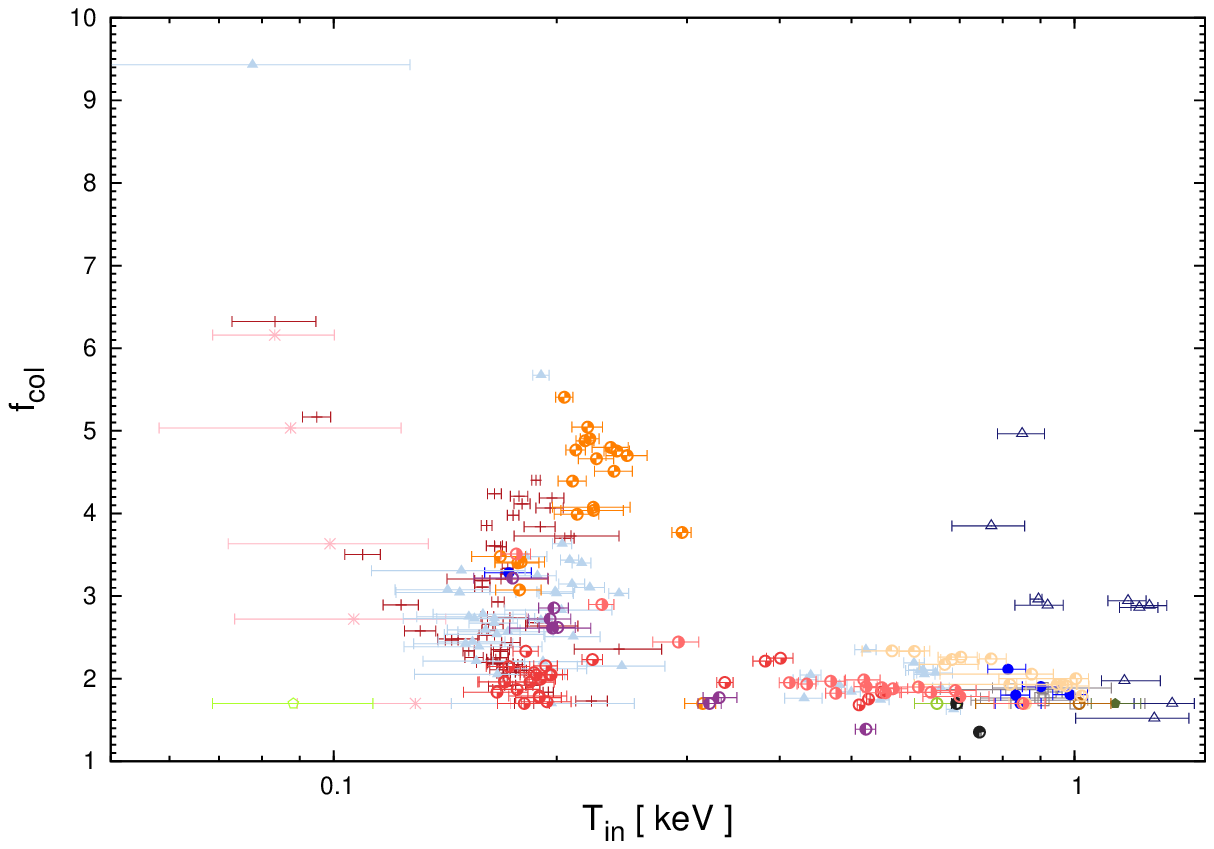}}
\subfigure{\includegraphics[height=0.17\textheight]{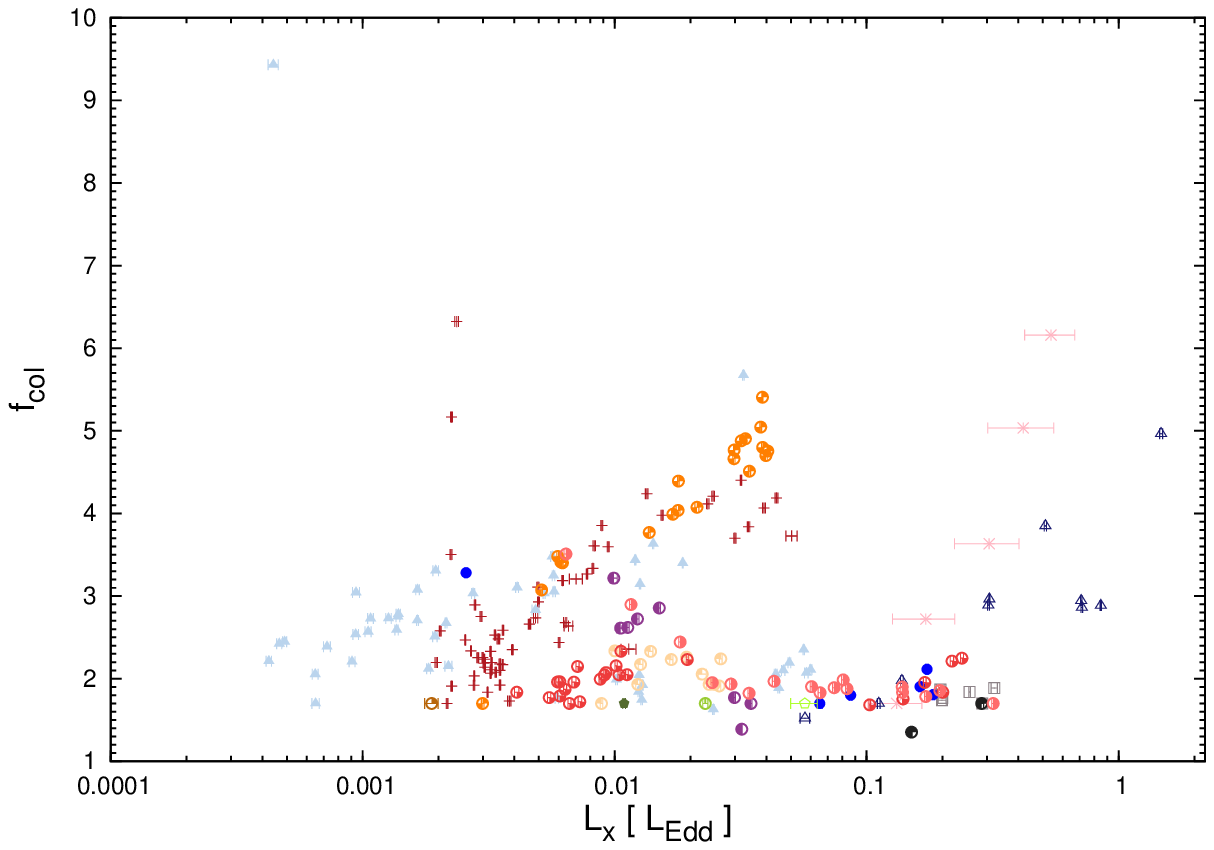}}
\subfigure{\includegraphics[height=0.17\textheight]{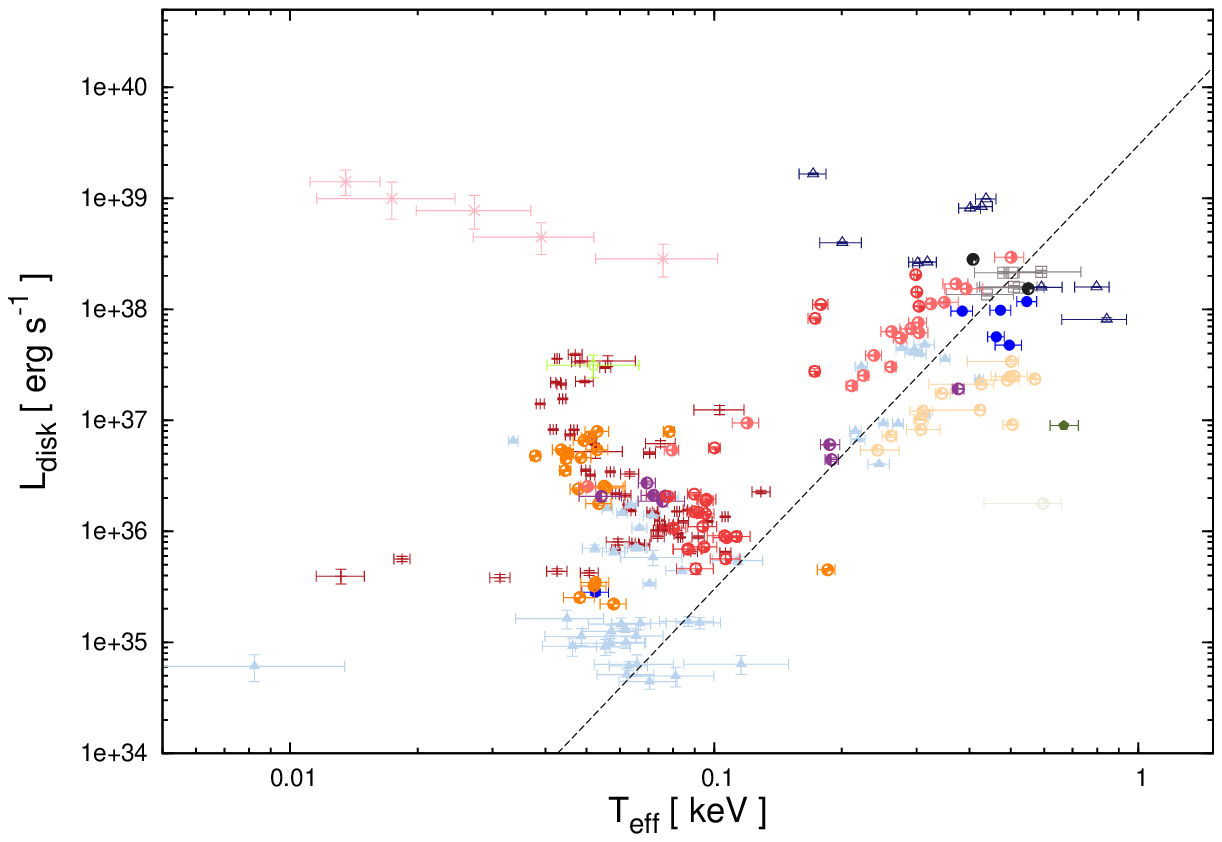}}
\caption{\textbf{Left:} Color correction factor versus disk
  temperature. \textbf{Middle:} Color correction factor versus Eddington scaled
  luminosity for the systems in our sample. The Color correction has been calculated
  via the \texttt{diskpn} model assuming a constant inner disk radius of 6 $\rm
  R_g$. \textbf{Right:} Luminosity versus accretion disk temperature $\rm T_{eff}$,
  i.e., the colour corrected temperature of the accretion disk. The solid line
  denotes $\rm L_x \propto T^4$. The symbol/color are as defined in
  Fig. \ref{BH_all_systems}.}
\label{fcol_plot}
\end{center}
\end{figure*}

In cases where the default value at $\rm norm_{min}$ of $f_c = 1.7$ is incorrect, for
example, in systems such as Swift J1753.5-0127 that remained in the hard state at all
times, a larger value of the spectral hardening factor may be warranted. This will not
affect the observed pattern, which remains approximately the same; however, the
amplitude will be reduced. Therefore our choice of $f_c = 1.7$ at $\rm norm_{min}$ is
conservative, and the derived maximum hardening factors should be considered as upper
limits in this case. In Fig. \ref{fcol_plot}, we illustrate the behavior of the
spectral hardening factor as a function of the disk temperature (left) and 0.6 -- 10
keV luminosity (middle). Finally, in the right hand panel we plot the corrected disk
luminosity versus disk effective temperature, i.e., $\rm L_x \propto T_{eff}^4$.

\subsubsection{Broadband accretion flow properties}
An important question when modelling the broadband emission from an XRB is to
determine whether the emission is direct viscous dissipation from the accretion disk,
reprocessed hard X-ray emission, or if it is a separate component entirely, i.e., a
jet. Previous work has demonstrated that irradiation of the disk is important during
the outburst of a black hole binary, e.g.,
\citet{vanparadijs94,dubus01,rykoff07}. \citet{russell06}, using a large sample of
black holes binaries, showed that the opt/nIR vs hard X-ray relationship was
consistent with either an irradiated disk or jet origin.\\

\noindent \textit{i. X-ray -- UV/optical flux relations}\\
The UVOT obtained simultaneous observations in the optical/UV spectral range (5000 \AA
-- 2000 \AA) for over half of our sample in at least a single filter (V, B, U, UVW1,
UVM2, UVW2). The lightcurves are morphologically similar to those observed at X-ray
wavelengths. Unfortunately due to the typical \textit{Swift} daily observing cadence,
and exposure time of $\sim$ 100s -- 1000s, we are unable to search for correlated
delays of the optical/UV emission with respect to that observed at X-ray
energies. Such features can be used to probe the response of the outer to the inner
flow, e.g., \citet{gandhi08}. Broadly characterizing the opt/UV photometry, we measure
the ratio of the flux at X-ray (0.6 -- 10 keV) to optical/UV-wavelengths ($\rm
f_{\nu}$) to be $f_{opt}/f_x \sim 10^{-5} - 10^{-6}$ in the V-band. The UV flux ratio
is similar, though we caution that the red leak in the UVW1 filter \citep{breeveld11}
may prohibit us from constraining the true UV flux.

\begin{figure*}[t]
\begin{center}
\subfigure{\includegraphics[height=0.25\textheight]{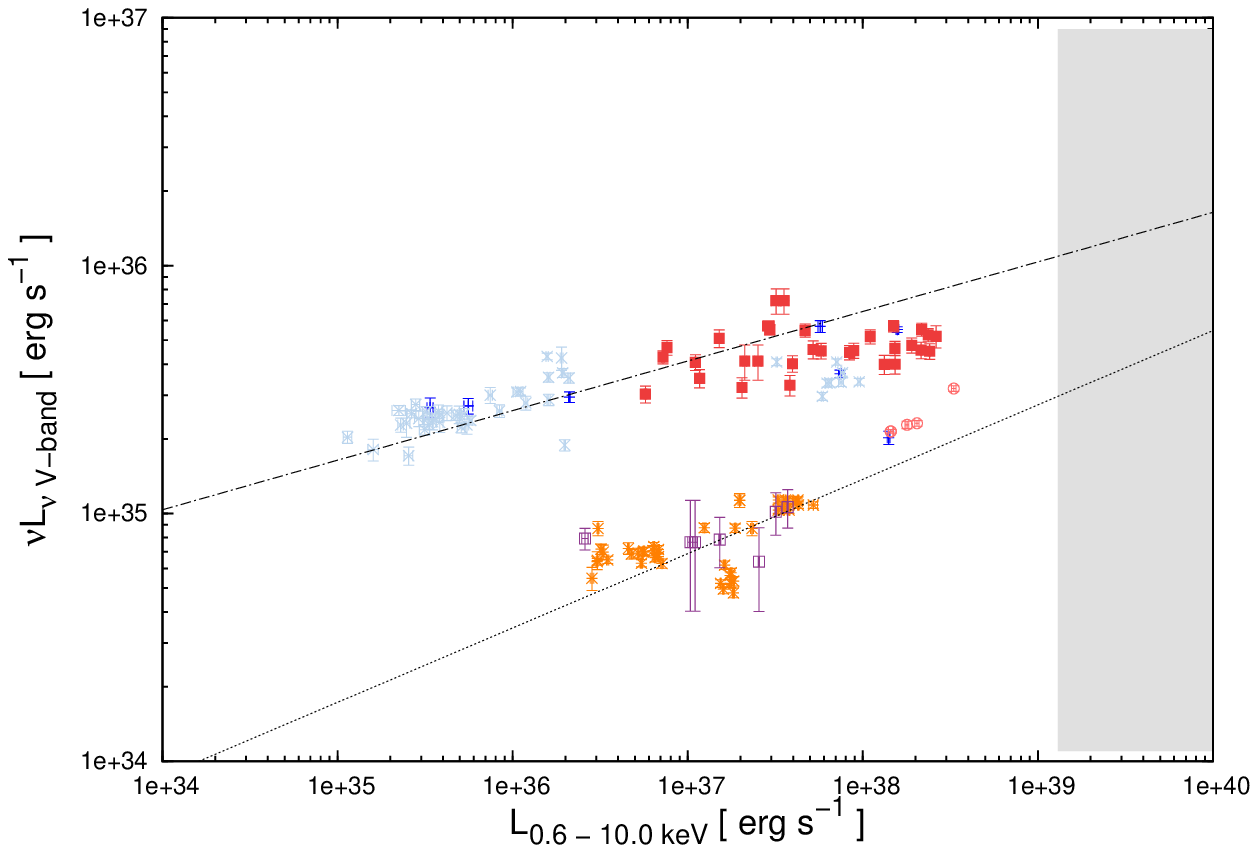}}
\subfigure{\includegraphics[height=0.25\textheight]{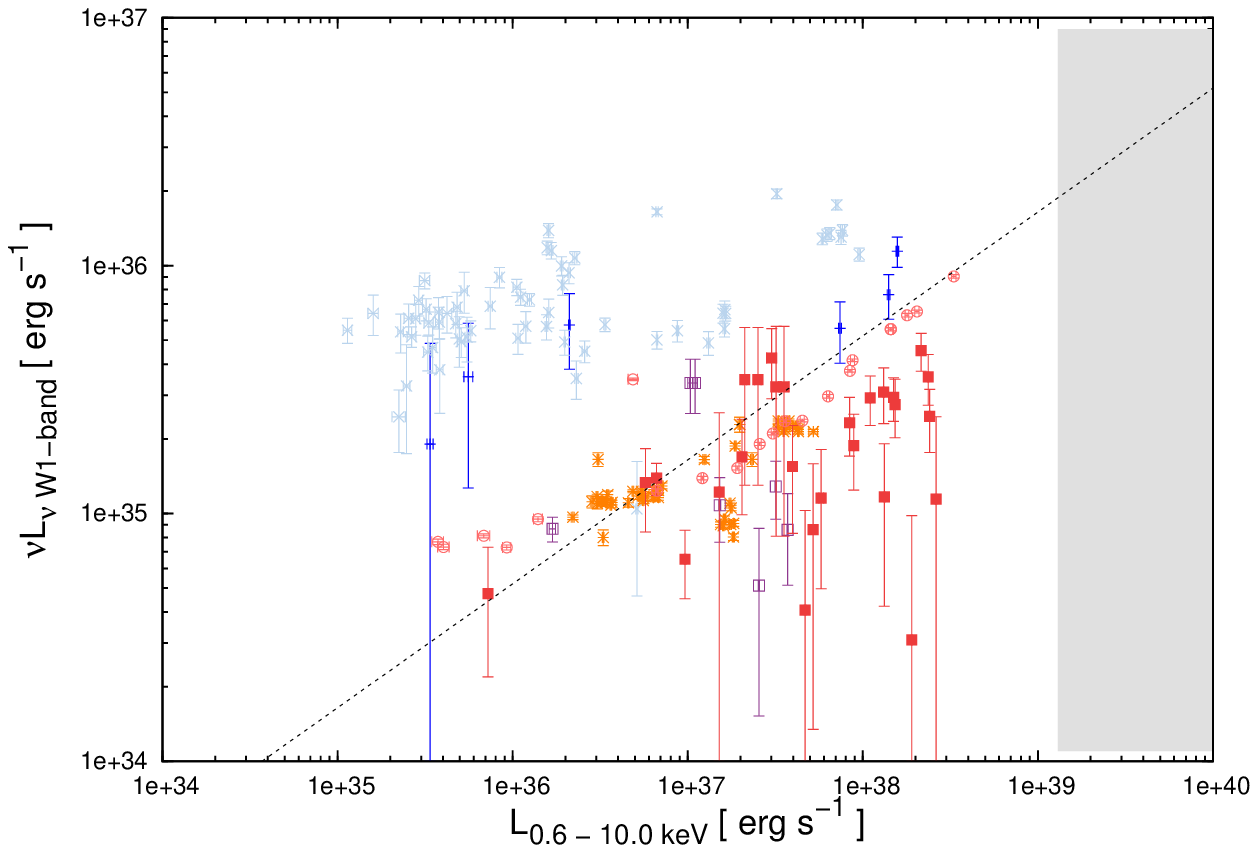}}
\caption{\textbf{Left:} Extinction corrected V-band versus X-ray flux. The optical
  data is consistent with expectations for viscous emission from the accretion disk
  ($\rm L_{V-band} \propto L_x^{0.2 - 0.3}$), indicated by the dot-dashed and dotted
  lines respectively. \textbf{Right:} Extinction corrected UVW1-band flux versus X-ray
  flux. The UV emission is displays evidence for irradiated disk emission ($\rm
  L_{V-band} \propto L_x^{0.5}$) at higher X-ray fluxes, as indicated by the dashed
  line. The shaded region denotes the Eddington limit for a 10 $\rm M_{\sun}$ black
  hole. The symbol/color are as defined in Fig. \ref{BH_all_systems}.}
\label{optuv_vs_xray_flux}
\end{center}
\end{figure*}

The relationship between the flux observed in the optical/UV versus that in the X-ray
can be used to constrain the physical processes contributing to the observed spectrum,
for example, a correlation of the form $\rm L_{opt} \propto L_x^{0.5}$ is expected
from the reprocessing of X-rays by the accretion disk \citep{vanparadijs94}.  In
Fig. \ref{optuv_vs_xray_flux}, we plot the relationship between the total unabsorbed
X-ray luminosity in the 0.6 -- 10 keV band and that detected in the optical/UV
bandpasses. The differing number of points in each plot reflects the differing
observing strategies undertaken for each system, i.e., one or multiple UVOT
filters. The optical emission is observed to be consistent with viscous emission from
the accretion disk ($\rm L_{opt/UV} \propto L_{X}^{0.2 - 0.3}$, e.g., see
\citealt{russell06}), while in the UV we find evidence for both viscous emission and
irradiated emission from the accretion disk ($\rm L_{opt/UV} \propto L_{X}^{0.5}$). In
the case of GX 339-4 (lightblue), we find an approximately flat relation between the
UV and the X-ray fluxes, particularly apparent for luminosities less than $\rm \sim
10^{37}~erg~s^{-1}$. This hints at the presence of an additional source of flux during
these observations, which we note are primarily from when the X-ray spectrum is
power-law dominated, i.e., in the low-hard state. In the optical, two distinct tracks
are observed likely due the nature of the observed emission (purely viscous or
contaminated) and to the varying characteristics of each system (e.g., disk size $\rm
\propto P_{orb}$).

Neither the optical or the UV flux is consistent with a significant contribution from
the jet, which would be expected to produce a relation $\rm L_{opt/UV} \propto
L_{X}^{0.7}$ \citep{russell06}. Previous observations of a number of black hole
binaries have revealed convincing evidence for the presence of a significant
contribution from the jet at nIR wavelengths, with a smaller contribution also in the
optical, e.g., \citet{coriat09,russell10}. However the analysis presented herein is
necessarily limited by the bandpass of our observations, i.e., $\lambda_{min} \approx
5000$ \AA, which is at a wavelength shorter than that at which the jet component has
been observed to dominate.

\vspace{5mm}
\noindent \textit{ii. Broadband spectroscopy}\label{bband_spec}\\ In order to
investigate the relations observed above in more detail, the UVOT data were converted
into \textsc{xspec} readable files and both the optical/UV and the X-ray data were fit
simultaneously.  To investigate the origin of the UV emission, three models were
considered (i) a standard steady state disk plus a power-law component, i.e.,
\texttt{diskbb+po}, (ii) variable temperature profile accretion disk plus a power-law,
i.e., \texttt{diskpbb+po}, and (iii) variable temperature profile accretion disk plus
Comptonization, i.e., \texttt{diskpbb+comptt}. In all cases, we utilized
\texttt{phabs} for the X-ray column \& \texttt{redden} for the optical/UV column
assuming the standard Galactic extinction curve \citep{cardelli89} and the dust to gas
ratio from \citet{predehl95}. The column density was fixed at the standard X-ray value
(Table \ref{BHC_table}, see appendix A\ref{appendix_param} for a discussion of the
uncertainties in the column density).  In those observation where we had multiple
opt/UV datapoints, it is possible to constrain the column density. The typical best
fit values were found to be consistent with the assumed X-ray column density within
the errors in all cases.

The results of the X-ray plus UV spectral fits may be summarised as follows: Model (i)
\texttt{diskbb+po}: 267 broadband spectra in total, and 196 of these are found to
require an accretion disk at greater than the 5$\sigma$ confidence level. The number
of broadband spectra which do not require a disk component are 71, 60 and 46 at the
5$\sigma$, 4$\sigma$, and 3$\sigma$ levels respectively. Model (ii)
\texttt{diskpbb+po}: 264 broadband spectra in total. Model (iii)
\texttt{diskpbb+comptt}: 257 broadband spectra in total. A similar number of models as
in (i) above were found not to require an accretion disk for model (ii). These spectra
are discussed in detail later. A disk component is required in all of the
\texttt{diskpbb+comptt} models as the corona cannot provide significant flux below the
temperature of the input seed photons. When considering the quality of the model fits
to the sample as a whole, they are found to be comparable, i.e., as measured by the
respective $\chi^2_{\nu}$ distributions. However, the physical differences in the
model components result in differing contributions from the accretion disk relative to
the power-law, for example, as a function of wavelength. In the appendix
A\ref{appendix_bband}, we examine the UV spectral models via a detailed look at a
hard \& soft state observation of the transient system GX 339-4. 

The basic properties of the accretion disks ($\rm T_{col},~r_{col}$) in these fits are
consistent with those measured in fits made to the X-ray data alone. However, it is
important to note that it is impossible to constrain the temperature profile using the
X-ray data alone, as the data will invariably be consistent with a steady state disk
within the statistical errors, e.g., \citet{kubota05}. For the variable temperature
profile model ($\rm T(r) \propto r^{-p}$) the distribution of the index differs
between the Comptonization and power-law models reflecting the contribution by the
hard component, in the power-law model case, to the optical/UV emission. The average
temperature profile ($\pm1\sigma$) we measure is $p \sim 0.59\pm0.06$ for the
\texttt{diskpbb+comptt} model and $p \sim 0.62\pm0.09$ for the \texttt{diskpbb+po}
model (see Fig. \ref{histogram_tprofile}), though in both cases we are biased by the
large number of hard state observations in our sample as may be seen in
Fig. \ref{disk_temp_histogram}. We emphasise, that irradiated disks are seen at all
luminosities, with hard state observations typically requiring a more irradiated disk
in comparison to soft state observations.

There is no evidence for a dominant jet contribution at UV wavelengths. The power-law
component is not dominant for any of the models in the hard state, and while it can
dominate in the case of the \texttt{diskbb+po} model in the soft state, we do not
consider this model to be valid, i.e., the power-law index would have to
be much harder than is typically observed in a soft state (e.g., see
Fig. \ref{gx339_bband_hss} \& \ref{gx339_bband_lhs} in the appendix). This is
consistent with the known absence of detectable jet emission in the soft state
\citep{fender06,russell11}. In all other cases the emission from the irradiated
accretion disk dominates at UV wavelengths at all times. Though it is important to
emphasise the fundamental difference implied by the two hard component choices. When
the hard X-ray emission is most successfully modelled by a Comptonization component
(i.e., soft state), then the UV emission is dominated by the accretion disc. In
contrast, during harder states, where both the Comptonization and power-law components
provide fits of comparable statistical quality, two different physical scenarios are
suggested. For the Comptonization model, again all of the UV emission originates in
the irradiated accretion disk, whereas for the power-law model the UV is dominated by
emission from the irradiated accretion disk \textit{but} with a contribution from the
power-law component that may be as large as 10\% of the total continuum flux in the
UV. Such a model may be consistent with timing observations that have found evidence
for non-thermal flux at UV wavelengths, e.g., \citet{gandhi08}. Ideally we would have
had access to simultaneous nIR observations that would allow us to break this
degeneracy and accurately constrain the possible power-law (jet) contribution at
optical/UV/X-ray wavelengths, e.g., \citet{russell10}.

As stated earlier, there are a number of the broadband spectra that are consistent
with a power-law alone, i.e, the addition of a disk component does not significantly
improve the fit. In the case of the \texttt{diskpbb+po} model 50/264 of the spectra
are consistent with a power-law alone when we only mark an accretion disk as detected
if it improves the fit statistic by greater than the 5$\sigma$ level as measured by an
\texttt{ftest}. If instead we lower the criteria to a 3$\sigma$ improvement only 28
spectra are consistent with a power-law alone. The number of spectra consistent with a
power-law alone is larger for the \texttt{diskbb+po} model; however, the excess
relative to the \texttt{diskpbb} model is due to the inability of a standard steady
state accretion disk to adequately reproduce the observed data. 

To examine these power-law spectra in more detail, we restrict ourselves to spectra
with greater than 1000 counts and a best fit $\chi^2_{\nu} \leq 1.5$, leaving us with
24 spectra distributed across 6 systems (GRO J1655-40, GX 339-4, SWIFT J1753.5-1027,
SWIFT J1842.5-1124, XTE J1752-223, XTE J1818-245). These systems broadly represent the
contents of our sample, with no obvious biases towards system parameters, e.g., $\rm
P_{orb}$. Though our knowledge of these parameters is limited in most of these
systems, their observed behavior to date provides no evidence of any fundamental
difference in comparison to the well known systems, i.e., GRO J1655-40.  The average
spectral index for this sample is found to be $\Gamma = 1.66 \pm 0.06~(1\sigma)$,
while the average luminosity is $\rm (1.5\pm0.7)\times10^{-3}~L_{Edd}(1\sigma)$, both
consistent with expectations for emission from a black hole in the low-hard state. The
absence of a detectable accretion disk in these spectra may point toward a truncated
accretion disk or conversely a disk that has cooled below our detection limit. An
alternative possibility, suggests that these spectra are dominated by emission form
the broadband jet, e.g., \citet{markoff01}.

\section{Discussion}
We have presented a comprehensive systematic analysis of all of the available
\textit{Swift} data on accreting stellar mass black hole binaries as of summer
2010. This is the most detailed study of accretion onto stellar mass black holes at
luminosities of $\rm \gtrsim 10^{-3}~L_{Edd}$. To summarize:
\begin{itemize}
\item{We analysed 476 CCD spectra of 27 systems, in the 0.6 -- 10 keV
  band. Approximately half of these had simultaneous coverage in the optical/UV band.}
\item{An accretion disk is required to successfully model the data in over half of the
  X-ray observations. The excellent sensitivity at low energies allows us to follow
  the evolution of the accretion disk to temperatures $\sim$ 0.2 keV. This is the
  largest sample of such cool accretion disks collected to date.}
\item{The observed accretion disk components are consistent with maintaining inner
  radii $\rm \lesssim 40~R_g$ at all times. More severe disk truncation must take
  place at luminosities $\rm \lesssim 10^{-3}~L_{Edd}$.}
\item{The evolution of the inner disk radius as a function of spectral hardness, i.e.,
  a decreasing radius as the source hardens, suggests a contribution from spectral
  hardening to the observed disk radius variations.}
\item{The simultaneous X-ray plus optical/UV data allow us to constrain the
  temperature profile of the accretion disk for a large sample of
  observations. Irradiation of the accretion disk is important at all times, likely
  even at X-ray wavelengths, i.e., $\rm T(r) \propto r^p,~p \neq 0.75$.}
\item{In the soft state, the broadband spectra are consistent with a spectrum
  dominated by emission from the accretion disk and an additional hard component
  consistent with Compton scattering of seed disk photons.}
\item{While an irradiated disk is preferred in the hard state, the data are unable to
  statistically distinguish between power-law and Comptonization models for the hard
  flux.}
\item{The relationship between the flux emitted by the accretion disk and that emitted
  by the corona is in broad agreement with the observed behavior in Seyfert galaxies,
  suggesting a scale invariant coupling between the accretion disk and the corona.}
\end{itemize}
We discuss these results and their implications in detail below.

\subsection{Accretion disk truncation}\label{disk_trunc_discuss}
The largest radii measured in our sample are $\sim$ 40 $\rm R_g$, which rules out any
large scale truncation of the accretion disk, for the luminosities probed in this
survey. The observed distribution of disk temperature and radius for those disks
detected at a luminosity $\rm \leq 0.01~L_{Edd}$ serves to emphasise this point
(Fig. \ref{disk_temp_histogram}).  

There is currently no conclusive observational evidence that the accretion disk is
truncated in the hard state, at least at the luminosities probed in this work ($\rm
L_X \gtrsim 10^{-3}~L_{Edd}$). This has long been known to be the case at the higher
luminosities observed in the soft state, where the inner edge of the accretion disk is
expected to lie at the ISCO, e.g., \citet{done03,gierlinski04}. It is this constancy
of radius that allows one to constrain the black hole spin via the accretion disk
continuum emission, e.g., \citet{steiner10}. In recent years evidence that the
accretion disk is not truncated in the hard state has also been accumulating, e.g.,
\citet{miller06a,reis10}. However, we acknowledge that the disk inner radius
distribution measured herein can be interpreted as evidence for small scale truncation
of the inner disk at lower luminosities. Clearly, further observational/theoretical
tests are required to break this degeneracy.

If the accretion disk does not truncate (e.g., \citealt{belo99,markoff01,merloni02}),
then we must re-consider the driving mechanism behind state transitions, in particular
the hard and soft states, and how this is related to the production of a relativistic
jet. The most promising means for powering these relativistic jets is the
`Blandford-Znajek' mechanism, where the luminosity of the jet is $\rm L_{BZ} \propto
B*r_g^2*f(a/M)$ \citep{bz77}. If the inner disk is present at the ISCO at all times as
the above data implies, then this suggests that the accretion disk itself is not
required to launch the jet. Indeed simulations have demonstrated the formation of jets
in the absence of a disk. For example, \citet{barkov11} analyze the formation of a jet
launched from a black hole accreting from the stellar wind of a companion star. 

The driving mechanism behind the state transitions is less clear, though it is clear
that a more detailed understanding of the evolution of, and interaction between, the
hard component and the accretion disk is key. The evidence presented herein points
towards a non-varying disk geometry during the hard to soft state transition, as such,
attention must turn to the corona and its coupling to the accretion disk as a function
of mass accretion rate.

\begin{figure}[t]
\begin{center}
\includegraphics[height=0.27\textheight]{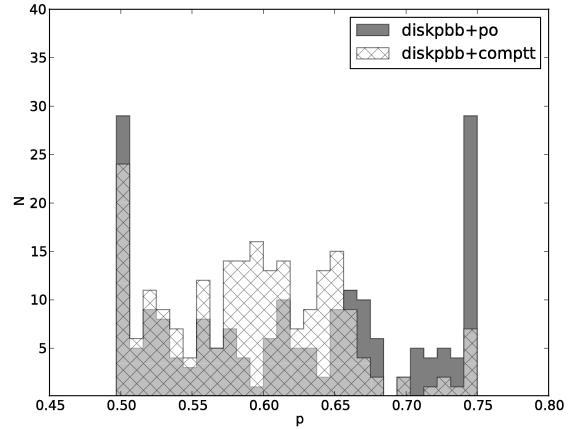}
\caption{Histogram of the best fit accretion disk temperature profile for the
  \texttt{diskpbb} model, i.e., $\rm T \propto r^{-p}$. An irradiated temperature
  profile is required in a large majority of the observations.}
\label{histogram_tprofile}
\end{center}
\end{figure}

\subsection{Spectral hardening}\label{spec_hard_discuss}
As discussed in \S\ref{spec_hard}, the accretion disk temperature measured from the
observed X-ray spectrum is not the true effective temperature of the disk, but is
instead the temperature of the disk modified as the emitted disk photons pass through
the disk atmosphere. This effect must be accounted for if we are to correctly
interpret the data. This is done via the spectral hardening factor, $\rm f_c$, i.e.,
$\rm T_{col} \equiv f_cT_{eff}$. Theory predicts the correction factor to be in the
range $\rm 1.4 \leq f_c \leq 3.0$ \citep{shimura95,merloni00,davis05,davis06}.
In Fig. \ref{rin_histogram}, it is clear that the behavior of the apparent disk radius
is consistent with spectral hardening: the apparent radius decreases as the source
hardens (e.g., \citealt{merloni00}).  This runs counter to the prediction of ADAF
models, which suggest that the disk should gradually recede in the hard state.
Indeed, fits with the `\texttt{diskpn}' model explicitly show that the spectra are
consistent with a constant radius while the correction factor changes in the manner
predicted by \citet{merloni00}.  This finding does not rule out ADAF models in a broad
sense; disk truncation at much lower fractions of the Eddington limit is expected and
may be partially confirmed by recent data, e.g., \citet{tomsick09} on GX 339-4.

For the first time, we present a large sample of empirically determined color
correction factors for a large sample of systems, see Fig. \ref{fcol_histogram} \&
\ref{fcol_plot}. The color correction factor is found to be consistent with the
canonical value ($\rm f_c \sim 1.7$) for the majority of our observations, with a tail
extending to larger values, i.e., 76\% (88\%) and 60\% (73\%) of the observations
require $f_c \leq 3.0~(4.0)$ for the \texttt{comptt} and \texttt{po} models
respectively. When the hard X-ray component is modelled using the \texttt{comptt}
model, the maximum value of the color correction factor is $\rm f_{c, max}\sim$ 5.
Figure \ref{fcol_plot}c plots the corrected disk temperature as a function of
luminosity. At larger temperatures (larger disk fractions) the standard $\rm L_x
\propto T_{eff}^4$ is recovered; however, as we move toward lower disk temperatures
(dominant hard component) a deviation is observed. We have demonstrated that a
significant fraction of this deviation can be accounted for with plausible variations
in the spectral hardening factor alone. Though the high $\rm f_c$ tail present in
Fig. \ref{fcol_histogram} does leave room for an additional process, i.e., is the disk
finally beginning to recede as we approach these luminosities ($\rm L_x \lesssim
10^{-3}~L_{Edd}$, also see \citealt{done07,tomsick09}) or are the limitations of our
spectral models, in the presence of a dominant hard component, being exposed, see
below.

Spectral hardening appears to contribute to the observed disk properties at
\textit{both} high and low disk temperatures/luminosities. At the highest
luminosities, we are likely beginning to see the re-emergence of the hard component as
the system moves to the very high state where in contrast to the LHS the disk is now
hot, i.e., $\rm T_{in} \gtrsim 0.7~keV$ \citep{mcclintockremillard06}. This is clearly
seen in the study of accreting black holes by \citet{dunn10a,dunn10b}, which due to
the low energy cut-off of \textit{RXTE} ($\rm \sim 3 keV$) was necessarily limited to
studying the brighter accretion phases. At low luminosities, a dominant power-law
component modifies the emitted disk spectrum.  In both cases, the color correction
factor naturally increases as the non-thermal emission begins to dominate (e.g.,
\citealt{merloni00,davis05}).

At the lowest measured disk temperatures, we begin to approach the lower energy limit
of the XRT used herein, i.e., 0.6 keV (remember that the disk spectrum peaks at $\rm
\sim 3kT_{in}$). In the limit of low S/N some of the low temperature disks we detect
could be systematics caused by an excess of flux in some bins at the low energy end of
our bandpass, which may mimic a low temperature disk component. Our decision to
categorize a disk as detected \textit{only} if it improves the fit by greater than
5$\sigma$ as measured by an \texttt{ftest}, should eliminate such errors in this
sample of low temperature disks. This has been verified by re-fitting a sample of the
data assuming a lower energy bound of 0.4, 0.5 keV respectively, where we find disk
components entirely consistent with those measured assuming a lower bound of 0.6 keV.
Similarly, we have no reason to believe there is an issue with the XRT calibration;
nonetheless, this is an area worthy of further detailed consideration, which we defer
to a future study.

Previous theoretical efforts to model the behavior of the spectral hardening factor
have primarily been focused on brighter more disk dominated states
\citep{shimura95,merloni00,davis05}. The study by \citet{merloni00} is the only one to
consider a situation in which the disk does not dominate the X-ray emission. These
simulations pointed towards an increase in $\rm f_c$ as the disk fraction decreased,
in agreement with the observed behavior of the accretion disks studied herein (see
Fig. \ref{rin_histogram}). However significant issues remained, including the use of a
number of simplifying assumptions regarding the structure of the accretion disk, and
the failure to account for the effect of the returning coronal radiation on the disk
itself \citep{merloni00}.

Subsequent work clearly illustrated how the assumption of a constant density for the
vertical structure of the accretion disk is inappropriate for a cold gas pressure
dominated disk in the presence of a powerful corona \citep{davis05}. While the study
of \citet{davis05} included a detailed treatment of the vertical structure of the
accretion disk in addition to opacity related effects, it focused on soft-state
accretion disks, i.e., kT $\gtrsim$ 0.5 keV, and as such, ignored the effect of the
returning radiation from the corona on the spectrum from the accretion disk, which
should be minimal in this regime. This reflected radiation can significantly modify
the observed spectral shape \citep{ross93,ross07} and dominate the spectrum in X-ray
bandpass, e.g., see \citet{miller12} for a detailed study of Cyg X-1. Finally, we note
that even in the limit of a disk dominated spectrum such as that observed from LMC
X-3, significant model uncertainties remain \citep{kubota10}.

Given the importance of this issue, more theoretical investigations, particularly in
states where the hard component is dominant over the emission from the disk, are
clearly warranted. The empirical distribution of the color correction factor provided
herein (see Fig. \ref{fcol_histogram} \& \ref{fcol_plot}) will provide a valuable
observational constraint for these studies.

\subsection{Accretion disk irradiation}
In this work, we have utilized the combination of simultaneous X-ray and UV/optical
data provided by \textit{Swift} to investigate the irradiation phenomenon in black
holes binaries.  In soft state observation, we find evidence for a mildly irradiated
disk, i.e., $\rm T(r) \propto r^{-0.6}$, whereas hard-state observations are
consistent with an increasingly irradiated disk, i.e., $\rm T(r) \propto r^{-0.5}$.
The change in the temperature profile relative to the soft state is consistent with an
increase in the irradiating hard X-ray flux, as expected in the hard state.  In
Fig. \ref{histogram_tprofile}, we plot the distribution of accretion disk temperature
profiles obtained from the \texttt{diskpbb} broadband spectral fits.  There is a
strong preference for a non-standard temperature profile irrespective of how the hard
spectral component is modelled, i.e., \texttt{po} or \texttt{comptt}. In the appendix
(A\ref{appendix_bband}), we discuss specific examples of broadband fits to an
observation of GX 339-4 in both the hard and soft states.

In the hard state both the \texttt{comptt}/corona and \texttt{po}/jet provide
acceptable fits to the data.  Fundamentally, we are unable to differentiate between
these 2 physical scenarios with the available data, though appealing to previous
timing observations (e.g., \citet{gandhi08}) leads us to favor the power-law model. In
addition, correlated multi-wavelength studies (radio -- X-ray) have revealed clear
evidence for a jet in the hard state \citep{fender06,coriat09,russell10}; however,
these studies suggest that it is unlikely to contribute significantly at UV
wavelengths, consistent with the analysis herein. The soft state spectra appear to
rule out a power-law origin for the observed hard X-ray component as this results in a
much harder spectral index than is normally measured in the soft state. This is in
agreement with the well known behavior of relativistic jets, whereby they are observed
to be present in the hard state, but appear to be absent in the soft state
\citep{fender06}.

In light of these results, it may be necessary to re-consider the effect of the
assumed temperature profile on the measured black hole spin, as even though the X-ray
spectrum may be consistent with a standard steady state disk, when one considers the
broadband spectrum deviations from this profile become apparent. This may materially
impact the derived disk parameters.  For example, a single observation of the black
hole binary 4U 1957+11 is included in this study. When the broadband data is fit with
the variable temperature profile disk model, the best fit temperature profile is
consistent with an irradiated accretion disk with a lower disk temperature in
comparison to the steady state case. Recently, \citet{nowak11} have measured the spin
of the black hole in this system via the continuum method using \textit{Suzaku}
spectra, where the spin was consistent with a near maximal value. Clearly, failure to
correctly account for disk irradiation could bias the measured spin. We defer detailed
consideration of this to a future study.

\section{Conclusions}
We have presented a comprehensive analysis of all of the data on stellar mass black
binaries in the \textit{Swift} public archive as of June 2010. This is the largest
collection of CCD spectra of accreting stellar mass black holes published to date,
amounting to 476 spectra of 27 binary systems. The excellent sensitivity and broad
soft X-ray bandpass of \textit{Swift} allows us to study the evolution of the
accretion disk down to temperatures as low as $\sim$ 0.2 keV for the first
time. Approximately half of the spectra are found to require the addition of an
accretion disk component at greater than the 5$\sigma$. 

There is no compelling evidence in the data presented herein, which samples the
accretion disk evolution down to luminosities of $\rm \sim 10^{-3}~L_{Edd}$, for a
large scale change in the inner disk radius during the transition from the hard to the
soft state, i.e., $\rm R_{in} \lesssim 40~R_g$ at all times. Consideration of the
evolution of the inner radius versus spectral hardness points towards spectral
hardening being a factor in the observed disk radius evolution.  Clearly, future
observations probing luminosities below $\rm 10^{-3}~L_{Edd}$ are necessary to provide
further constraints on the evolution of the accretion flow geometry.

Irradiation is found to be present in the majority of observed spectra where
simultaneous optical/UV and X-ray data are available. This irradiated disk component
is found to dominate the optical/UV emission in both hard and soft states. The soft
state spectrum is found to be consistent with a combination of an irradiated accretion
disk plus corona, whereas for the hard state a significant contribution from a
power-law component (i.e., a jet) cannot be ruled out at optical/UV energies.

We find the relationship between the flux emitted by the accretion disk and that
emitted by the corona observed herein is in broad agreement with the observed behavior
in Seyfert galaxies, suggesting a scale invariant coupling between the accretion disk
and the corona. Though, further study is warranted.

The data and analysis presented herein clearly demonstrates the value of the broadband
observational capabilities of the \textit{Swift} observatory, in the study of
accretion physics.  An analysis of the neutron star XRBs observed by \textit{Swift},
and a comparison with the black hole binary systems presented herein will be presented
in a companion paper (Reynolds et al., 2012 in prep.).  Finally, we would like to
re-emphasise the importance of the broadband coverage provided by \textit{Swift} ($\rm
\sim 1eV - 10~keV$) in constraining the properties of the observed accretion disk
emission, which is key if we are to understand the relationship between the accretion
disk inflow and the jet outflow.

\acknowledgements It is a pleasure to acknowledge the excellent contribution made to
the study of accretion physics by the \textit{Swift} observatory, we extend our thanks
to all those on the \textit{Swift} team who made this possible. We thank the referee
Andrea Merloni for an insightful report, which improved the content and clarity of
this paper. We also thank Rubens Reis, Dipankar Maitra and David Russell for useful
discussions. We acknowledge the use of public data from the Swift data archive. This
research made extensive use of data obtained from the High Energy Astrophysics Science
Archive Research Center (HEASARC), provided by NASA's Goddard Space Flight Center, of
the \textit{SIMBAD} database, operated at CDS, Strasbourg, France and NASA's
Astrophysics Data System.


\newpage
\begin{appendix}

\begin{figure*}[t]
\begin{center}
\subfigcapskip=-5mm
\subfigure[Cyg X-1: \texttt{diskbb+po}]{\includegraphics[height=0.26\textheight]{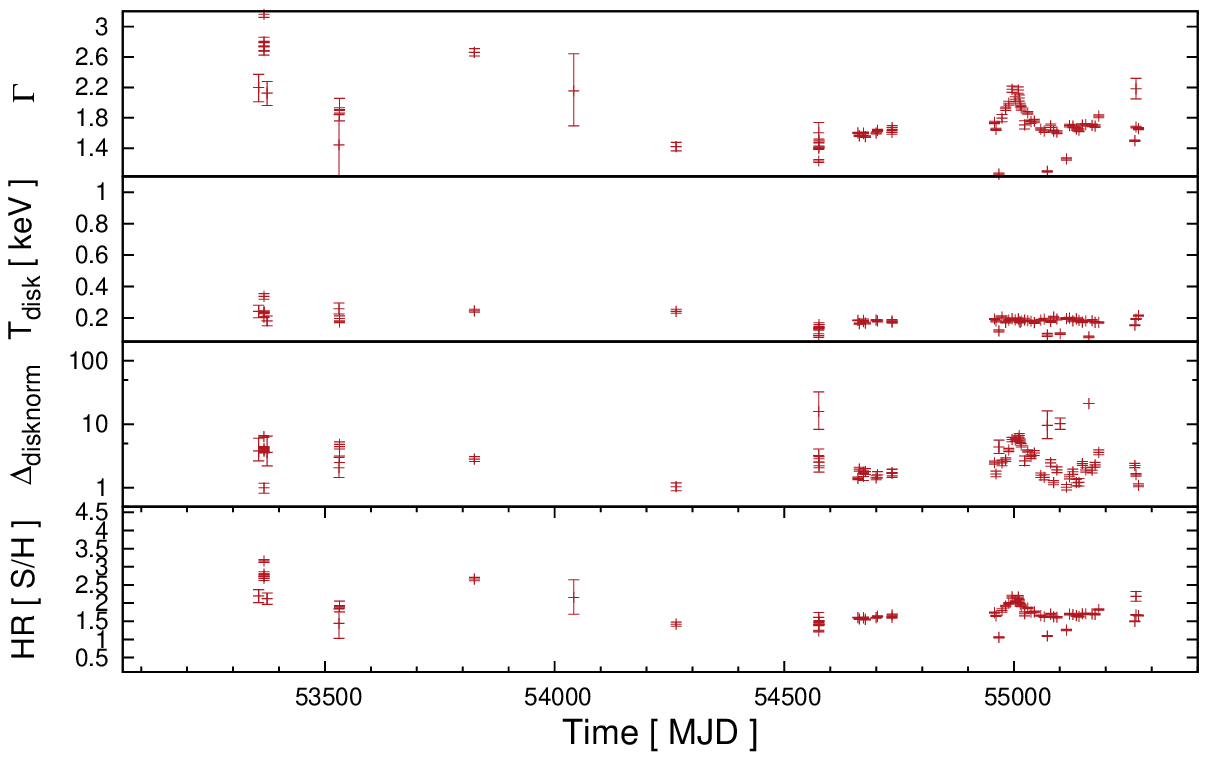}}
\subfigure[GX 339-4: \texttt{diskbb+po}]{\includegraphics[height=0.26\textheight]{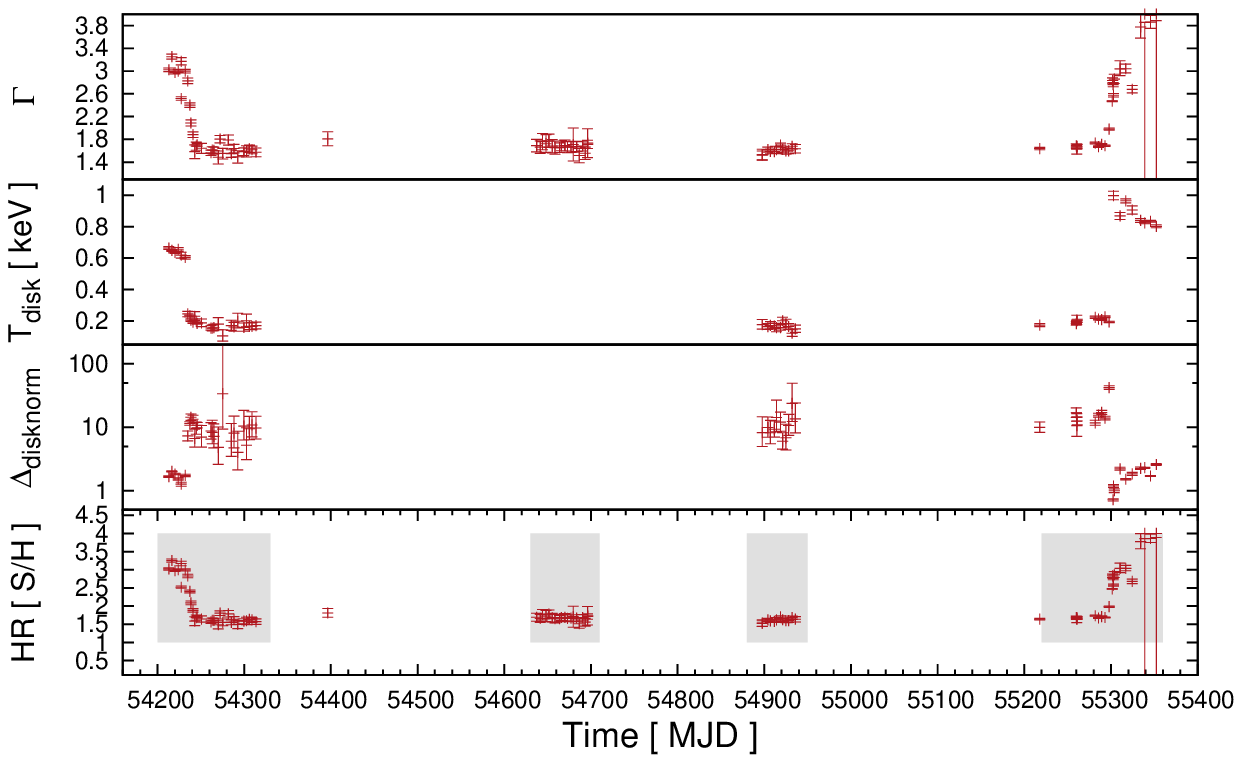}}
\subfigure[Cyg X-1: \texttt{diskbb+comptt}]{\includegraphics[height=0.26\textheight]{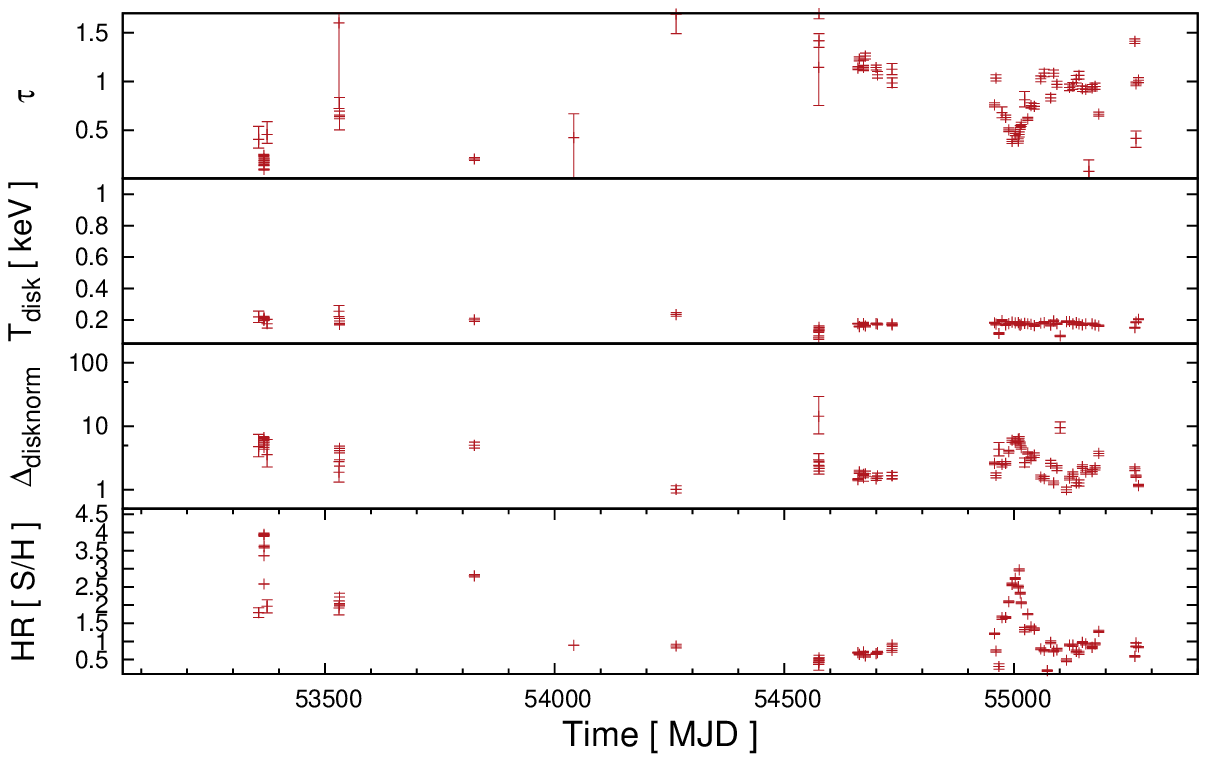}}
\subfigure[GX 339-4: \texttt{diskbb+comptt}]{\includegraphics[height=0.26\textheight]{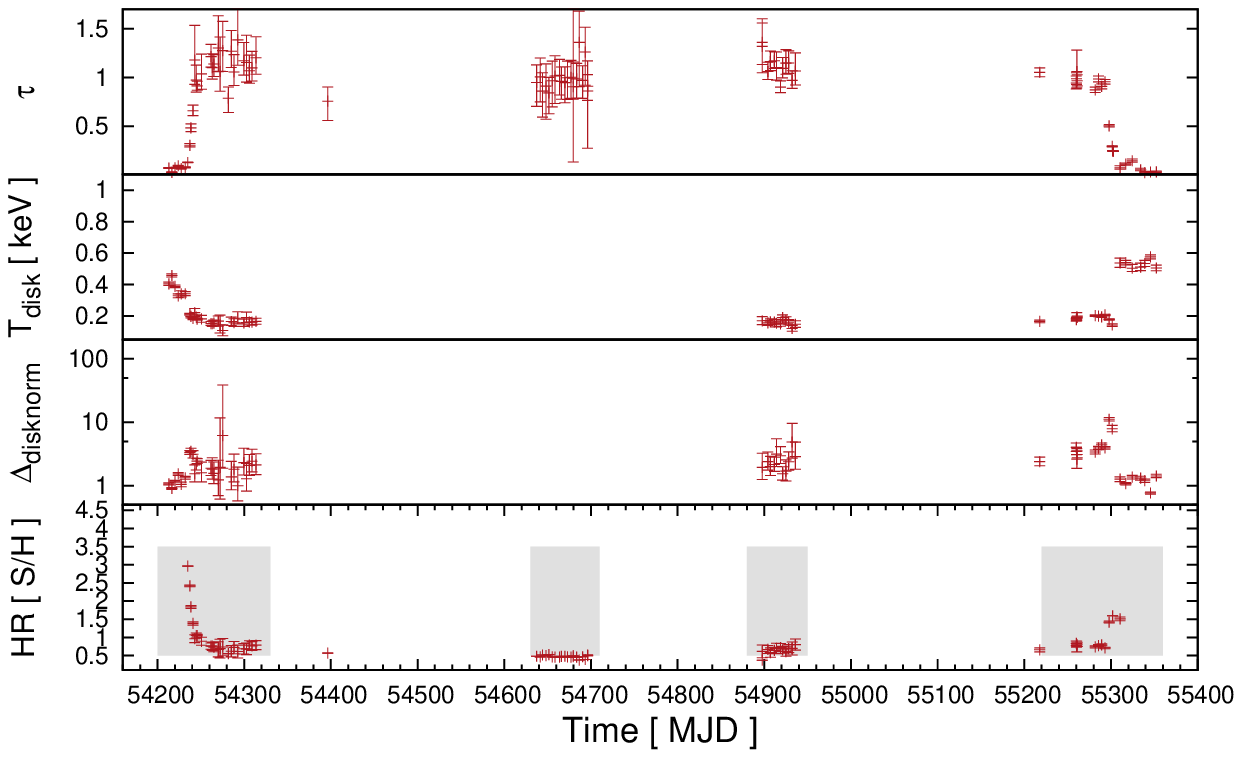}}
\caption{Accretion disk properties for Cyg X-1 (left) \& GX 339-4 (right) taken from
  the best fit \texttt{diskbb+po} (top) and \texttt{diskbb+comptt} (bottom)
  models. The electron temperature ($\rm kT_e$) of the \texttt{comptt} component has
  been frozen at 50 keV. Displayed from top to bottom are the power-law index $\rm
  \Gamma$ (or optical depth, $\tau$), disk temperature $\rm T_{in}$, change in the
  disk normalization relative to the minimum $\rm \Delta_{disknorm}$, and the hardness
  ratio defined as $\rm f_{0.6 - 2.0 keV}/f_{2.0 - 10.0 keV}$. Four outbursts were
  detected from GX 339-4, indicated by the grey highlighting in the hardness ratio
  plot, whereas Cyg X-1 is detected at all times.}
\label{disk_properties}
\end{center}
\end{figure*}

\subsection{A.1) A detailed look at the spectral evolution of Cyg X-1 \& GX 339-4 in X-rays}\label{appendix_cyg}
In Fig. \ref{disk_properties}, we plot the primary disk properties as a function of
time for the best fit \texttt{diskbb+po} models in the case of the HMXB Cyg X-1 and
the LMXB GX 339-4. These systems are chosen as examples of the accretion flow
behaviors observed from a persistent and transient black hole binary system
respectively.

Cyg X-1 is a persistent X-ray source where the $\sim$ 14.8 $\rm M_{\sun}$ black hole
accretes matter from the O-type secondary star \citep{orosz11b}. It is normally
detected at a flux of $\sim$ 0.2 $\rm ct~cm^{-2}~s^{-1}$ by Swift BAT in the 15 -- 50
keV band, while it is in the spectrally hard state. In Fig. \ref{disk_properties}, we
see that Cyg X-1 was observed sparingly prior to the spectral softening which occurred
in late 2010. The disk normalization (where the color radius, $\rm r_{col} \propto
norm^{0.5}$) and spectral index are observed to increase as the spectrum softens while
the temperature of the disk component remains fairly constant. For the Comptonization
model the optical depth is observed to behave in the opposite sense to the power-law
spectral index. We remind the reader that we have fixed the electron temperature to be
50 keV due to the lack of data above 10 keV. Even though Cyg X-1 is a HMXB with a
massive O-type donor, in contrast to the the low mass main sequence stars typically
detected in the transient black hole binaries, the basic X-ray continuum behavior
described above is also observed from the persistent LMXB Swift J1753.5-0127. This
suggests that the behavior we observe is driven by the inner accretion flow and not by
the nature of the mass transfer, i.e., wind vs Roche lobe overflow.

In contrast to the persistent systems, the transients as represented by GX 339-4
display discrete changes in measured disk temperature, consistent with the well known
transitions between the low temperature hard state and the high temperature soft
state, after brightening from the quiescent state \citep{mcclintockremillard06}. Since
the launch of the \textit{Swift} mission there have been 4 outbursts detected from GX
339-4 (Fig. \ref{disk_properties} -- right). Two of these outbursts are consistent
with the system remaining in the hard-state for the duration of the outburst. In fact,
a statistically significant (5$\sigma$) accretion disk component was not detected
during the second outburst. Such hard-state outbursts have been detected from
numerous black hole systems, e.g., GRO J0422+32, XTE J1118+480, GS 1354-64 among
others \citep{brocksopp04}. The other two outbursts are observed to show a clear
transition to the accretion disk dominated soft-state. Here, we see that the
normalization (color radius) of the accretion disk increases as the disk temperature
decreases, though this effect essentially disappears when the hard component is
modelled using the Comptonization model. The power-law index is observed to decrease
in concert with the decreasing disk temperature, while again the optical depth behaves
in the opposite sense, i.e., $\tau$ increases as the disk temperature decreases.

Finally, we note that this change in the disk temperature and any related
changes in the structure/geometry of the accretion flow takes place on timescales of
days (e.g., see Fig. \ref{disk_properties}), consistent with evolution of the inner
accretion disk on the viscous timescale.

\subsection{A.2) Caveats}\label{appendix_param} 
The analysis presented in the preceding sections is dependent on a number of
assumptions regarding the properties of the individual X-ray binary systems, see Table
\ref{BHC_table}. In particular, the inner radius of the accretion disk depends on the
line of sight column density and the assumed distance and inclination of the source
($\rm R_{in} \propto norm_{diskbb}^{0.5} * cos\theta^{-0.5} * d$, see
\S\ref{accretion_disk}). Likewise, the broadband spectral fitting (see
\S\ref{bband_spec}) will be sensitive to the column density as it is the data at
optical and UV wavelengths that provides the constraints on deviations of the
accretion disk temperature profile from that expected for a steady state disk. We
discuss these points in further detail below.\\

\textit{1) Column density -- $N_H$:}\\ There are a number of different issues
regarding the column density, which affect different areas of the paper. The assumed
value of the column density will effect the low energy region of the X-ray spectrum,
and hence the normalization of the measured accretion disk component.  The assumed
abundance model will alter the magnitude of flux absorbed by the intervening gas for a
given column density. A different gas/dust ratio will effect the disk temperature
profile for the joint X-ray/UV fits.

\citet{miller09} have demonstrated, via \textit{Chandra} CCD grating observations of a
sample of black hole binaries, that the column density is consistent with remaining
constant for those binary systems dominated by disk accretion. This is in contrast to
the HMXBs (e.g., Cyg X-1, LMC X-1), where a large stellar wind may be present. We
emphasise here that the winds detected from accretion disks during certain stages of
an XRB outburst \citep{miller06c} are different to the absorbing line of sight column,
i.e., density/temperature, making it possible to separate the local/non-local
absorption component. The column density is held fixed at a value consistent with
previous observations, see Table \ref{BHC_table} for details. This is in agreement
with the observed absence of significant local absorption in these systems
\citep{miller09}, with obvious exceptions for the case of the wind-accreting HMXBs.

Care must also be taken with the chosen abundance model, for example, the metal
abundances in the latest solar model differs from the \citet{angr89} abundances used
in the absorption model herein (\texttt{phabs}). Updated models reflecting our
improved knowledge of the solar abundance contain lower metal abundances, which imply
decreased absorption for a constant column density, i.e., inferring larger disk
components via the extra soft excess. Such models are available in \textsc{xspec} for
usage with the \texttt{phabs} model via the ``wilms" or ``aspl" abundance models
\citep{wilms00,aspl09}.  We have repeated the X-ray fits assuming the ``aspl"
abundances and find that while the measured disk properties may change during
individual observations at the 5\% -- 10\% level, with disk temperature increasing and
inner radius decreasing, the disk temperature and inner radius distributions remain
consistent with those plotted in Fig. \ref{disk_temp_histogram}.

The accretion disk contribution to the optical/UV portion of the spectrum is
especially susceptible to the chosen gas to dust ratio. Throughout the broadband
fitting a value of $\rm N_H = 5.3\times 10^{21} * E(B-V)$ was assumed to link the
optical/UV (\texttt{redden}) to the X-ray (\texttt{phabs}) absorption
\citep{predehl95}. However, different values have been measured previously e.g., $\rm
N_H = 5.8\times 10^{21} * E(B-V)$ \citep{bohlin78}, $\rm N_H = 6.6\times 10^{21} *
E(B-V)$ \citep{guver09}. The broadband fits in \S\ref{bband_spec} were repeated using
both gas/dust ratios above in order to determine if the observed accretion disk
temperature profile behavior is real or simply an artifact due to the assumed law. As
noted earlier for the disk temperature and radius, the temperature profile is observed
to vary with respect to the values determined in \S \ref{bband_spec} for individual
observations. These deviations are typically at less than the 5\% level, but the
overall distribution is similar to that presented in Fig. \ref{histogram_tprofile},
e.g., 68\% of the observation have p $\lesssim$ 0.64 and 90\% have p $\lesssim$ 0.68
when using the \citet{guver09} gas/dust ratio in comparison to 62\% and 66\% for the
\citet{predehl95} model respectively.\\

\textit{2) Mass, distance \& inclination -- $\rm M_x,~d,~\theta$:} \\ 
Less than half the systems in our sample have dynamical constraints on the black hole
mass. For the other systems, we assume a fiducial mass of $\rm M_x = 8~M_{\sun}$, in
agreement with the peak in the mass distribution for the stellar mass black holes with
dynamical mass estimates \citep{ozel10,kreidberg12}. This mass will be a good estimate
for the majority of the systems, though we might expect systems such as IGR
J17091-3624, GRS 1758-258 \& 1E 1740.7-2942 to be outliers in the mass distribution in
analogy with GRS 1915+105. However, the mass of the black hole is only important in
those plots where luminosities are plotted in Eddington units. The measured accretion
disk properties are independent of mass, at least as characterized by the models used
herein. 

The distances to the majority of these systems are unknown, and are instead estimates
based on a number of factors (e.g., the observed column density, observed outburst
flux, magnitude of the optical counterpart in outburst/quiescence), and comparison
with those systems whose properties have been constrained via detailed observations,
see Table \ref{BHC_table}. The distance enters the inner radius calculation as a
linear factor and it is highly unlikely that any of the distances are over/under
estimated by a factor of $\sim$ 2. Crucially, we have no reason to believe that we are
systematically over or underestimating the distance and as such these errors should
tend towards canceling each other out, for the sample as a whole.

As with distance, the inner radius scales proportional to $\rm
cos(\theta)^{-0.5}$. This amounts to a factor of 1.7 for an inclination of
70$\degr$. To date, no eclipses have been observed from any of these systems
indicating that typical inclinations are less than 70$\degr$. For more typical
inclinations of 30$\degr$ -- 60$\degr$ (see \citet{charles06}), the inclination will
increase the inner radius by between 7\% -- 40\% percent.

Bearing these caveats in mind, we believe that the distributions of accretion disk
properties presented herein radii (e.g., $R_{in},~T_{in}$, see
Fig. \ref{disk_temp_histogram}) are valid to within the limitations of the currently
available data.

\begin{figure*}[t]
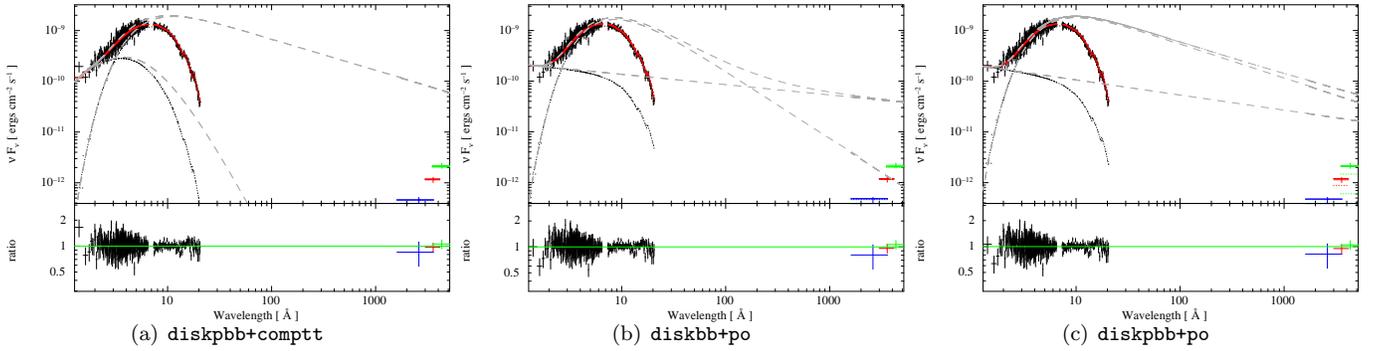

\begin{center}
\subfigure[\texttt{diskpbb+comptt}]{\includegraphics[height=0.25\textheight,angle=-90]{fig20a.eps}}
\subfigure[\texttt{diskbb+po}]{\includegraphics[height=0.25\textheight,angle=-90]{fig20b.eps}}
\subfigure[\texttt{diskpbb+po}]{\includegraphics[height=0.25\textheight,angle=-90]{fig20c.eps}}
\caption{Simultaneous XRT \& UVOT observation of GX 339-4 in the soft state, i.e., the
  accretion disk temperature is $\sim$ 0.6 keV. The fits return $\chi^2_{\nu}$ of
  1.23, 1.28 \& 1.25 for models (a), (b) and (c) respectively. While the fits are
  statistically comparable, the power-law fits requires $\Gamma \sim 1.8$, in contrast
  to the expected value for a black hole in the soft state, i.e., $\Gamma \gtrsim
  2$. The models also attribute differing amounts of the optical/UV flux to the
  accretion disk component, ranging from 100\% in model (a) to less than 10\% in model
  (b).  The preferred model (\texttt{diskpbb+comptt} -- a) requires an irradiated
  accretion disk, i.e., $T(r) \propto r^{-0.6}$ (see text). The unabsorbed model
  components are indicated by the thick grey dashed lines. The residuals are plotted
  with respect to the absorbed model.}
\label{gx339_bband_hss}
\end{center}
\end{figure*}

\begin{figure*}[t]
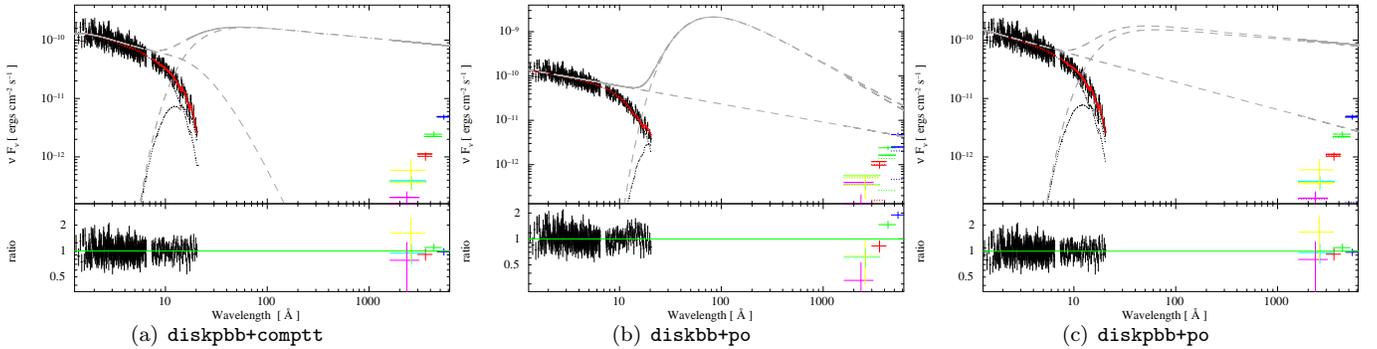

\begin{center}
\subfigure[\texttt{diskpbb+comptt}]{\includegraphics[height=0.25\textheight,angle=-90]{fig21a.eps}}
\subfigure[\texttt{diskbb+po}]{\includegraphics[height=0.25\textheight,angle=-90]{fig21b.eps}}
\subfigure[\texttt{diskpbb+po}]{\includegraphics[height=0.25\textheight,angle=-90]{fig21c.eps}}
\caption{Simultaneous XRT \& UVOT observation of GX 339-4 in the hard state, i.e., the
  accretion disk temperature is $\lesssim$ 0.2 keV. The fits return $\chi^2_{\nu}$ of
  1.02, 1.27, \& 1.02 for models (a), (b) and (c) respectively. In this case the steady
  state accretion disk model in unable to reproduce the observed optical/UV spectral
  slope (panel b). The optical/UV emission is dominated by the accretion disk
  component in each model. The preferred models (those with the \texttt{diskpbb}
  component -- a, c) both require an irradiated accretion disk, i.e., $T(r) \propto
  r^{-0.5}$ (see text). The unabsorbed model components are indicated by the thick
  grey dashed lines. The residuals are plotted with respect to the absorbed model.}
\label{gx339_bband_lhs}
\end{center}
\end{figure*}

\subsection{A.3) A detailed look at broadband spectral fits to GX 339-4}\label{appendix_bband}
As an illustrative test case for the broadband fits, we plot the results of
these fits to representative XRT+UVOT spectra of GX 339-4. In
Fig. \ref{gx339_bband_hss}, we plot fits to a soft state observation (obsid:
00030919005). In fact, this observation is more accurately categorized as being a soft
intermediate state, see \citet{mcclintockremillard06}. The unabsorbed model components
are indicated by the thick grey dashed lines. All 3 models return a similar
temperature for the accretion disk of $\rm T_{in} \sim 0.6~keV$, and disk
normalization of $\lesssim$ 1000, i.e, $\rm R_{in} \lesssim ~40~d_{8
  kpc}~\theta_{45}~km$. As can be seen in the figure, there is a clear difference
between the model with the steady state accretion disk where the optical emission is
dominated by the power-law component and those models with an irradiated disk, where
the optical emission is dominated by the disk component. However, the power-law index
is poorly constrained by the X-ray data, which is dominated by the disk emission. Both
of the variable temperature profile models return non-standard profiles, i.e., $T(r)
\propto r^{-0.6}$; however, the standard disk model provides a statistically
comparable fit to the data. We must now consider the returned parameters and their
implications. In particular, we note that both of the models that utilize a power-law
to generate the hard X-ray emission, return hard photon indices ($\Gamma \sim
1.8$). Such hard power-law emission is in conflict with the well known soft power-laws
normally observed in the soft disk dominated states, i.e., $\Gamma \gtrsim 2.2$
\citep{mcclintockremillard06}. Hence for the soft state the data actually favour a
model consisting of an irradiated accretion disk and a corona, i.e.,
\texttt{diskpbb+comptt}.

In Fig. \ref{gx339_bband_lhs}, we plot fits to a hard state observation of GX 339-4
(obsid: 00030943002), where the unabsorbed model components are again indicated by the
thick grey dashed lines. In contrast to the soft state observation above, here the
\texttt{diskbb+po} model returns a significantly poorer fit in comparison to the
\texttt{diskpbb} models, i.e., $\chi^2_{\nu} \sim 1.3~vs~1.0$. This is due to an
inability of this model to re-produce the observed optical/UV slope as may be clearly
seen in the residuals (middle panel). The models with the \texttt{diskpbb} component
are both observed to provide fits of a statistically similar quality. The returned
accretion disk parameters are also similar, e.g., $\rm T_{in} \sim 0.17~keV, R_{in}
\sim 100~d_{8 kpc}~\theta_{45}~km$. Crucially in both models, a non steady state
accretion disk is required. The measured temperature profile for the accretion disk is
$\rm p = 0.52\pm0.01$. This temperature profile is consistent with that expected from
an accretion disk that is being irradiated. Such irradiation is expected to occur in
the hard state, when the X-ray emission is dominated by the hard X-ray component,
i.e., $\rm E \gtrsim 2~keV$.  Finally, we note the crucial role played by the
broadband coverage provided by the XRT plus UVOT combination. In particular, the
coverage from optical to UV wavelengths is crucial in allowing us to constrain the
temperature profile of the accretion disk, which in this case is consistent with an
irradiated accretion disk, i.e., $\rm T \propto r^{-0.5}$.


\begin{table*}[p]
\caption{Confirmed \& Candidate black hole binaries: primary properties.}\label{BHC_table}
\begin{center}
\begin{tabular}{lcccccccccc}
\tableline\tableline\\ [-2.0ex] 
Source & $\alpha$ & $\delta$ & P$\rm_{orb}$ & Spec & d & f(M) & M$\rm_x$ & $\rm N_H$ & Ref\\ [0.5ex]
 -- & -- & -- & [ hrs ] & -- & [ kpc ] & [ $\rm M_{\sun}$ ] & [ $\rm M_{\sun}$ ] & [ $\rm  10^{21}~cm^{-2}$] & -- \\ [0.5ex]
\tableline\tableline\\ [-2.0ex] 
NGC 300 X-1        & 00:55:09.00 & -37:42:12.16 & 32.8  & WR          & 1800     & 2.6
$\pm$ 0.3     & 12 -- 24 & 0.6 & $^{1}$ \\ [0.5ex]
IC 10 X-1          & 00:20:29.05 & +59:16:52.3  & 35.6  & WR          & 660     & 7.4 $\pm$ 1.25    & $\geq$ 23.1 & 12.5 & $^{2}$\\ [0.5ex]
LMC X-3            & 05:38:56.63 & -64:05:03.2  & 40.9  & B3V         & 50.0 $\pm$ 2.3
& 2.3 $\pm$ 0.3     & 5.9 -- 9.2 & 0.6 & $^{3, 4}$\\ [0.5ex]
GX 339-4            & 17:02:49.5 & -48:47:23    & 42.1  & --          & 4.0 $\pm$ 2.0
& 5.8 $\pm$ 0.5     & -- & 6.0 & $^{5, 6}$\\ [0.5ex]
GRO J1655-40       & 16:54:00.14 & -39:50:44.9  & 62.9  & F3 -- F5IV  & 3.2 $\pm$ 0.2
& 2.73 $\pm$ 0.09   & 6.0 -- 6.6 & 7.4 & $^{7, 8, 9}$\\ [0.5ex]
SAX J1819.3-2525   & 18:19:21.58 & -25:24:25.1  & 67.6  & B9III       & 9.8 $\pm$ 2.5  & 3.13 $\pm$ 0.13   & 6.8 -- 7.4 & 1.7 & $^{10}$\\ [0.5ex]
M33 X-7            & 01:33:34.12 & +30:32:11.6  & 82.8  & O7 -- O8III & 840     & 0.48 $\pm$ 0.08   & 15.65 $\pm$ 1.45 & 1.0 & $^{11}$\\ [0.5ex]
LMC X-1            & 05:39:38.7 & -69:44:36     & 101.5 & O7III       & 50.0 $\pm$ 2.3
& 0.14 $\pm$ 0.05   & 4.0 -- 10.0 & 11.0 & $^{12, 13, 14}$\\ [0.5ex]
Cyg X-1            & 19:58:21.68 & +35:12:05.78 & 134.4 & O9.7Iab     & 2.1 $\pm$ 0.1
& 0.244 $\pm$ 0.005 & 14.8 $\pm$ 1.0 & 6.6 & $^{15}$\\ [0.5ex]
GRS 1915+105       & 19:15:11.55 & +10:56:44.8  & 804.0 & K -- MIII   & 11.5 $\pm$ 0.5
& 9.5 $\pm$ 3.0     & 10.1 -- 13.4 & 43.5 & $^{16, 17, 18, 19}$\\ [0.5ex]
 -- & -- & -- & -- & -- & -- & -- & -- &-- \\ [0.5ex]
1E 1740.7-2942	   & 17:43:54.88 & -29:44:42.5   & 12.7 & -- & 8.5 & --
& -- & 100 & $^{20}$\\ [0.5ex]
H1743-322	   & 17:46:15.61 & -32:14:00.6   & --   & -- & 10 & -- & -- &
18 & $^{21, 22}$\\ [0.5ex] 
4U 1630-47	   & 16:34:01.61 & -47:23:34.8   & --   & -- & 10 & -- &
-- & 82 & $^{23, 24}$ \\ [0.5ex]
GRS 1758-258	   & 18:01:12.67 & -25:44:26.7	 & 18.5 & -- & 8.5 & --
& -- & 15 & $^{25}$\\ [0.5ex]
4U 1957+11	   & 19:59:24.0  & +11:42:30     & 9.3 & -- & 10 & -- &
-- & 1.5 & $^{26, 27}$ \\ [0.5ex]
SLX 1746-331	   & 17:49:48.94 & -33:12:11.6   & -- & -- & 7.0 & -- &
-- & 7.0 & $^{28, 29}$ \\ [0.5ex]
 -- & -- & -- & -- & -- & -- & -- & -- &-- \\ [0.5ex]
IGR J17091-3624	   & 17:09:07.6  & -36:24:24.9   & --   & -- & 10 & -- & -- & 8 & $^{30}$\\ [0.5ex]  
SWIFT J1539.2-6277 & 15:39:11.96  & -62:28:02.3   & -- & -- & 10 & -- & -- & 4.5 & $^{31}$\\ [0.5ex]
XTE J1652-453      & 16:52:20.33  & -45:20:39.6   & -- & -- & 10 & -- & -- & 49 & $^{32}$\\ [0.5ex]
IGR J17098-3628	   & 17:09:45.93  & -36:27:57.30  & -- & -- & 10 & -- & -- & 9.0 &
$^{30, 33}$\\ [0.5ex]
IGR J17497-2821	   & 17:49:38.1   & -28:21:17	  & -- & -- & 10 & -- & -- & 50 &
$^{34, 35}$\\  [0.5ex]
XTE J1752-223	   & 17:52:15.095 & -22:20:32.782 & -- & -- & 8.5 & -- & -- & 4.7 &
$^{36, 37}$\\ [0.5ex]
SWIFT J1753.5-0127 & 17:53:28.29  & -01:27:06.22  & -- & -- & 6.0 & -- & -- & 2.0 & $^{38}$\\ [0.5ex]	
XTE J1817-330 	   & 18:17:43.526 & -33:01:07.47  & -- & -- & 8.5 & -- & -- & 1.5 & $^{39}$\\ [0.5ex]	
XTE J1818-245	   & 18:18:24.43  & -24:32:17.96  & -- & -- & 3.5 & -- & -- & 5.4 & $^{40}$\\ [0.5ex]
SWIFT J1842.5-1124 & 18:42:17.44  & -11:25:03.95  & -- & -- & 8.5 & -- & -- & 3.9 &
$^{41, 42}$\\ [0.5ex]  
XTE J1856+053	   & 18:56:42.92  & +05:18:34.3   & -- & -- & 10 & -- & -- & 44 &
$^{43, 44}$\\ [0.5ex]
\tableline\tableline\\ [-2.0ex]
\end{tabular}
\tablecomments{Parameters for the 27 black hole binary systems in our sample. These
  include 10 dynamically confirmed systems (NGC 300 X-1 -- GRS 1915+105) and another
  17 systems that are strong candidate black hole binaries, e.g., see
  \citet{mcclintockremillard06}.  For those systems that have not had their mass
  dynamically estimated we assume a fiducial mass of 8.0 M$_{\sun}$ with a 20\%
  uncertainty, consistent with the known distribution of masses for dynamically
  measured black holes \citep{ozel10,kreidberg12}. Please see the following references
  for the additional system parameters:\\ $^{1}$ \citealt{crowther10}; $^{2}$
  \citealt{I213}; $^{3}$ \citealt{I39}; $^{4}$ \citealt{I40}; $^{5}$ \citealt{I41};
  $^{6}$ \citealt{I42}; $^{7}$ \citealt{I43}; $^{8}$ \citealt{I44}; $^{9}$
  \citealt{I45}; $^{10}$ \citealt{I46}; $^{11}$ \citealt{I212}; $^{12}$ \citealt{I39};
  $^{13}$ \citealt{I47}; $^{14}$ \citealt{I48}; $^{15}$ \citealt{orosz11b}; $^{16}$
  \citealt{I54}; $^{17}$ \citealt{I55}; $^{18}$ \citealt{I56}; $^{19}$ \citealt{I57};
  $^{20}$ \citealt{reynolds10b}; $^{21}$ \citealt{atel314}; $^{22}$ \citealt{prat09};
  $^{23}$ \citealt{hjellming99}; $^{24}$ \citealt{kubota07}; $^{25}$
  \citealt{pottschmidt06}; $^{26}$ \citealt{thorstensen87}; $^{27}$ \citealt{nowak11};
  $^{28}$ Torres priv. comm.; $^{29}$ \citealt{atel1237}; $^{30}$ \citealt{atel1140};
  $^{31}$ \citealt{atel1893}; $^{32}$ \citealt{atel2120}; $^{33}$ \citealt{atel490};
  $^{34}$ \citealt{atel900}; $^{35}$ \citealt{walter07}; $^{36}$ \citealt{atel2278};
  $^{37}$ \citealt{atel2261}; $^{38}$ \citealt{reynolds10a}; $^{39}$
  \citealt{rykoff07}; $^{40}$ \citealt{cadollebel09}; $^{41}$ \citealt{atel1720};
  $^{42}$ \citealt{atel1706}; $^{43}$ \citealt{atel1072}; $^{44}$ \citealt{atel1062}}
\end{center}
\end{table*}

\end{appendix}

\vspace{1cm}
\footnotesize{This paper was typeset using a \LaTeX\ file prepared by the 
author.}



\begin{thebibliography}{}  

\bibitem[\protect\citeauthoryear{Abbey et al.}{2006}]{abbey06} Abbey T., Carpenter J.,
  Read A. et al., 2006, in Proceedings of 'The X-ray Universe 2005', ed. A. Wilson
  (ESA SP 604; Noordwijk: ESA), 943 

\bibitem[Akritas \& Siebert(1996)]{akritas96} Akritas M.G., \& Siebert J., 1996,
  MNRAS, 278, 919

\bibitem[\protect\citeauthoryear{Anders \& Grevesse}{1989}]{angr89} Anders E.,
  Grevesse N., 1989, GeCoA, 53, 197

\bibitem[Asplund et al.(2009)]{aspl09} Asplund M., Grevesse N., Sauval A.J.,  Scott
  P., 2009, ARA\&A, 47, 481



\bibitem[\protect\citeauthoryear{Balucinska-Church \& McCammon}{1992}]{bcmc92}
  Balucinska-Church M., McCammon D., 1992, ApJ, 400, 699

\bibitem[\protect\citeauthoryear{Barkov \& Khangulyan}{2012}]{barkov11} Barkov M.V.,
  Khangulyan D.V., 2012, MNRAS, 421, 1351

\bibitem[\protect\citeauthoryear{Barret et al.}{1996}]{I21} Barret D., McClintock
  J.E., Grindlay J.E., 1996, ApJ, 473, 963

\bibitem[Barthelmy et al.(2005)]{barthelmy05} Barthelmy, S.~D., 
Barbier, L.~M., Cummings, J.~R., et al.\ 2005, \ssr, 120, 143

\bibitem[\protect\citeauthoryear{Belloni}{2004}]{belloni04} Belloni T., 2004,
  Nucl. Phys. B, 132, 337

\bibitem[\protect\citeauthoryear{Beloborodov}{1999}]{belo99} Beloborodov A. 1999, ApJ,
  510, 123 

\bibitem[\protect\citeauthoryear{Blandford \& Znajek}{1977}]{bz77} Blandford R.D.,
  Znajek R.L., 1977, MNRAS, 179, 433 



\bibitem[Bohlin et al.(1978)]{bohlin78} Bohlin R.C., Savage B.D., Drake, J.F., 1978,
  ApJ, 224, 132

\bibitem[\protect\citeauthoryear{Bradt, Rothschild, \& Swank}{1993}]{bradt93} Bradt
  H.~V., Rothschild R.~E., Swank J.~H., 1993, A\&AS, 97, 355

\bibitem[\protect\citeauthoryear{Breeveld et al.}{2010}]{uvot_calII} Breeveld A.A.,
  Curran P.A., Hoversten E.A. et al., 2010, MNRAS, 406, 1687

\bibitem[\protect\citeauthoryear{Breeveld et al.}{2011}]{breeveld11} Breeveld A.~A.,
  Landsman W., Holland S.~T., Roming P., Kuin N.~P.~M., Page M.~J., 2011, AIPC, 1358,
  373

\bibitem[Brocksopp et al.(2004)]{brocksopp04} Brocksopp, C., 
Bandyopadhyay, R.~M., \& Fender, R.~P.\ 2004, NewA, 9, 249

\bibitem[\protect\citeauthoryear{Brocksopp et al.}{2009}]{atel2278} Brocksopp C. et
  al., 2009, ATel \#2278

\bibitem[\protect\citeauthoryear{Burrows et al.}{2005}]{burrows05} Burrows D.N. et
  al., 2005, SSRev, 120, 165 

\bibitem[Cadolle Bel et al.(2009)]{cadollebel09} Cadolle Bel, M., Prat, L., Rodriguez,
  J., et al.\ 2009, \aap, 501, 1

\bibitem[\protect\citeauthoryear{Cardelli et al.}{1989}]{cardelli89} Cardelli J.A.,
  Clayton G.C., Mathis J.S., 1989, ApJ, 345, 245

\bibitem[\protect\citeauthoryear{Casares et al.}{1993}]{I53} Casares J., Charles P.A.,
  Naylor T., Pavlenko E.P., 1993, MNRAS, 265, 834

\bibitem[\protect\citeauthoryear{Casares \& Charles}{1994}]{I52} Casares J., Charles
  P.A., 1994, MNRAS, 271, 5

\bibitem[\protect\citeauthoryear{Casares et al.}{1995}]{b19} Casares J., Martin A.C.,
  Charles P.A., Martin E.L., Rebolo R., Harlaftis E.T., Castro-Tirado A.J., 1995,
  MNRAS, 276, 35

\bibitem[\protect\citeauthoryear{Cash et al.}{1979}]{cash79} Cash W., 1979, ApJ 228, 939

\bibitem[Charles \& Coe(2006)]{charles06} Charles P.A., Coe M.J., 2006, Compact
  stellar X-ray sources, 215

\bibitem[\protect\citeauthoryear{Chen et al.}{1997}]{chen97} Chen W., Shrader C.R.,
  Livio M., 1997, ApJ, 491, 312

\bibitem[Cheng et al.(1992)]{cheng92} Cheng, F.~H., Horne, K.,  Panagia,
  N., Shrader, C.~R., Gilmozzi, R., Paresce, F.,  \& Lund, N.\ 1992, \apj, 397, 664


\bibitem[\protect\citeauthoryear{Coppi}{1999}]{coppi99} Coppi P.S., 1999, in ASP
  Conf. Ser. 161, High Energy Processes in Accreting Black Holes, Poutanen J.,
  Svensson R., eds (San Francisco, CA: ASP), 375

\bibitem[\protect\citeauthoryear{Corbel et al.}{2006}]{corbel06} Corbel S., Tomsick
  J.A., Kaaret P., 2006, ApJ, 636, 971

\bibitem[\protect\citeauthoryear{Coriat et al.}{2009}]{coriat09} Coriat M., Corbel S., 
Buxton M.M., et al., 2009, MNRAS, 400, 123

\bibitem[\protect\citeauthoryear{Cowley et al.}{1983}]{I40} Cowley A.P., Crampton D.,
  Hutchings J.B., et al., 1983, ApJ, 272, 118

\bibitem[\protect\citeauthoryear{Cowley et al.}{1987}]{I41} Cowley A.P., Crampton D.,
  Hutchings J.B., 1987, AJ, 92, 195

\bibitem[\protect\citeauthoryear{Cowley et al.}{1995}]{I48} Cowley A.P., Schmidtke
  P.C., Anderson A.L., McGrath T.K., 1995, PASP, 107, 145

\bibitem[\protect\citeauthoryear{Croton et al.}{2006}]{croton06} Croton D.J. et al.,
  2006, MNRAS, 365, 11 

\bibitem[\protect\citeauthoryear{Crowther et al.}{2010}]{crowther10} Crowther P.~A.,
  Barnard R., Carpano S., Clark J.~S., Dhillon V.~S., Pollock A.~M.~T., 2010, MNRAS,
  403, L41

\bibitem[\protect\citeauthoryear{Davis et al.}{2005}]{davis05} Davis S.W., Blaes
  O.M., Hubeny I., Turner N.J., 2005, ApJ, 621, 372

\bibitem[\protect\citeauthoryear{Davis et al.}{2006}]{davis06} Davis S.W., Done C.,
  Blaes O.M., 2006, ApJ, 647, 525

\bibitem[\protect\citeauthoryear{Done \& Gierlinski}{2003}]{done03} Done C.,
  Gierlinski M., 2003, MNRAS, 342, 1041

\bibitem[\protect\citeauthoryear{Done et al.}{2007}]{done07} Done C., Gierli{\'n}ski
  M., Kubota A., 2007, A\&ARv, 15, 1

\bibitem[\protect\citeauthoryear{Done et al.}{2008}]{done08} Done C., Davis S.W.,
  2008, ApJ, 683, 389

\bibitem[\protect\citeauthoryear{Done \& Diaz Trigo}{Done et
    al.}{2010}]{done10} Done C., Diaz Trigo M., 2010, MNRAS, 407, 2287

\bibitem[\protect\citeauthoryear{Dubus et al.}{1999}]{dubus99} Dubus G., Lasota J.P.,
  Hameury J.M., Charles P., 1999, MNRAS, 303, 139

\bibitem[\protect\citeauthoryear{Dubus et al.}{2001}]{dubus01} Dubus G., Hameury J.M.,
  Lasota J.P., 2001, A\&A, 373, 251

\bibitem[\protect\citeauthoryear{Dunn et al.}{2010}]{dunn10a} Dunn R.J.H., Fender
  R.P., Kording E.G., Belloni T., Cabanac C., 2010, MNRAS, 403, 61

\bibitem[\protect\citeauthoryear{Dunn et al.}{2011}]{dunn10b} Dunn R.J.H., Fender
  R.P., Kording E.G., Belloni T., Merloni A., 2011, MNRAS, 411, 337

\bibitem[\protect\citeauthoryear{Durant et al.}{2008}]{durant08} 
Durant M., Gandhi P., Shahbaz T., Fabian A.~P., Miller J., Dhillon V.~S., 
Marsh T.~R., 2008, ApJ, 682, L45 

\bibitem[\protect\citeauthoryear{Ebisawa et al.}{1996}]{ebisawa96} Ebisawa K., Ueda
  Y., Inoue H., Tanaka Y., White N.E., 1996, ApJ, 467, 419

\bibitem[\protect\citeauthoryear{Esin et al.}{1997}]{I17} Esin A.A., McClintock J.E.,
  Narayan R., 1997, ApJ, 489, 865

\bibitem[\protect\citeauthoryear{Fabbiano}{2006}]{fabbiano06} Fabbiano G., 2006,
  ARA\&A, 44, 323

\bibitem[\protect\citeauthoryear{Falcke et al.}{2004}]{falcke04} Falcke H., Kording
  E., Markoff S., 2004, A\&A, 414, 895

\bibitem[\protect\citeauthoryear{Fender et al.}{1999}]{I55} Fender R.P., Garrington
  S.T., McKay D.J., Muxlow T.W.B., Pooley G.G., Spencer R.E., Stirling A.M., Waltman
  E.B., 1999, MNRAS, 304, 865

\bibitem[\protect\citeauthoryear{Fender et al.}{2004}]{fender04} Fender R.P., Belloni
  T.M., Gallo E., 2004, MNRAS, 355, 1105

\bibitem[\protect\citeauthoryear{Fender}{2006}]{fender06} Fender 
R., 2006, Compact stellar X-ray sources, 381

\bibitem[\protect\citeauthoryear{Ferrarese \& Merritt}{2000}]{ferraresemerritt00}
  Ferrarese F., Merritt D. 2000, ApJ, 539, 9  






\bibitem[\protect\citeauthoryear{Freedman et al.}{2001}]{I39} Freedman W.L., Madore
  B.F., Gibson B.K., 2001, ApJ, 553, 47

\bibitem[\protect\citeauthoryear{Gallo et al.}{2004}]{gallo04} Gallo E., Corbel S.,
  Fender R.P., Maccarone T.J., Tzioumis A.K., 2004, MNRAS, 347, 52

\bibitem[\protect\citeauthoryear{Gallo}{2010}]{gallo10} Gallo E., 2010, Lecture Notes 
in Physics, Berlin Springer Verlag, 794, 85

\bibitem[\protect\citeauthoryear{Gandhi et al.}{2008}]{gandhi08} 
Gandhi P., et al., 2008, MNRAS, 390, L29

\bibitem[\protect\citeauthoryear{Gebhardt et al.}{2000}]{gebhardt00} Gebhardt K. et
  al., 2000, ApJ, 539, 13 

\bibitem[\protect\citeauthoryear{Gehrels et al.}{2004}]{gehrels04} Gehrels N. et al.,
  2004, ApJ, 611, 1005



\bibitem[\protect\citeauthoryear{George \& Fabian}{1991}]{george91} George I.~M.,
  Fabian A.~C., 1991, MNRAS, 249, 352


\bibitem[\protect\citeauthoryear{Gierlinski et al.}{1999}]{diskpn_gierlinski99}
  Gierlinski M., Zdziarski A.A., Poutanen J., Coppi P.S., Ebisawa K., Johnson W.N.,
  1999, MNRAS, 309, 496

\bibitem[\protect\citeauthoryear{Gierlinski \& Done}{2004}]{gierlinski04} Gierlinski
  M., Done C., 2004, MNRAS, 347, 885

\bibitem[\protect\citeauthoryear{Gilfanov}{2004}]{gilfanov04}
  Gilfanov M., 2004, MNRAS, 349, 146

\bibitem[\protect\citeauthoryear{Greiner et al.}{2001a}]{I56} Greiner J., Cuby J.G.,
  McCaughrean M.J., 2001a, Nature, 414, 522

\bibitem[\protect\citeauthoryear{Greiner et al.}{2001b}]{I57} Greiner J., Cuby J.G.,
  McCaughrean M.J., et al., 2001b, A\&A, 373, 37

\bibitem[\protect\citeauthoryear{Grimm et al.}{2002}]{grimm02} Grimm H.J., Gilfanov
  M., Sunyaev R., 2002, A\&A, 391, 923

\bibitem[\protect\citeauthoryear{Grimm et al.}{2003}]{grimm03} Grimm H.J., Gilfanov
  M., Sunyaev R., 2003, MNRAS, 339, 793


\bibitem[\protect\citeauthoryear{Grove et al.}{1998}]{grove98} Grove J.E., Johnson
  W.N., Kroeger R.A., McNaron-Brown K., Skibo J.G., Phlips B.F., 1998, ApJ, 500, 899

\bibitem[\protect\citeauthoryear{Gultekin et al.}{2009}]{gultekin09} Gultekin K. et
  al., 2009, ApJ, 706, 404 

\bibitem[G{\"u}ver \& {\"O}zel(2009)]{guver09} G{\"u}ver T., {\"O}zel F., 2009,
  MNRAS, 400, 2050

\bibitem[\protect\citeauthoryear{Hands et al.}{2004}]{hands04} Hands A.D.P., Warwick
  R.S, Watson M.G., Helfand D.J., 2004, MNRAS, 351, 31


\bibitem[\protect\citeauthoryear{Hasinger \& van der Klis}{1989}]{hasinger89} Hasinger
  G., van der Klis M., 1989, A\&A, 225, 79

\bibitem[\protect\citeauthoryear{Hjellming \& Rupen}{1995}]{I43} Hjellming R.M., Rupen
  M.P., 1995, Nature, 375, 464

\bibitem[\protect\citeauthoryear{Hjellming et al.}{1999}]{hjellming99} Hjellming
  R.~M., et al., 1999, ApJ, 514, 383

\bibitem[\protect\citeauthoryear{Homan et al.}{2001}]{homan01} Homan J., Wijnands R.,
  van der Klis M., Belloni T., van Paradijs J., Klein-Wolt M., Fender R. Mendez M.,
  2001, ApJS, 132, 1377

\bibitem[\protect\citeauthoryear{Homan \& Belloni}{2005}]{homanbelloni05} Homan J.,
  Belloni T., 2005, Ap\&SS, 300, 107.

\bibitem[\protect\citeauthoryear{Hutchings et al.}{1987}]{I47} Hutchings J.B.,
  Crampton D., Cowley A.P., et al., 1987, AJ, 94, 340

\bibitem[\protect\citeauthoryear{Hynes et al.}{1998}]{hynes98} Hynes R.I. et al.,
  1998, MNRAS, 300, 64 

\bibitem[\protect\citeauthoryear{Hynes et al.}{2003a}]{I42} Hynes R.I., Steeghs D.,
  Casares J., Charles P.A., O'Brien K., 2003a, ApJ, 583, 95

\bibitem[\protect\citeauthoryear{Hynes et al.}{2006}]{hynes06} Hynes R.I. et al.,
  2006, ApJ, 651, 401 

\bibitem[Isobe et al.(1990)]{isobe90} Isobe T., Feigelson E.D., Akritas M.G., Babu
  G.J., 1990, ApJ, 364, 104

\bibitem[\protect\citeauthoryear{Kalemci et al.}{2004}]{kalemci04} Kalemci E., Tomsick
  J.A., Rothschild R.E., Pottschmidt K., Kaaret P., 2004, ApJ, 603, 231 

\bibitem[\protect\citeauthoryear{Kalemci et al.}{2006}]{kalemci06} Kalemci E., Tomsick
  J.A., Rothschild R.E., Pottschmidt K., Corbel S., Kaaret P., 2006, ApJ, 639, 340 

\bibitem[\protect\citeauthoryear{Kanbach et al.}{2001}]{kanbach01} Kanbach G.,
  Straubmeier C., Spruit H.~C., Belloni T., 2001, Natur, 414, 180

\bibitem[\protect\citeauthoryear{Kennea et al.}{2006}]{atel900} Kennea J. et al.,
  2006, ATel \#900

\bibitem[\protect\citeauthoryear{Kennea et al.}{2007a}]{atel1140} Kennea J. et al.,
  2007a, ATel \#1140

\bibitem[\protect\citeauthoryear{Kennea et al.}{2007b}]{atel1237} Kennea J. et al.,
  2007b, ATel \#1237

\bibitem[\protect\citeauthoryear{Kim \& Fabbiano}{2004}]{kim04} Kim D., Fabbiano G.,
  2004, ApJ, 611, 846

\bibitem[\protect\citeauthoryear{King \& Ritter}{1998}]{kingritter98} King A.R.,
  Ritter H., 1998, MNRAS, 293, 42 

\bibitem[\protect\citeauthoryear{Kong et al.}{2002}]{kong02} Kong A.K.H., McClintock
  J.E., Garcia M.R., Murray S.S., Barret D., 2002, ApJ, 570, 277

\bibitem[\protect\citeauthoryear{Kording et al.}{2006}]{kording06} Kording E.G.,
  Jester S., Fender R.P., 2006, MNRAS, 372, 1366 

\bibitem[\protect\citeauthoryear{Kording et al.}{2007}]{kording07} Kording E.G.,
  Migliari S., Fender R., Belloni T., Knigge C., McHardy I., 2007, MNRAS, 380, 301 

\bibitem[\protect\citeauthoryear{Kording et al.}{2008}]{kording08} Kording E.G.,
  Jester S., Fender R.P., 2008, MNRAS, 383, 277 


\bibitem[Kreidberg et al.(2012)]{kreidberg12} Kreidberg L., Bailyn C.D., Farr W.M.,
  Kalogera V., 2012, ApJ, 757, 36

\bibitem[\protect\citeauthoryear{Krimm et al.}{2008}]{atel1706} Krimm H.A. et al.,
  2008, ATel \#1706 

\bibitem[\protect\citeauthoryear{Krimm et al.}{2009}]{atel1893} Krimm H.A. et al.,
  2009, ATel \#1893 

\bibitem[\protect\citeauthoryear{Kubota et al.}{1998}]{kubota98} Kubota A., Tanaka Y.,
  Makishima K., Ueda Y., Dotani T., Inoue H., Yamaoka K., 1998, PASJ, 50, 667

\bibitem[\protect\citeauthoryear{Kubota et al.}{2005}]{kubota05} Kubota A., Ebisawa
  K., Makishima K., Nakazawa K., 2005, ApJ, 631, 1062

\bibitem[\protect\citeauthoryear{Kubota et al.}{2007}]{kubota07} Kubota A., et al.,
  2007, PASJ, 59, 185

\bibitem[\protect\citeauthoryear{Kubota et al.}{2010}]{kubota10} Kubota A., Done C., 
Davis S.W. et al., 2010, ApJ, 714, 860

\bibitem[\protect\citeauthoryear{Kuulkers et al.}{2007}]{kuulkers07} Kuulkers E., Shaw
  S.E., Paizis A. et al., 2007, A\&A, 466, 595

\bibitem[\protect\citeauthoryear{Lasota}{2001}]{lasota01} Lasota J.P., 2001, NewAR,
  45, 449

\bibitem[\protect\citeauthoryear{Li et al.}{2005}]{li05} Li L., Zimmerman E.R.,
  Narayan R., McClintock J.E., 2005, ApJS, 157, 335



\bibitem[\protect\citeauthoryear{Makishima et al.}{1986}]{diskbb_makishima86}
  Makishima K., Maejima, Y., Mitsuda, K., Bradt H.V., Remillard R.A., Tuohy I.R.,
  Hoshi R., Nakagawa M., 1986, ApJ, 308, 635

\bibitem[\protect\citeauthoryear{Makishima et al.}{2000}]{makishima00} Makishima, K.,
  et al., 2000, ApJ, 535, 632

\bibitem[\protect\citeauthoryear{Makishima et al.}{2008}]{makishima08} Makishima K.,
  Takahashi H., Yamada S. et al., 2008, PASJ, 60, 585

\bibitem[\protect\citeauthoryear{Markoff et al.}{2001}]{markoff01} Markoff S., Falcke
  H., Fender R., 2001, A\&A, 372, 25

\bibitem[\protect\citeauthoryear{Markoff et al.}{2005}]{markoff05} Markoff S., Nowak
  M.A., Wilms J., 2005, ApJ, 635, 1203

\bibitem[\protect\citeauthoryear{Marsh et al.}{1994}]{I23} Marsh T.R., Robinson E.L.,
  Wood J.H., 1994, MNRAS, 266, 137

\bibitem[\protect\citeauthoryear{Markwardt et al.}{2009a}]{atel2120} Markwardt C.B. et
  al., 2009a, ATel \#2120

\bibitem[\protect\citeauthoryear{Markwardt et al.}{2009b}]{atel2261} Markwardt C.B. et
  al., 2009b, ATel \#2261

\bibitem[\protect\citeauthoryear{Merloni et al.}{2000}]{merloni00} Merloni A., Fabian
  A.C., Ross R.R., 2000, MNRAS, 313, 193

\bibitem[\protect\citeauthoryear{Merloni et al.}{2002}]{merloni02} Merloni A., Fabian,
  A. 2002, MNRAS, 332, 165 

\bibitem[\protect\citeauthoryear{Merloni et al.}{2003}]{merloni03} Merloni A., Heinz
  S., di Matteo T., 2003, MNRAS, 345, 1057 

\bibitem[\protect\citeauthoryear{McClintock et al.}{1995}]{mcclintock95} McClintock
  J.E., Horne K., Remillard R.A., 1995, ApJ, 442, 358 

\bibitem[\protect\citeauthoryear{McClintock \& Remillard}{2000}]{I24} McClintock J.E.,
  Remillard R.A., 2000, ApJ, 531, 956


\bibitem[\protect\citeauthoryear{McClintock \&
    Remillard}{2006}]{mcclintockremillard06} McClintock J.E., Remillard R.A., 2006,
  Compact stellar X-ray sources, 157

\bibitem[\protect\citeauthoryear{McClintock et al.}{2011}]{mcclintock11} McClintock
  J.E., Narayan R., Davis S.W., et al., 2011, arXiv:1101.0811

\bibitem[McHardy et al.(2004)]{mchardy04} McHardy I.M., Papadakis I.E.,
  Uttley P., Page M.J., Mason K.O., 2004, MNRAS, 348, 783 

\bibitem[Miller et al.(2004)]{miller04} Miller, J.~M., Fabian, A.~C., \& Miller,
  M.~C.\ 2004, \apjl, 614, L117

\bibitem[\protect\citeauthoryear{Miller, Homan, 
\& Miniutti}{2006a}]{miller06a} Miller J.~M., Homan J., Miniutti G., 2006a,
  ApJ, 652, L113 

\bibitem[\protect\citeauthoryear{Miller et al.}{2006b}]{miller06b} 
Miller J.~M., Homan J., Steeghs D., Rupen M., Hunstead R.~W., Wijnands R., 
Charles P.~A., Fabian A.~C., 2006b, ApJ, 653, 525

\bibitem[\protect\citeauthoryear{Miller et al.}{2006c}]{miller06c} Miller J.M., Raymond
  J., Fabian A., Steeghs D., Homan J., Reynolds C., van der Klis M., Wijnands R.,
  2006c, Natur, 441, 953

\bibitem[\protect\citeauthoryear{Miller}{2007}]{miller07} Miller J.M., 2007,
  ARA\&A, 45, 441

\bibitem[\protect\citeauthoryear{Miller et al.}{2009}]{miller09} Miller J.M., Cackett
  E.M., Reis R.C., 2009, ApJ, 707, 77

\bibitem[\protect\citeauthoryear{Miller et al.}{2010}]{miller10} Miller J.M., D'Aì A.,
  Bautz M.W., Bhattacharyya S., Burrows D.N., Cackett E.M., Fabian A.C., Freyberg
  M.J., Haberl F., Kennea J., Nowak M.A., Reis R.C., Strohmayer T.E., Tsujimoto M.,
  2010, ApJ, 724, 1441

\bibitem[\protect\citeauthoryear{Miller, Miller, \& Reynolds}{2011}]{miller11} Miller
  J.~M., Miller M.~C., Reynolds C.~S., 2011, ApJ, 731, L5

\bibitem[Miller et al.(2012)]{miller12} Miller J.M., Pooley G.G., Fabian A.C., et al.,
  2012, ApJ, 757, 11

\bibitem[\protect\citeauthoryear{Mineshige et al.}{1994}]{mineshige94} Mineshige S.,
  Hirano A., Kitamoto S., Yamada T.T., Fukue J., 1994, ApJ, 426, 308

\bibitem[\protect\citeauthoryear{Mirabel \& Rodriguez}{1994}]{I54} Mirabel I.F.,
  Rodriguez L.F., 1994, Nature, 371, 46

\bibitem[\protect\citeauthoryear{Mirabel \& Rodriguez}{2003}]{I49} Mirabel I.F.,
  Rodriguez L.F., 2003, Science, 300, 1119

\bibitem[\protect\citeauthoryear{Mirabel et al.}{2011}]{mirabel11} Mirabel I.F.,
  Dijkstra M., Laurent P., Loeb A., Pritchard J.R., 2011, A\&A, in press
  (arXiv:1102.1891) 

\bibitem[\protect\citeauthoryear{Mitsuda et al.}{1984}]{diskbb_mitsuda84} Mitsuda, K.,
  et al., 1984, PASJ, 36, 741

\bibitem[\protect\citeauthoryear{Monet et al.}{2003}]{usnob1} Monet D.G. et al., 2003,
  AJ, 125, 984

\bibitem[\protect\citeauthoryear{Morley et al.}{2001}]{morley01} Morley J.E., Briggs
  K.R., Pye J.P., Favata F., Micela G., Sciortino S., 2001, MNRAS, 326, 1161

\bibitem[\protect\citeauthoryear{Motch et al.}{1998}]{motch98} Motch C., Guillout
  P., Haberl F., Krautter J., Pakull M.W., Pietsch W., Reinsch K., Voges W., Zickgraf
  F.J., 1998, A\&AS, 132, 341



\bibitem[\protect\citeauthoryear{Narayan \& Yi}{1994}]{narayan94} Narayan R., Yi I.,
  1994, ApJ, 428, 13

\bibitem[\protect\citeauthoryear{Narayan \& McClintock}{2008}]{narayan08} Narayan R.,
  McClintock J.E., 2008, NewAR, 51, 733



\bibitem[\protect\citeauthoryear{Nowak et al.}{2011}]{nowak11} Nowak
  M.~A., Wilms J., Pottschmidt K., Schulz N., Maitra D., Miller J., 2011, arXiv,
  arXiv:1109.6008

\bibitem[\protect\citeauthoryear{Orosz et al.,}{1996}]{I29} Orosz J.A., Bailyn C.D.,
  McClintock J.E., Remillard R.A., 1996, ApJ, 468, 380

\bibitem[\protect\citeauthoryear{Orosz \& Bailyn}{1997}]{I44} Orosz J.A., Bailyn C.D.,
  1997, ApJ, 477, 876

\bibitem[\protect\citeauthoryear{Orosz et al.,}{1998}]{I36} Orosz J.A., Rain J.K.,
  Bailyn C.D., McClintock J.E., Remillard R.A., 1998, ApJ, 499, 375

\bibitem[\protect\citeauthoryear{Orosz et al.}{2001}]{I46} Orosz J.A., Kuulkers E.,
  van der Klis M., McClintock, J.E., Garcia M.R., Callanan P.J., Bailyn C.D., Jain
  R.K., Remillard R.A., 2001, ApJ, 555, 489

\bibitem[\protect\citeauthoryear{Orosz et al.}{2002a}]{I38} Orosz J.A., Groot P.J.,
  van der Klis M., McClintock J.E., Garcia M.R., Zhao P., Jain R.K., Bailyn C.D.,
  Remillard R.A., 2002a, ApJ, 568, 845

\bibitem[\protect\citeauthoryear{Orosz et al.}{2002b}]{I35} Orosz J.A., Polisensky
  E.J., Bailyn C.D., Tourtellotte S.W., McClintock J.E., Remillard R.A., 2002b, AAS,
  201, 1511

\bibitem[\protect\citeauthoryear{Orosz et al.}{2007}]{I212} Orosz J.A., McClintock
  J.E., Narayan R., Bailyn C.D., Hartman J.D., Macri L., Liu J., Pietsch W., Remillard
  R.A., Shporer A., Mazeh T., 2007, Nature, 449, 872

\bibitem[\protect\citeauthoryear{Orosz et al.}{2011a}]{orosz11a} Orosz J.A., Steiner
  J.F., McClintock J.E., Torres M.A.P., Remillard R.A., Bailyn C.D., Miller J.M.,
  2011a, ApJ, 730, 75

\bibitem[\protect\citeauthoryear{Orosz et al.}{2011b}]{orosz11b} Orosz, J.~A.,
  McClintock, J.~E., Aufdenberg, J.~P., Remillard, R.~A., Reid, M.~J., Narayan, R., \&
  Gou, L.\ 2011b, ApJ, 742, 840

\bibitem[{\"O}zel et al.(2010)]{ozel10} {\"O}zel F., Psaltis D., Narayan R.,
  McClintock J.~E., 2010, ApJ, 725, 1918 

\bibitem[\protect\citeauthoryear{Pakull et al.}{2010}]{pakull10} Pakull M.W., Soria
  R., Motch C., 2010, Natur, 466, 209

\bibitem[\protect\citeauthoryear{Poole et al.}{2008}]{uvot_calI} Poole T.S.,
  Breeveld A.A., Page M.J. et al., 2008, MNRAS, 383, 627 

\bibitem[\protect\citeauthoryear{Pottschmidt et al.}{2006}]{pottschmidt06} Pottschmidt
  K., Chernyakova M., Zdziarski A.~A., Lubi{\'n}ski P., Smith D.~M., Bezayiff N.,
  2006, A\&A, 452, 285

\bibitem[\protect\citeauthoryear{Poutanen \& Svensson}{1996}]{ps96} Poutanen J.,
  Svensson R., 1996, ApJ, 470, 249


\bibitem[\protect\citeauthoryear{Prat et al.}{2009}]{prat09} Prat L., et al., 2009,
  A\&A, 494, L21

\bibitem[\protect\citeauthoryear{Predehl 
\& Schmitt}{1995}]{predehl95} Predehl P., Schmitt J.~H.~M.~M., 1995, A\&A,
  293, 889


	

\bibitem[\protect\citeauthoryear{Psaltis}{2008}]{psaltis08} Psaltis D., 2008, LRR, 11,
  9 
\bibitem[\protect\citeauthoryear{Reis, Miller, \& Fabian}{2009}]{reis09} Reis R.~C.,
  Miller J.~M., Fabian A.~C., 2009, MNRAS, 395, L52

\bibitem[\protect\citeauthoryear{Reis, Fabian, \& Miller}{2010}]{reis10} Reis R.~C.,
  Fabian A.~C., Miller J.~M., 2010, MNRAS, 402, 836

\bibitem[\protect\citeauthoryear{Reis et al.}{2011a}]{reis11a} Reis R.C., Miller J.M.,
  Fabian A.C., Cackett E.M., Maitra D., Reynolds C.S., Rupen M., Steeghs D.T.H.,
  Wijnands R., 2011a, MNRAS, 410, 2497

\bibitem[\protect\citeauthoryear{Reis et al.}{2011b}]{reis11b} 
Reis R.~C., Miller J.~M., Reynolds M.~T., Fabian A.~C., Walton D.~J., 2011b, 
arXiv, arXiv:1111.6665

\bibitem[\protect\citeauthoryear{Remillard et al.}{1996}]{I32} Remillard R.A., Orosz
  J.A., McClintock J.E., Bailyn C.D., 1996 ApJ, 459, 226


\bibitem[\protect\citeauthoryear{Revnivtsev et al.}{2011a}]{revnivtsev11a} Revnivtsev
  M., Sazonov S., Forman W., Churazov E., Sunyaev R., 2011, MNRAS, 414, 495

\bibitem[\protect\citeauthoryear{Revnivtsev et al.}{2011b}]{revnivtsev11b} Revnivtsev
  M., Postnov K., Kuranov A., Ritter H., 2011b, A\&A, 526, 94



\bibitem[\protect\citeauthoryear{Reynolds et al.}{2010a}]{reynolds10a} Reynolds
  M.~T., Miller J.~M., Homan J., Miniutti G., 2010a, ApJ, 709, 358

\bibitem[\protect\citeauthoryear{Reynolds \& Miller}{2010b}]{reynolds10b} Reynolds M.T.,
  Miller J.M., 2010b, ApJ, 716, 1431

\bibitem[\protect\citeauthoryear{Reynolds \& Miller}{2011}]{reynolds11} Reynolds,
  M.~T., \& Miller, J.~M.\ 2011, \apjl, 734, L17

\bibitem[\protect\citeauthoryear{Romano et al.}{2006}]{romano06} Romano P. et al.,
  2006, A\&A, 456, 917

\bibitem[\protect\citeauthoryear{Roming et al.}{2005}]{roming05} Roming P.W.A. et al.,
  2005, SSRev, 120, 95

\bibitem[\protect\citeauthoryear{Rupen et al.}{2004}]{atel314} Rupen M.P. et al.,
  2004, ATel \#314

\bibitem[\protect\citeauthoryear{Rupen et al.}{2005}]{atel490} Rupen M.P. et al.,
  2005, ATel \#490

\bibitem[Ross \& Fabian(1993)]{ross93} Ross R.R., Fabian, A.C., 1993, MNRAS, 261, 74

\bibitem[Ross \& Fabian(2007)]{ross07} Ross, R.R., \& Fabian, A.C., 2007, MNRAS, 381,
  1697

\bibitem[\protect\citeauthoryear{Russell et al.}{2006}]{russell06} Russell D.M.,
  Fender R.P., Hynes R.I., Brocksopp C., Homan J., Jonker P.G., Buxton M.M., 2006,
  MNRAS, 371, 1334

\bibitem[\protect\citeauthoryear{Russell et al.}{2010}]{russell10} Russell D.M.,
  Maitra D., Dunn R.J.H., Markoff S., 2010, MNRAS, 405, 1759

\bibitem[\protect\citeauthoryear{Russell et al.}{2011}]{russell11} Russell D.M., 
Miller-Jones J.C.A., Maccarone T.J. et al., 2011, ApJ, 739, L19

\bibitem[\protect\citeauthoryear{Rykoff et al.}{2007}]{rykoff07} Rykoff E.S., Miller
  J.M., Steeghs D., Torres M.A.P., 2007, ApJ, 666, 1129 

\bibitem[\protect\citeauthoryear{Sakano et al.}{2002}]{sakano02} Sakano M., Koyama
  K., Murakami H., Maeda Y., Yamauchi S., 2002, ApJSS, 138, 19

\bibitem[\protect\citeauthoryear{Sala et al.}{2007}]{atel1062} Sala G. et al.,
  2007, ATel \#1062 

\bibitem[Sazonov et al.(2012)]{sazonov12} Sazonov S., Willner S.P., Goulding A.D. et
  al., 2012, ApJ, 757, 181

\bibitem[\protect\citeauthoryear{Schrader et al.}{1993}]{schrader93} Schrader C. et
  al., 1993, A\&A, 276, 373 

\bibitem[\protect\citeauthoryear{Schrader et al.}{1994}]{schrader94} Schrader C. et
  al., 1994, ApJ, 434, 698 

\bibitem[\protect\citeauthoryear{Shahbaz et al.,}{1994}]{I51} Shahbaz T., Ringwald
  F.A., Bunn J.C., 1994, MNRAS, 271, 10

\bibitem[\protect\citeauthoryear{Shahbaz et al.,}{1997}]{I31} Shahbaz T., Naylor T.,
  Charles P.A., 1997, MNRAS, 285, 607

\bibitem[\protect\citeauthoryear{Shahbaz et al.,}{1999}]{I45} Shahbaz T., van der
  Hooft F., Casares J., 1999, MNRAS, 306, 89

\bibitem[\protect\citeauthoryear{Shakura \& Sunyaev}{1973}]{ss73} Shakura N.I.,
  Sunyaev R.A., 1973, A\&A, 24, 337

\bibitem[\protect\citeauthoryear{Shimura \& Takahara}{1995}]{shimura95} Shimura T.,
  Takahara F., 1995, ApJ, 445, 780

\bibitem[\protect\citeauthoryear{Silverman \& Filippenko}{2008}]{I213} Silverman J.M.,
  Filippenko A.V., 2008, ApJ, 678, 17

\bibitem[\protect\citeauthoryear{Steiner et al.}{2009}]{steiner09} Steiner J.F., 
Narayan R., McClintock J.E., Ebisawa K., 2009, PASP, 121, 1279 

\bibitem[\protect\citeauthoryear{Steiner et al.}{2010}]{steiner10} Steiner J.F.,
  McClintock J.E., Remillard R.A., Gou L., Yamada S., Narayan R., 2010, ApJ, 718,
  L117

\bibitem[\protect\citeauthoryear{Sugizaki et al.}{2001}]{sugizaki01} Sugizaki M.,
  Mitsuda K., Kaneda H., Matsuzaki K., Yamauchi S., Koyama K., 2001, ApJSS, 134, 77 


\bibitem[\protect\citeauthoryear{Tanaka \& Shibazaki}{1996}]{tanakashibazaki96} Tanaka
  Y., Shibazaki N., 1996, ARA\&A, 34, 607

\bibitem[\protect\citeauthoryear{Thorstensen}{1987}]{thorstensen87} Thorstensen J.~R.,
  1987, ApJ, 312, 739

\bibitem[\protect\citeauthoryear{Titarchuk}{1994}]{titarchuk94}
  Titarchuk L., 1994, ApJ, 434, 570

\bibitem[\protect\citeauthoryear{Tomsick et al.}{2005}]{tomsick05} Tomsick J.A.,
  Corbel S., Goldwurm A., Kaaret P., 2005, ApJ, 630, 413

\bibitem[\protect\citeauthoryear{Tomsick et al.}{2008}]{tomsick08} Tomsick J.~A., et al., 2008, ApJ, 680, 593

\bibitem[\protect\citeauthoryear{Tomsick et al.}{2009}]{tomsick09} Tomsick J.A.,
  Yamaoka K., Corbel S., Kaaret P., Kalemci E., Migliari S., 2009, ApJ, 707, 87

\bibitem[\protect\citeauthoryear{Torres et al.}{2007}]{atel1072} Torres M.A.P. et al.,
  2007, ATel \#1072 


\bibitem[\protect\citeauthoryear{Torres et al.}{2008}]{atel1720} Torres M.A.P. et al.,
  2008, ATel \#1720 


\bibitem[\protect\citeauthoryear{van Paradijs \& McClintock}{1994}]{vanparadijs94} van
  Paradijs \& McClintock, 1994, A\&A, 290, 133 

\bibitem[\protect\citeauthoryear{van der Klis}{2006}]{vanderklis06} van der Klis M.,
  2006, Compact stellar X-ray sources, 39

\bibitem[\protect\citeauthoryear{Vierdayanti et al.}{2010}]{vierdayanti10} Vierdayanti
  K., Mineshige S., Ueda Y., 2010, PASJ, 62, 239

\bibitem[\protect\citeauthoryear{Voges et al.}{1999}]{voges99} Voges W.,
  Aschenbach B., Boller T. et al., 1999, A\&A, 349, 389


\bibitem[Walter et al.(2007)]{walter07} Walter, R., Lubi{\'n}ski, P., Paltani, S., et
  al.\ 2007, \aap, 461, L17

\bibitem[Walton et al.(2012)]{walton12} Walton D.J., Reis R.C., Cackett E.M., Fabian
  A.C., Miller J.M., 2012, MNRAS, 422, 2510



\bibitem[\protect\citeauthoryear{Wilkinson \& Uttley}{2009}]{wilkinson09} Wilkinson T., Uttley P., 2009, MNRAS, 397, 666

\bibitem[Wilms et al.(2000)]{wilms00} Wilms J., Allen A., McCray R., 2000, ApJ, 542, 914

\bibitem[\protect\citeauthoryear{Woods et al.}{1996}]{woods96} Woods D.T., Klein R.I.,
  Castor J.I., McKee C.F., Bell J.B., 1996, ApJ., 461, 767


\bibitem[\protect\citeauthoryear{Zhang et al.}{1997}]{zhang97} Zhang S.N., Cui W.,
  Chen W., 1997, ApJ, 482, 155

\bibitem[\protect\citeauthoryear{Zurita et al.}{2002}]{I37} Zurita C.,
  Sanchez-Fernandez C., Casares J., Charles P.A., Abbott T.M., Hakala P.,
  Rodríguez-Gil P., Bernabei S., Piccioni A., Guarnieri A., Bartolini C., Masetti N.,
  Shahbaz T., Castro-Tirado A., Henden A., 2002, MNRAS, 334, 999

\end{thebibliography}
\end{document}